\documentclass{aastex62}
\usepackage{color}
\usepackage{natbib}
\usepackage{subfloat}

\pdfoutput=1

\newcommand\ergcms{erg/cm$^2$/s}
\usepackage{mathtools}

\accepted{by ApJ, September 5, 2019}

\shorttitle{Ro-Vibrational H$_2$ Emission from the Disks of Young Stars}
\shortauthors{Beck \& Bary}

\begin{document}

\title{A Search for Spatially Resolved Infrared Ro-Vibrational Molecular Hydrogen Emission from the Disks of Young Stars}

\correspondingauthor{Tracy L. Beck}

\email{tbeck@stsci.edu, bary@colgate.edu}

\author{Tracy L. Beck}
\affiliation{The Space Telescope Science Institute, 3700 San Martin Dr. Baltimore, MD 21218, USA}

\author{Jeffrey S. Bary}
\affiliation{Colgate University, Department of Physics \& Astronomy, 13 Oak Drive, Hamilton, NY 13346, USA \\}



\begin{abstract}

We present results from a survey searching for spatially resolved near-infrared line emission from molecular hydrogen gas in the circumstellar environments of nine young stars: AA~Tau, AB~Aur, DoAr~21, GG~Tau, GM~Aur, LkCa~15, LkH$\alpha$~264, UY~Aur, and V773~Tau. Prior high-resolution spectra of these stars showed the presence of ro-vibrational H$_2$ line emission at 2.12~$\mu$m with characteristics more typical of gas located in proto-planetary disks rather than outflows. In this study, we spatially resolve the H$_2$ emission in the eight stars where it is detected. LkCa~15 is the only target that exhibits no appreciable H$_2$ despite a prior detection. We find an anti-correlation between H$_2$ and X-ray luminosities, likely indicating that the X-ray ionization process is not the dominant H$_2$ excitation mechanism in these systems. AA~Tau, UY~Aur, and V773~Tau show discrete knots of H$_2$, as typically associated with shocks in outflowing gas. UY~Aur and V773~Tau exhibit spatially resolved velocity structures, while the other systems have spectrally unresolved emission consistent with systemic velocities. V773~Tau exhibits a complex line morphology indicating the presence of multiple excitation mechanisms, including red and blue-shifted bipolar knots of shock-excited outflowing gas. AB~Aur, GM~Aur, and LkH$\alpha$~264 have centralized, yet spatially resolved H$_2$ emission consistent with a disk origin. The H$_2$ images of AB~Aur reveal spiral structures within the disk, matching those observed in ALMA CO maps. This survey reveals new insights into the structure and excitation of warm gas in the circumstellar environments of these young stars.

  
\end{abstract}



\section{Introduction} \label{sec:intro}

Emission from the {\it v}~=~1-0~S(1) ro-vibrational transition of molecular hydrogen at 2.12~$\mu$m was discovered in the spectra of a range of astronomical objects during the rapid expansion of infrared (IR) astronomy capabilities in the 1970s.  In the context of star formation science, this emission was first detected from shock-excited outflows in the Orion nebula \citep{gaut76}. Since this time, molecular hydrogen emission from electronic transitions in the ultraviolet (UV), ro-vibrational transitions in the near-IR, and pure rotational transitions in the mid-IR have been studied in the spectra of young stars and their environs \citep{ardi02, sauc03, walt03, herc04, herc06, davi01a, davi02a, bary03, taka04, duch05, carm08a, carm08b, beck08, thi99, thi01, rich02, sher03, sako05}.  More recently, the detection of {\it v}=1-0 S(1) H$_2$ emission in Classical T Tauri Stars (CTTSs), embedded proto-stars, and Herbig-Haro (HH) energy sources is commonplace.

Molecular hydrogen is the primary constituent of cool gas in the circumstellar disks of young stars.  As the sites of planet formation, a detailed understanding of the evolution of the gas, namely the H$_2$, will lead to a clearer picture of how circumstellar disks evolve into planetary systems.  Due to the lack of a dipole moment, the hydrogen molecule emits radiation comparably weakly from quadrupolar transitions.  In cool circumstellar environments, the H$_2$ molecules require a stimulation mechanism to increase the overall flux to detectable levels. Hence, many studies of molecular hydrogen emission in the inner $\sim$200~AU regions of CTTSs are for systems that are known to drive HH outflows, which produce shock-excited H$_2$ emission \citep{sauc03,taka04,kasp02,duch05,beck08}.  At UV wavelengths, the H$_2$ emission is typically stimulated in low-density gas by non-thermal excitation of the H$_2$ Lyman and Werner electronic bands by strong Ly~$\alpha$ emission from the stellar chromosphere, and in the case of an accreting TTS, from accretion hot spots \citep{walt03,herc04,herc06}. Pure rotational transitions in the mid-IR arise from thermally excited dense H$_2$ material closer to the disk mid-plane \citep{bitn08}.  Understanding the excitation mechanism of the range of molecular hydrogen features is especially of interest in the inner $\sim$50~AU where emission from H$_2$ in the planet-forming regions of circumstellar disks is believed arise \citep{malo96,bary03,nomu05,nomu07}.  

In addition to being excited by shocks, the ro-vibrational transitions of H$_2$ observed in the near-IR spectra of young stars can be excited directly and indirectly by the absorption of high-energy photons (e.g., UV fluorescence, stellar UV or X-ray ionization and heating).  Shock-excited H$_2$ typically shows thermal level populations (T$_{\rm{ex}}\sim$1500-2200~K) in ro-vibrational features detected in K-band spectra (2.0-2.4~$\mu$m), sometimes with high and low H$_2$ velocity features arising from different environments in the outflow or wind \citep{burt89a,eisl00a,eisl00b,davi01a,davi02a,taka04,taka07,beck08}. In the case of UV fluorescence, molecular hydrogen is pumped into an electronically-excited state usually by the stellar Ly$\alpha$ flux \citep{ardi02,walt03,herc04,herc06}, and lines in the near IR arise from the corresponding cascade through the vibration-rotation transitions \citep{sauc03}.  As a result of the photo-excitation, high-$v$  and electronically-excited states are populated quite differently from thermal excitation in a medium density gas \citep{blac87,hase87,ardi02}.  However, at gas densities higher than $\sim$a few times 10$^4$~cm$^{-3}$, the UV stimulated H$_2$ will quickly thermalize.  Stellar UV and X-ray fluxes also heat the ambient circumstellar gas and dust.  The H$_2$ molecules are excited into thermal equilibrium and level populations of low-$v$ states can be similar to those observed in shock-excited regions \citep{malo96,tine97,nomu05,nomu07}.   However, H$_2$ emission excited due to heating by the high energy photons generated near the central star is not expected to extend beyond $\sim$30-50~AU from the star, such emission should be centered on the systemic radial velocity with narrower line widths than emission excited by shocks in the outflows. 

In the UV, {\it HST} STIS and {\it FUSE} observations of TW~Hya, a nearby TTS (D$\sim$60.1~pc), were used to constrain the location of the emitting H$_2$ gas to within 2~AU of the central star.  Ly$\alpha$ photons produced by accretion activity in this source provides a natural stimulation mechanism for the H$_2$ molecules and the many ro-vibrational electronic transitions that align energetically with the Ly$\alpha$ transition.  \citet{herc04}, using relative line strengths and reconstructing the Ly$\alpha$ profile, constrained both temperature and column density of the emitting gas to $T$~=~2500~K and $\log N$(H$_2$)~=~18.5.  More recently, the DAO (Disks Accretion and Outflows) of Tau guest program on {\it HST} observed 33 additional TTSs with the {\it HST} Cosmic Origins Spectrograph (COS), successfully detecting H$_2$ emission from 27 TTSs or 100\% of the accreting sources in their sample.  The COS observations spectrally resolved the emission features providing kinematic constraint on the location of the H$_2$ gas.  While the gas remained constrained to the inner regions of the disks (0.1~$\le$~R$_{\rm{H_2}}$~$\le$~10~AU) across the entire sample, \citet{fran12} find that the average radial position of the H$_2$ emission increases as the system evolves.  For systems with near-IR and UV H$_2$ emission, the UV H$_2$ features are significantly broader locating the gas at smaller disk radii.

The development of high-resolution near-IR spectrometers (R$\sim$35,000-60,000) on moderate aperture telescopes allowed for the detection of narrow 2.12~$\mu$m emission lines with line centers within a few km/s of the systemic velocities of a handful of TTSs \citep{wein00,bary02,bary03,bary08,itoh03,rams07,carm08a}.  On average, these lines were comparatively weaker than the more obvious shock-excited lines from outflowing gas and represented H$_2$ gas masses in the range of 10$^{-10}$ to 10$^{-12}$~M$_\odot$.  The narrow line profiles were interpreted as evidence that the gas resides in the disks, and may be in Keplerian motion about the stars at distances on the order of 10~AU.  Upper limits to the flux of the $v$~=2-1~S(1) 2.24~$\mu$m line for a couple of sources suggested that UV fluorescence was not the most likely stimulation mechanism unless the gas was sufficiently dense to thermalize.  Most authors concluded that the gas was likely confined to the upper atmospheres of the disk at intermediate orbital radii (R$_{\rm{H_2}}$~=~10-30~AU).  In addition to the near-IR detections, H$_2$ emission from purely rotational transitions was detected from one TTS and a couple of Herbig AeBe stars, using high-resolution mid-infrared spectrometers \citep{mart07,bitn07,bitn08}.  The mid-IR lines were also weak, narrow, and centered within a few km/s of the systemic stellar velocities.  Line profile fitting for AB~Aur suggested that the rotational H$_2$ emission originates at a distance of 18~AU \citep{bitn07}.  Likewise, one- and two-temperature local thermodynamic equilibrium (LTE) models of the H$_2$ line fluxes pointed towards extremely small masses for the emitting gas and temperatures that required an additional source of heating for gas located at intermediate orbital radii.

Despite the high-spectral resolution, reasonable kinematic arguments, and sophisticated models of the line emission, spectroscopic detections of H$_2$ gas in the circumstellar environments of protoplanetary disks provide limited methods for constraining locations of the emitting gas. Adaptive optics-fed integral field units, such as NIFS on the Gemini North 8-meter telescope, provide an opportunity to spatially resolve H$_2$ emission structures with angular sizes of less than 0.\arcsec1 with exceptional contrast \citep{mcgr03}.  \citet{beck08} presented the first IFU survey of K-band H$_2$ emission from the near environments of six TTSs, all known to drive Herbig-Haro outflows. Using NIFS, the study found spatially-extended shock-excited H$_2$ emission arising from the jets and winds associated with these stars and little evidence for quiescent emission originating in the circumstellar disks. \cite{gust08} modeled a portion of the near-IR IFU data on the T Tauri system as a disk encircling T Tau North, and \cite{beck12} revealed the structure of emitting gas in the environment of GG Tau A.  These studies suggest that warm H$_2$ emission from inner disks in CTTS systems may be well characterized using IFU techniques.  Following on these studies and focusing specifically on sources that may possess disk-like H$_2$ emission, we acquired AO-fed NIFS images of nine additional TTSs with high-resolution spectroscopic detections of H$_2$:  AA Tau, AB Aur, DoAr 21, GG Tau A, GM Aur, LkCa 15, LkH$\alpha$ 264, UY Aur and V773 Tau.  Here we present the results from our study and place them in context to constrain the nature and evolution of gas in protoplanetary disks through their planet-building phase.

\subsection{The Excitation of H$_2$ in CTTS Environments}

The bulk gas constituent in circumstellar disks around young sun-like stars is in the form of molecular hydrogen.   Table~1 presents the excitation and origins of the UV, near-IR and mid-IR molecular hydrogen features from the environments of CTTSs. This table serves as a summary of key H$_2$ features that have been detected or interpreted in CTTSs; it is not meant to be a complete listing of all theorized excitation mechanisms for H$_2$ from young stars.   Figure~1a shows a schematic cross-cut of an edge on disk system highlighting the location of UV, near-IR and Mid-IR H$_2$ emission from the inner $\sim$100~AU of a young single star system.  The structure in this figure is based on the density, temperature and chemical models of circumstellar disks developed in the past two decades \cite{nomu05, nomu07,dull07, woit09,wals10,woit18}.  The UV H$_2$ traces hot low density gas predominantly excited by Ly$\alpha$ pumping in the inner 10AU disk regions and the inner outflows.  The UV H$_2$ can also arise from non-thermal Ly$\alpha$ pumping in regions of the collimated jets and inner winds where the H$_2$ is not dissociated by shocks or high temperatures.  In regions with significant UV H$_2$ emission the gas temperatures are typically in the 2000-3000K range, but non-thermal pumping of the H$_2$ by Ly$\alpha$ is the dominant excitation mechanism over collisional excitation.  Near-IR H$_2$ measures denser warm gas in LTE the upper disk layers excited by central stellar flux, and in the shock excited inner outflows.  The mid-IR H$_2$ traces cooler, denser gas closer to the disk mid-plane within $\sim$30~AU of the central star.  Figure~1b is a representation of the H$_2$ emission from an equal mass 100AU separation binary star.  The disk emissions shown in Figure~1a are also present here, though each circumstellar disk is dynamically truncated at $\sim$30~AU \citep{arty94}.  The system is encompassed by a circumbinary ring with 300~AU radius, and the inner circumstellar disks are fed by accretion streamers that funnel material into the inner regions \citep{arty94,arty96}.  Shocked ro-vibrational H$_2$ emission from material in accretion infall is also postulated as a viable excitation mechanism in dynamically complex young star multiples.

At UV wavelengths, the electronic transitions of H$_2$ are fluorescently pumped by the strong L$\alpha$ emission from young stars and trace very warm ($>$2000K) low density ($<$10$^3$) gas in the central disk.  In \cite{fran12}, UV H$_2$ emission was measured in 100\% of the actively accreting stars surveyed with the HST COS.  They found a correlation between H$_2$ flux and stellar Ly$\alpha$ and accretion generated CIV luminosity, as well as a decreasing trend of UV H$_2$ emission with age from $\sim$1 to 10 Myr.  Analysis of the UV H$_2$ emission profiles reveals an origin within the inner $\sim$3~AU inner disk from interaction between the strong Ly$\alpha$ emission with the molecular disk surface.  Only three stars overlap between our present study and the \cite{fran12} sample:  AA Tau, GM Aur and LkCa 15.  Conversely, the pure rotational H$_2$ transitions in the mid-infrared trace cooler, dense molecular hydrogen at deeper depths near the circumstellar disk mid-plane.  The {\it v}=0-0 S(2), S(1) and S(0) lines at 12.28, 17.03 and 28.0$\mu$m are used to quantify the molecular hydrogen component in dense extended disks.  The {\it Spitzer Space Telescope} IRS instrument with the R$\sim$600 high resolution mode was not quite sensitive enough to detect these very weak quadrupolar rotational H$_2$ transitions above bright mid-IR continuum emission in most of the systems studied in the CTTS surveys \citep{bald11}.  From our present survey sample, only AA Tau, UY Aur and AB Aur have measured rotational H$_2$ transitions from Spitzer or high spectral resolution ground-based measurements\citep{carr11, bitn07}.  In AB Aur, the measured mid-IR rotational transitions indicate a gas temperature of 670K and an H$_2$ emission location within 18~AU in the circumstellar disk \citep{bitn07}.  \cite{woit18} found that strong rotational H$_2$ emission requires very large column densities and a large temperature contrast between the gas and dust at deeper layers inside disks, which they do not see in their models.

The near-IR ro-vibrational H$_2$ transitions provide an important cooling mechanism in the inner $<$3~AU terrestrial regions of circumstellar disks \citep{woit09}.   While the ro-vibrational H$_2$ emission has been studied in the spectra of young stars for decades, clear identification of the spatially resolved H$_2$ arising from circumstellar disks has proven difficult.  In many of the youngest systems, these near-IR H$_2$ features are stronger or more readily detected from shock-excited emission associated with or encompassing the outflows from the stars \citep{beck08}.  Additionally, disk excited ro-vibrational H$_2$ emission arises predominantly from the low to moderate density heated disk surface layer (10$^3$ - 10$^5$ molecules cm$^{-3}$; Figure~1a), and it is not expected to extend beyond $\sim$ 50~AU from the central star \citep{nomu05}.   

\begin{deluxetable}{cccc|ccc}
\tabletypesize{\scriptsize}
\tablecaption{UV, Near-IR and Mid-IR H$_2$ Emissions in CTTS Environments}
\tablewidth{0pt}
\tablehead{
\colhead{H$_2$ Emission} & \colhead{Wavelength}   &  \colhead{Gas Density }  &  \colhead{Temperature}  & \colhead {Excitation}  & \colhead {Emission} & \colhead {References} \\  
\colhead{Diagnostic} & \colhead{Range}   &  \colhead{ (cm$^{-3}$) }  &  \colhead{(K)}  & \colhead {Mechanism}  & \colhead {Location} & \colhead {} }
\startdata
\hline
\hline
UV Electronic  & 1100-1600A & $<$10$^3$ & 2000-3000  & (1) Non-thermal pumping & R$<\sim$10AU, & (1) \\
Dipole  &  &  &   & by stellar Ly$\alpha$ & Disk Surface &  \\
Transitions &  &  &  & (2) Non-thermal pumping  &  Extended Outflows & (2) \\
 &  &  &  & by Ly$\alpha$ &   and Winds &  \\
\hline
\hline
IR Ro-Vibrational  & 1-6$\mu$m &  10$^3$-10$^6$ & 1500-3000 & (1) Collisional excitation in  & Extended outflows and & (3) \\
Quadrupole &  &  &  & shock heated material & winds to $>>$100AU &  \\
Transitions &  &  & $<$2000 & (2) Collisional excitation through & R$<\sim$50AU, & (4)  \\
  &  &  &  &  stellar X-Ray ionization+heating &  disk surface &  \\
  &  &  &  & (3) Collisional excitation through & R$<\sim$50AU, &  (5) \\
  &  &  &  &   stellar UV heating &  disk surface &  \\
  &  &  &  & (4) IR cascade from non-thermal & R$<\sim$30AU, & (6) \\
  &  &  &  &  pumping by Ly$\alpha$ &  disk surface &  \\
  &  &  &  & (5) Shock excitation from & Regions that & (7) \\
  &  &  &  &  mass accretion infall in  &  should be &  \\
  &  &  &  &  multiple stellar systems  &  dynamically cleared &  \\
 \hline
\hline
Mid-IR Rotational  & 6-28$\mu$m & $>\sim$10$^6$ & $<$1000K & (1) Collisional Excitation  & R$<\sim$30AU,  & (9) \\
Quadrupole &  &  &  & through Stellar heating & Closer to the &  \\
Transitions &  &  &  & & Disk Mid-Plane &  \\
\hline
\enddata
\tablecomments{References - (1): \cite{blac87,ardi02,herc02,herc04,fran12}, (2): \cite{ardi02,sauc03,walt03,fran12},  (3): \cite{burt89b,eisl00a,eisl00b,taka04,taka07,beck08} (4): \cite{tine97,malo96,bary03,nomu05,nomu07,bary08}, (5): \cite{bary03,nomu05,nomu07}, (6): \cite{blac87,bary03,nomu05,nomu07,bary08} (7): \cite{beck08,dutr16} (8): this study; (9) \cite{thi99,thi01,sher03,bitn07,bitn08} }
\end{deluxetable}

\begin{figure}
\plottwo{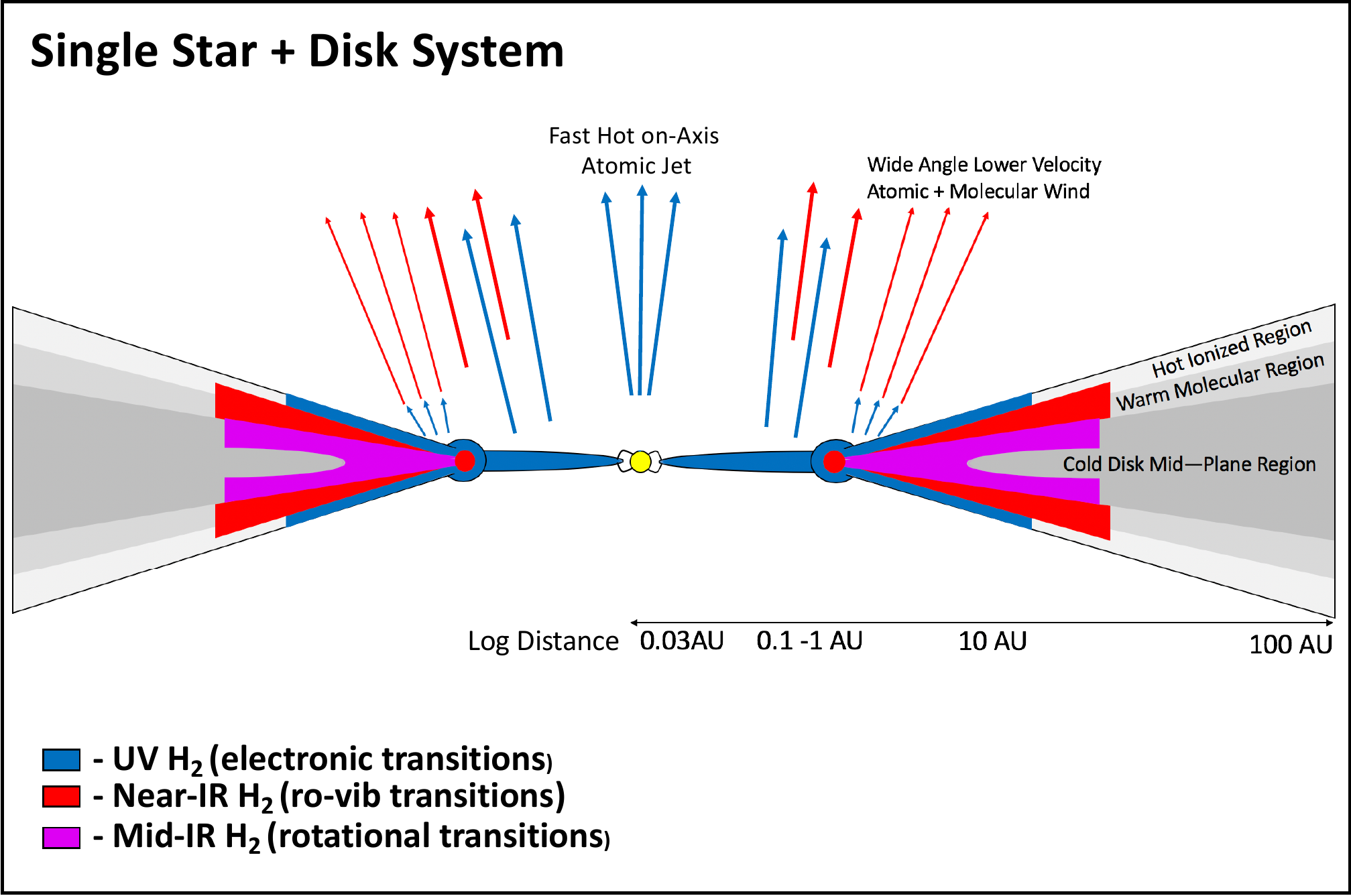}{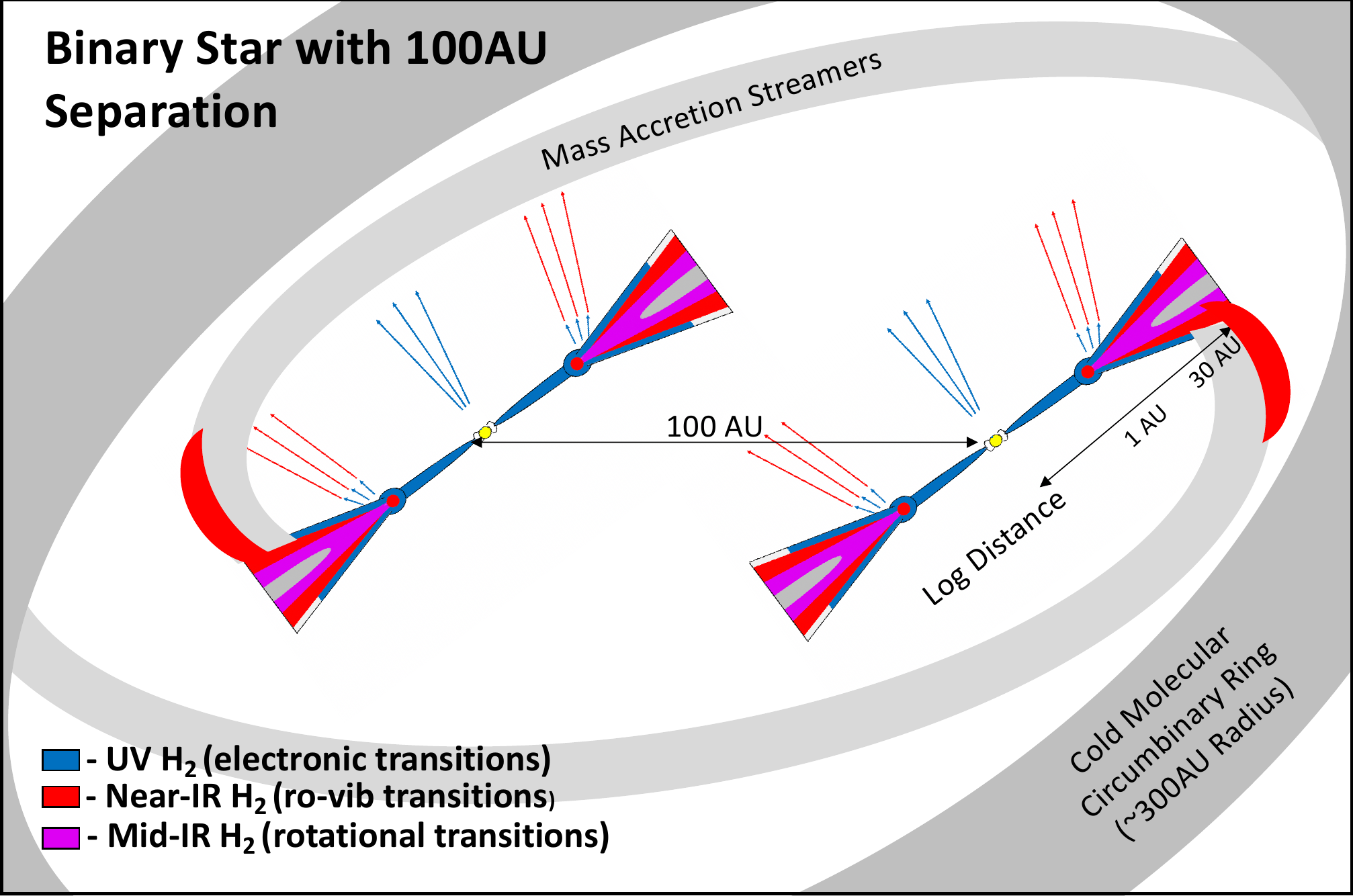}
\caption{Cartoon depictions of the possible locations of UV (blue), near-IR (red) and mid-IR (violet) molecular hydrogen emitting gas from the disks and outflows in a single CTTS system (a), and from an equal mass 100AU separation CTTS binary system (b).  The single star diagram presents and labels material in the disks and outflows and the approximate locations on a logarithmic distance scale from the star.  All H$_2$ emissions shown in (a) are also seen in the multiple star system (b).  The disks are truncated at 30~AU because of the binary, and additional material in the circumbinary ring and mass accretion streamer distributions is identified \citep{arty94,arty96}. \label{fig:h2singlemultiple}}
\end{figure}

\section{Observations and Data Reduction} \label{sec:obs}

\subsection{High Resolution Spectroscopy for H$_2$ Line Detection}

On UT 2001 November 15 and 16 high-resolution spectroscopy was acquired for using the CGS4 infrared spectrograph in echelle mode at the United Kingdom Infrared Telescope (UKIRT) on Mauna Kea, Hawaii (Table~1).  Using that 1-pixel wide slit ($\sim$~0.\arcsec61), long-slit observations were made using a standard ABBA nod pattern for the efficient removal of sky emission features, dark current, and sky background.  The velocity sampling is 3.7~km/s per pixel in the wavelength calibrated data, yielding an instrumental two-pixel resolving power of R~=~40,000.  The data were reduced using the standard packages in STARLINK software package CGS4DR and Figaro.  These data were significantly affected by fringing, most of which was successfully removed during the reduction process which necessitated a top-hat data smoothing to an effective resolving power of R~$=$~18000 and a two pixel velocity resolution of $v$~$=$~35~km/s.  

On UT 2010 September 25 a follow-up high-resolution spectrum of one target, GM~Aur, was obtained with the CSHELL spectrograph at the Infrared Telescope Facility on Mauna Kea, Hawaii (Table~1).  Using the five-pixel wide, 1.\arcsec0 slit, we observed the 2.12~$\mu$m emission from GM~Aur in a total of 42 minutes on source.  These data were reduced with a CSHELL specific IDL package called CSHELLEXT (C. Bender, private communication).  Alternatively, the data were also reduced with standard IRAF routines. A comparison of the two procedures found them to be similar within the uncertainties of the processed data.  The instrumental spectral sampling gave a two pixel resolving power of R~=~21,000 for this observation, or $v$~$=$~28~km/s.  The spectra presented here are the product of the IDL package.  

All of the H$_2$ emission line velocities are corrected for heliocentric motion at the time of the observation and placed into the stellar rest frame by removing the radial velocity of the star \citep{nguy12,hart86}.  All velocities presented in this project are v$_{Helio}$ rather than v$_{LSR}$.

\subsection{Integral Field Spectroscopy to Characterize Extended H$_2$}

Integral Field spectroscopic observations of nine CTTSs were obtained using the Near IR Integral Field Spectrograph (NIFS) at the Gemini North Frederick C. Gillette Telescope on Mauna Kea, Hawaii (Table~1).  NIFS is an image slicing IFU fed by Gemini's Near IR adaptive optics system, Altair, that is used to obtain integral field spectroscopy at spatial resolutions of $\le$0.$''$1 with a two pixel spectral resolving power of R$\sim$5300 at 2.2~$\mu$m \citep{mcgr03}.  The NIFS field is 3$''\times3''$ in size and the individual IFU pixels are 0.$''$1$\times$0.$''$04 on the sky.  Data were obtained at the standard K-band wavelength setting for a spectral range of 2.010-2.455~$\mu$m.  Each of the stars observed for this program was bright enough to serve as their own wavefront reference stars for the adaptive optics, and observations were typically acquired in natural seeing better than $\sim$0.$''$85 for excellent AO correction. 

The NIFS data for this project was acquired during three observing terms at Gemini under program IDs GN-2007B-Q-40, GN-2009A-Q-100, and GN-2009B-Q-40.  All nine sources observed possess detectable levels of $\it v$~$=$~1-0~S(1) ro-vibrational H$_2$ line emission in long slit high-resolution data such as those presented here and in our previous work (AA~Tau, GG~Tau~A, GM~Aur, UY~Aur, DoAr-21, GG~Tau~A, LkCa~15) or reported in the literature (LkH$\alpha$~264). One source, AB~Aur, had a published detection of mid-infrared rotational emission \citep{bitn08}.  The IFU data obtained is summarized in Table~1.  Each star was observed with many short exposures to avoid saturation on the bright central source.  In columns 3 and 4 of Table~2 the exposure times and number of individual exposures are listed.  Column 5 presents the final on-source median exposure time for each star.  All NIFS data on the young stars was acquired by offsetting to a blank sky field to measure background flux brightness every $\sim$third image, or approximately every 2-3 minutes.  For each observation, a standard set of calibrations was acquired using the Gemini facility calibration unit, GCAL.  The data for most stars was obtained over several nights, and associated supporting calibration data was acquired for each night.  Column 6 of Table~1 presents the adopted K-band magnitude for each star. In most cases, the K-band magnitudes are adopted from data acquired with WHIRC and column 7 presents the source for these measurements.  The supporting photometry was acquired in September 2009 with the 3.5-meter WIYN telescope and the WIYN High-Resolution Infrared Camera (WHIRC) using the K$'$ filter and the UKIRT faint standard FS~18 for flux calibration.    

\begin{deluxetable}{lcccccc}
\tablecaption{Log of Observations \label{tbl-2}}
\tablehead{
\colhead{Star} &  \colhead{Obs. Date}   &  \colhead {Exp Time(s)} & \colhead{\# Exp} & \colhead{Total On-Source } & \colhead {K-band} &  \colhead{Magnitude }  \\
\colhead{Name} & \colhead{} & \colhead{/ \# coadds}   &  \colhead {}  &  \colhead{Exp. Time (s)} & \colhead {Magnitude} &  \colhead{Source } } 
\startdata
  &    &  & UKIRT CGS4  &  & & \\
 \hline
 AA Tau &  2001 Nov 15  & 30.0  & 16 & 480  & --- & --- \\
GG Tau A &  2001 Nov 15  & 15.0  & 16 & 240  & --- & ---    \\
 GM Aur &  2001 Nov 16 & 30.0  & 8 & 240   & --- & ---  \\
UY Aur & 2001 Nov 16 & 30.0 & 9 &  270    & --- & --- \\
\hline
 &    & & IRTF CSHELL   &   \\
 \hline
  GM Aur &  2010 Sep 25 & 180.0 & 14 &  2520.0 &  --- & ---   \\
 \hline
 &    &  & Gemini NIFS  &   \\
 \hline
 AA Tau &  2007 Oct 14, Nov 11,  & 40.0 / 1 & 31 & 1240  & 7.9 & WIYN+WHIRC  \\
 &    2008 Jan 4,5  &  &  &   &  &   \\
AB Aur$^1$ &  2009 Nov 09, 19 & 40.0 /1  & 41 & 1640  & 4.23 & 2MASS \\
   &  2009 Dec 13, 26 &   &  &  &  &  \\
DoAr 21 &  2009 Sep 4,5  & 40.0 /1 & 76 & 3040  & 6.23 & 2MASS   \\
GG Tau A &  2009 Aug 14, Dec 01  & 40.0 / 1 & 78 & 3120 & 7.7$^2$ & WIYN+WHIRC  \\
GM Aur &  2009 Oct 18, 20, 24  & 40.0 / 1 & 123 & 4920 & 8.5 & WIYN+WHIRC   \\
LkCa 15 & 2007 Nov 12, 15, 24  & 40.0 /1 & 54 &  2160  & 8.2 & 2MASS   \\
LkH$\alpha$ 264 & 2009 Aug 10,11  & 40.0 / 1  & 105 & 4200  & 9.05 & WIYN+WHIRC \\
UY Aur & 2009 Aug 22  & 40.0 / 1 & 104 &  4160 & 7.6$^2$ & WIYN+WHIRC  \\
V773 Tau & 2009 Aug 11,13,22  & 20.0 / 2 & 119 & 4760  & 6.7$^2$ & WIYN+WHIRC   \\
 \enddata
 \tablenotetext{1}{AB Aur was observed using the 0.$\arcsec$2 diameter occulting disk in the NIFS to attenuate the continuum flux from the bright star.}
\tablenotetext{2}{The presented K-band magnitude values for GG Tau A, UY Aur and V773 Tau include flux contributions from all stars in these multiple systems.}
\end{deluxetable}

The raw NIFS IFU frames were reduced using the NIFS tasks in the Gemini IRAF package\footnote{Information on the Gemini data reduction package and pipeline results is available at http://www.gemini.edu/sciops/data-and-results}.  The basic reduction steps for processing NIFS raw data into flat fielded, sky subtracted, rectified pixels in a 3-D datacube is described in detail in \citet{beck08}. We refer to this manuscript for description of the main processing steps; {\bf nfreduce}, {\bf nffixbad}, {\bf nffitcoords} and {\bf nftransform}, which execute the basic NIFS IFU reductions.  For each of the science targets, we observed an A0 spectral type star with NIFS for removal of telluric absorption features.  These data were also processed using the above tasks in the Gemini IRAF package.  The A0 type stars were chosen based on their spectral types defined in the Hipparchos catalog and because they provided a good match to the airmass of the science target.  A pseudo long-slit spectrum of each calibration star was extracted from the data using a 1.$''$0 radius circular aperture using the IRAF task {\bf nfextract}.  The spectral continuum shape and atomic H~{\scshape i} absorption features were removed using a task {\bf nffixa0}, which was developed for this purpose.  The science datacubes of the YSOs were divided by the corrected spectra of the A0 calibrators (in the wavelength dimension) using the task {\bf nftelluric}.  

After the data were corrected for telluric absorption, they were built into individual datacubes using the {\bf nifcube} task with an 0.$''$04 square pixel spatial sampling.  The data for each source was collected using a small dither pattern to allow for the effective removal of hot/cold pixels, to even out the pixel-to-pixel sensitivity variations, and to improve the overall pixel sampling in the spatial dimensions. The resulting data cubes were registered and coadded using a 3-D shift-and-add routine, which determines the stellar central position in each data cube, shifts the spatial axes to a common, central location, and then median combines all the spectral data in $\lambda$-dimension corresponding to each spaxel. For optimal signal-to-noise on faint extended H$_2$ emission, individual NIFS data frames that were observed with instantaneous measured natural seeing of worse than 0.''85 were excluded from the final coadded datacubes.  As a result, the processed data had better spatial resolution ($\sim$0."1) and rounder PSF shapes than if all of the raw NIFS data had been included.  The final data products demonstrate this observing strategy and data reduction method to be particularly powerful, providing high contrast images that permit detection of low-level line emission in the environments of bright young stars.


After the cubes were combined into a single 3-D product, they were flux calibrated as a last step in the reduction.  This was accomplished by extracting a 1-D spectrum over the full spatial field, convolving this 1-D spectrum with a K-band filter function (filter function depends on the source of photometry, see column 7 in Table~2), and deriving and applying a relative flux calibration scaling using the adopted K-band magnitude (column 6 of Table~2) and the instrument flux sensitivity function.  Whenever possible, we used our own photometry taken nearby in time to calibrate the fluxes of these young stars, since YSOs are known to be variable in infrared flux. Typical uncertainties in our observed K-band magnitudes are at the 0.1 mag level.  For the sources that we were not able to acquire nearly simultaneous photometry for, we used magnitudes from the Two Micron All Sky Survey (2MASS) \citep{2mass}.  We conservatively estimate that the uncertainties in the flux calibrations to be $\sim$5-10\% when using the WHIRC photometry and $\sim$15\% for sources where we relied on 2MASS magnitudes -- AB~Aur, GM~Aur and DoAr~21.
 
AB~Aur was observed with the NIFS 0.$"2$ occulting disk in the entrance pupil wheel to block the flux from the bright central star in this system.  The continuum emission from AB~Aur would have saturated in the minimum available exposure time if the occulting disk was not used.  The occulting disk is a dark obscuring spot on a transmissive substrate. The light that passes through the substrate creates a fringing-like modulation that imprints a sinusoidal pattern on the continuum flux data at the $\sim$3-4\% level.  This sinusoidal modulation of the stellar flux can be mitigated by acquiring an occulting disk flat, which can lessen this continuum structure to the $\sim$0.5\% level.  It is also important to note that the occulting disk is slightly transmissive, attenuating the central flux by a factor of 10$^{4}$.  The central flux was recovered by correcting the AB~Aur data using a flat-field exposure taken with the occulting disk in the beam simultaneously with the observed data.  Given the attenuation and fringing associated with the occulting disk, the value of central flux measured for AB~Aur is very uncertain.  The approximate absolute flux calibration for AB~Aur was performed using the target acquisition images acquired with a neutral density filter. The flux calibration sensitivity of the AB~Aur data is correspondingly less certain than the other sources, and all line fluxes determined for AB~Aur are lower limits due to the attenuation of the central H$_2$ emission.

By inspecting the wavelength calibration and accuracy of sky emission line positions in raw exposures, we estimate that the absolute velocity accuracy of the individual IFU cubes is $\sim$9-12~km/s or roughly one third of a spectral pixel for a given night.  However, over multiple nights, the repositioning accuracy of the NIFS IFU grating wheel is repeatable only to $\sim\pm$0.75 pixels. Therefore, data acquired for the same source, but on different nights was not shifted and coadded in the wavelength dimension because of very weak signal from H$_2$ emission and to preserve velocity resolution.  It is important to note that the accuracy of the velocity centroids of the emission lines in the co-added multi-night data is clearly affected by the grating repositioning and the resulting shifts in the wavelength dimension.  However, all of the NIFS data on UY~Aur was acquired during the same night with no movement of the grating wheel.  As a result, the UY~Aur data has the single-night kinematic absolute accuracy of 9-12~km/s and relative in-field kinematic resolution near this level.  For all other sources, the estimated velocity accuracy of our IFU data is at the level of 1.0 to 1.2 pixels, or 28-35~km/s.  In these cases, the relative kinematic shifts within the spatial IFU field are likely accurate to approximately half of this value, or $\sim$14-17~km/s. In the following sections, aside from UY~Aur, only emission line velocity shifts of greater than the absolute accuracy level are taken to be genuine.  Additionally, all of the discussed H$_2$ emission line velocities are corrected for heliocentric motion at the time of the observation and placed into the stellar rest frame by removing the radial velocity of the star \citep{nguy12,hart86}.  All velocities presented in this project are v$_{Helio}$ rather than v$_{LSR}$.

While investigating the IFU data cubes for this project it became apparent that we discovered and characterized a previously unknown optical ghost in the NIFS instrument.  This ghost is seen in all of the K-band IFU spectral imaging cubes of our target CTTSs.  This ghost is presented and described in more detail in Appendix 1.

\section{Results} \label{sec:results}

\subsection{High Resolution Spectra}

Figure~2 presents the H$_2$ emission line detections in AA~Tau, GG~Tau~A, GM~Aur and UY~Aur from the high resolution spectroscopy.   Table~3 summarizes the velocity properties of the H$_2$ line emission.  Line profile velocity centroids and velocity widths were derived by fitting Gaussians to the observed features.  To place a measure on uncertainties in this process, the fits were carried out using three different tools and slightly different parameters for continuum level identification.  This investigation showed that the measured velocity values are accurate to about 3~km/s (0.2 pixels) for both CGS4 and CSHELL.  AA~Tau shows a moderate blueshifted H$_2$ line centroid, and the other three systems are largely consistent within 1-2$\sigma$ of the stellar rest velocity, which is at 0~km/s in Figure~2.  The line profile width for AA~Tau is perhaps marginally resolved, the other three systems are consistent with spectrally unresolved H$_2$ emission in our datasets.  The purpose of these high resolution measurements was to identify clear detection of the {\it v}~$=$~1-0~S(1) ro-vibrational H$_2$, and confirm existence of low velocity spectrally unresolved features indicative of potentially disk-bearing molecular hydrogen gas.



\begin{figure}[ht!]
\plotone{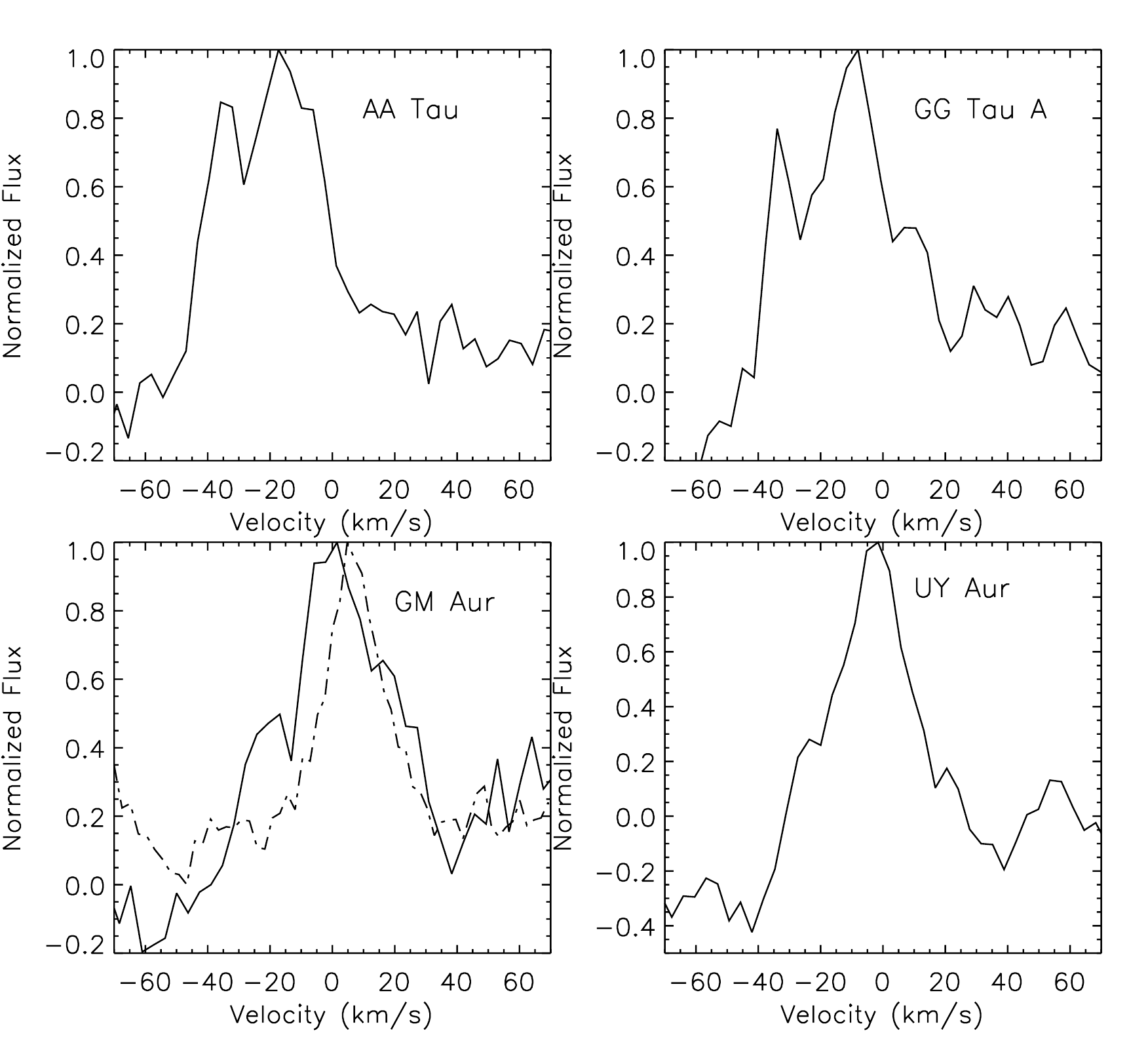}
\caption{High resolution spectroscopy of AA Tau, GG Tau A, GM Aur and UY Aur.   The flux is normalized to a value of 1.0 of the peak in the H$_2$ profile and plotted versus velocity.  The H$_2$ emission is blue shifted in AA Tau (a) and GG Tau A (b) with respect to the central stellar velocity at 0~km/s.  GM Aur shows the two measurements from UKIRT CGS4 (solid line) and IRTF CSHELL (dashed line) with a very slight velocity centroid shift of 6~km/s (2$\sigma$ significance) between them.  The H$_2$ emission from UY~Aur (d) is consistent with the stellar rest velocity.  \label{fig:hires}}
\end{figure}


\begin{deluxetable}{lccc}
\tabletypesize{\scriptsize}
\tablecaption{H$_2$ Velocity Structure from High Resolution Spectra\label{tbl-3}}
\tablewidth{0pt}
\tablehead{
\colhead{Star}  & \colhead{H$_2$ Line Center Velocity}  & \colhead{H$_2$ Velocity Width} & \colhead{Radial Velocity$^a$}  \\
\colhead{   } & \colhead{(km/s)}  & \colhead{(km/s)}  & \colhead{(km/s)}  }
\startdata
AA Tau & -16$\pm$3 &  35.3$\pm$3 & +16.9$^1$ \\
GG Tau A   & -10$\pm$3 &  23.6$\pm$3 & +13.7$^2$ \\
GM Aur (CGS 4)   & -1$\pm$3 &  29$\pm$3 & +15.2$^1$ \\
GM Aur (CSHELL) & 6$\pm$3 &  23$\pm$3  & +15.2$^1$ \\
UY Aur & -2.5$\pm$3 &  29.0$\pm$3  & +13.9$^1$ \\
\enddata
\tablenotetext{a}{Radial velocities are from 1 - \cite{nguy12} and 2 - \cite{hart86}}
\end{deluxetable}

\subsection{IFU Imaging Spectroscopy}

The goal of this study is to characterize the spatial extent of the H$_2$ emission previously detected in these systems and to isolate emission arising from the inner circumstellar disks.  Figure~3 presents the 2.12~$\mu$m continuum images of the three young multiple star systems that we spatially resolved in this study. The point-like continuum images of the single stars are not included.  These images were derived by fitting the average continuum level on the blue-ward and red-ward side of the H$_2$ line using a second order polynomial, and integrating in wavelength across the emission feature.  Our AO-fed imaging spectroscopy coupled with the multi-slice dithered observing strategy results in nicely rounded spatial point spread functions (PSFs) typically with spatial full-width half maximum (FWHM) values of 0.$\arcsec$08-0.$\arcsec$10.  DoAr~21, GG~Tau~Ab and V773~Tau have unresolved companions that were below the spatial sensitivity limit in our data \citep{bode12,loin08,difo14}.

\begin{figure}[ht!]
\plotone{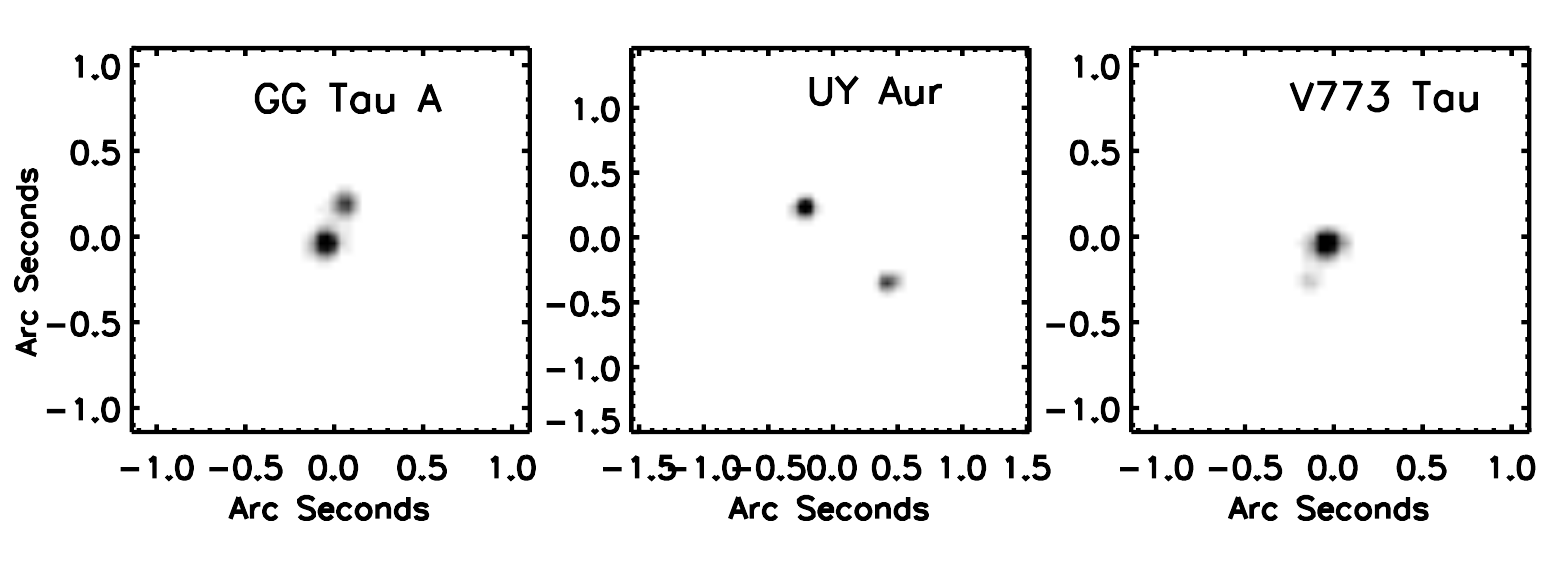}
\caption{ Maps of the continuum emission at 2.12~$\mu$m for GG Tau A, UY Aur and V773 Tau determined by fitting a line to the continuum emission blueward and redward of the {\it v}~$=$~1-0 S(1) line.  Images are scaled from +10\% to +70\% of the peak flux. In all panels, north is up and east is to the left. \label{fig:cont}}
\end{figure}

\subsection{Spatially Extended H$_2$ Emission}
 
Of the nine stars observed, seven have H$_2$ emission that was clearly spatially resolved in the IFU data.  Figure~4 presents continuum subtracted maps of the 2.12~$\mu$m flux showing the distribution of the H$_2$ emitting gas in AA~Tau, DoAr~21, GG~Tau~A, GM~Aur, LkH$\alpha$~264, UY~Aur and V773~Tau.  The H$_2$ images were made by first fitting the continuum level on either side of the emission line and interpolating between these points to estimate the continuum flux at the position of the feature. The continuum flux was then subtracted from the spectrum over the wavelength range covered by the line and the remaining line flux was integrated over the entire feature. These steps were performed for the $\lambda$-dimension of each pixel in the data cube. The final two-dimensional images presented in Figure~4 isolate the 2.12~$\mu$m line emission from the H$_2$ gas in seven of the systems. 


\begin{figure}[ht!]
\plotone{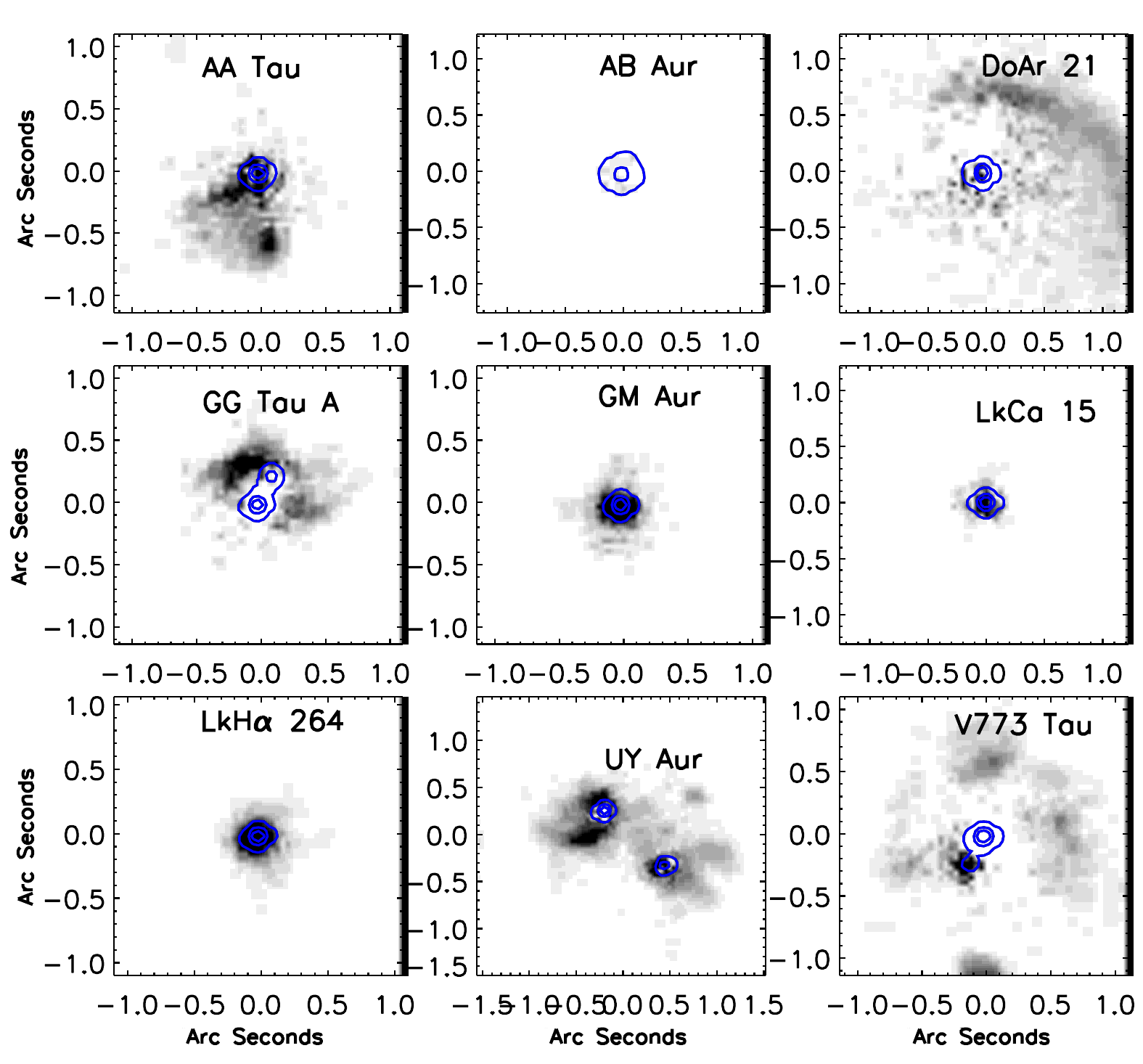}
\caption{Maps of the continuum subtracted, spatially extended {\it v}~$=$~1-0 S(1) H$_2$ line emission for each of the nine stars, the images are scaled from from +7.5\% to +70\% of the peak line flux.  Overplotted in blue are three contours of the continuum emission from +10\% to +50\% of the peak continuum level.  AB~Aur and LkCa~15 exhibit no appreciable H$_2$ emission in this view (see Figures~6 and 8 for continued analysis of these systems).  The H$_2$ from LkH$\alpha$~264 and GM~Aur is marginally extended and centered on the star (see Figure~5).  All other stars have spatially extended H$_2$ emission with significant spatial structure. In all panels, north is up and east is to the left. \label{fig:h2}}
\end{figure}

AA~Tau, UY~Aur and V773~Tau all have discrete knots of H$_2$ emission at extended distances from the stars.  The detected emission from AA~Tau shows an arc extending to the east-south east, and a bright knot of emission at a position angle of $\sim$185$\degr$, measured east of north.  The extension of emission is aligned with the known near edge-on disk encircling AA~Tau, and may arise from the disk surface in the inner 0$\farcs$3 regions ($<$50~AU) from the star.  The bright knot of emission is aligned with the direction of the known micro-jet from the AA~Tau system \citep{cox13}.  The H$_2$ emission detected in DoAr~21 is entirely spatially extended, starting at angular distances of greater than 0.$\arcsec$5 ($\sim$65~AU) from the central star.  The H$_2$ exhibits a smooth, arc-like morphology and is spatially consistent with the extended dust emission discovered in resolved mid-infrared maps of the system \citep{jens09}.  The H$_2$ emission in GG~Tau~A shows a bright arc to the north-east of the stars, with arc-like streamers that extend away from the central system.  An analysis of the observed H$_2$ emission from the interesting GG~Tau~A system has been published in a previous article \citep{beck12,dutr16}.

UY~Aur exhibits the strongest integrated H$_2$ flux of all the stars studied in this survey (Table~4).  The flux shows an extended distribution of H$_2$ gas entirely encompassing the stars in the binary with several extended knots of discrete emission.  There is also a bright arc of emission to the south of UY~Aur~A that extends to more than 0$\farcs$7 (100~AU) from the star toward the east, in the direction opposite from the companion, UY~Aur~B.  Several other arcs also extend to the south from UY Aur A and B.  UY~Aur is also known to drive an outflow as suggested by prior investigations \citep{hirt97,pyo14}.  The H$_2$ emission from the V773~Tau multiple also shows an extended distribution of flux to more than 140~AU from the brightest central star, and knots of emission.  The brightest H$_2$ emission in V773~Tau arises from the position of the 0$\farcs$2 separation companion.  We detect significant H$_2$ emission in the vicinity of both of the known infrared luminous companions in the V773~Tau and UY~Aur systems.  The H$_2$ emission velocity structure observed in V773~Tau reveals, for the first time, a bipolar outflow in this system, which will be discussed more thoroughly in Section 3.6.  Hence, an appreciable amount of the extended H$_2$ emission we observe in the AA~Tau, UY~Aur and V773~Tau systems seems to be stimulated by shocks in the outflows.  

Of the nine stars, the detected H$_2$ emission presented in Figure~4 for GM~Aur and LkH$\alpha$~264 is most similar to that expected to arise from the inner disks of T Tauri stars.  The emission from these systems appears centrally compact, spatially resolved and centered approximately on the stellar radial velocity (to our limited accuracy).  The measured gas is in agreement with high-resolution detections of the 2.12~$\mu$m lines in GM Aur and LkH$\alpha$ 264 \citep[Figure~2;][]{itoh03,carm08b}.  As an illustration of the extended nature of the emitting H$_2$ gas, we present comparisons of the encircled energy in the PSF and the PSF subtracted noise energy to that of the continuum subtracted H$_2$ in Figure~5.  The PSF encircled energy is essentially a sum of the flux as a function of distance from the central star, normalized to a value of 1.0 at a 1.$\arcsec$2 distance.  The encircled energy for the PSF subtraction noise was estimated by executing a continuum PSF subtraction at a wavelength of 2.18 $\mu$m, in a region that is largely devoid of spectral features, summing the PSF subtraction residuals into an image, calculating the noise energy level as a function of distance from the star, and normalizing in the same manner to 1.0 at a 1.$\arcsec$2 distance.  The difference between the observed PSF energy and the PSF subtraction noise profiles and the H$_2$ profile indicates that H$_2$ emission is spatially resolved and extended in GM~Aur and LkH$\alpha$~264.   The fact that the H$_2$ encircled energy curves for both GM Aur and LkHa 264 lie significantly below the summed PSF energy and noise curves means that the radial H$_2$ emission energy is spatially extended and resolved. 
 
 
\begin{figure}
\plottwo{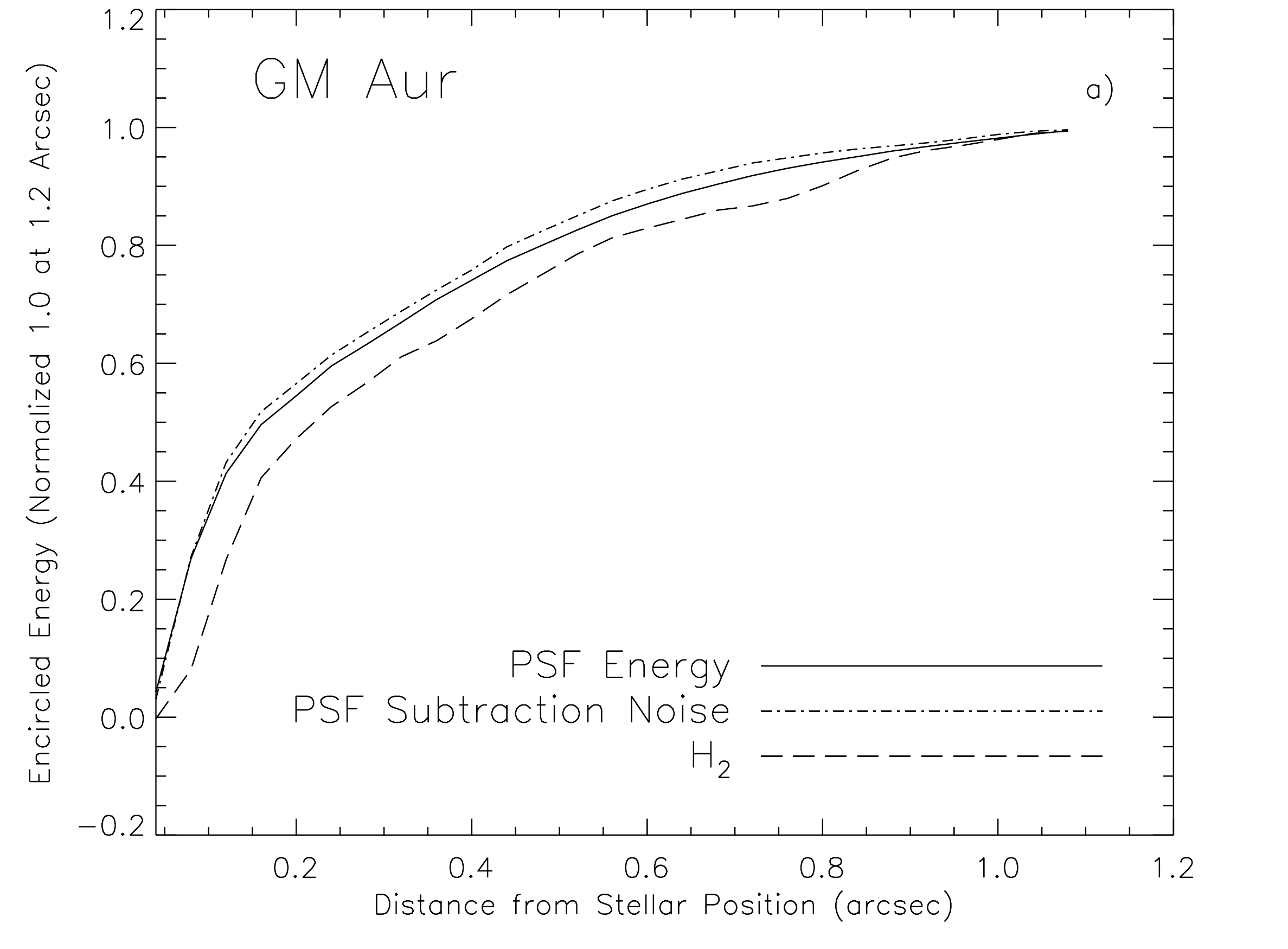}{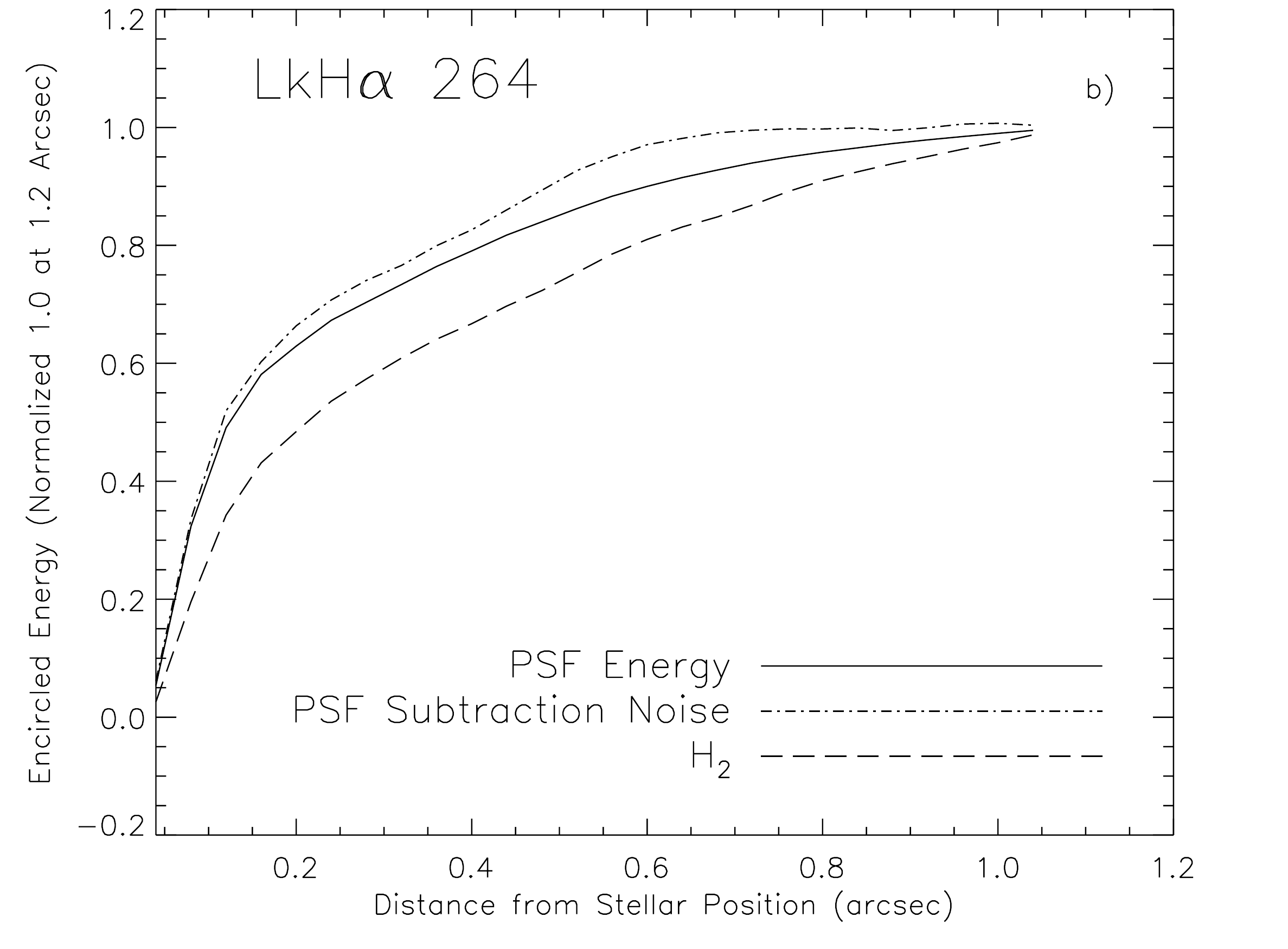}
\caption{Plots of the encircled energy versus position with increasing distance for the single stars GM Aur and LkH$\alpha$ 264.  Shown are curves for the PSF energy, the continuum subtracted H$_2$ emission, and a measure of the PSF subtraction noise.  The encircled energies are normalized to a value of 1.0 at a distance of 1.$\arcsec$2 from the central stars.   In both cases, the encircled H$_2$ energy for these stars is more extended from the stars compared to the PSF encircled energy shape and the subtraction noise residuals. \label{fig:ee}}
\end{figure}

Figure~6 shows the encircled energy profiles for LkCa~15, the PSF subtraction noise, and the continuum subtracted residuals at the wavelength of the H$_2$ feature.  The PSF subtraction noise is calculated in the same way as described for Figure~5.  The PSF noise and H$_2$ encircled energy curves deviate from the PSF shape by at most $\sim$3-4\% at all distances from the star.  Hence, the analysis at the H$_2$ wavelength shows no statistically significant difference from pure PSF subtraction residuals.  We conclude that there is no appreciable H$_2$ emission toward LkCa~15 that is measured in the IFU data.  This result is discussed in more detail in \S 4.0.1.

\begin{figure}[ht!]
\plotone{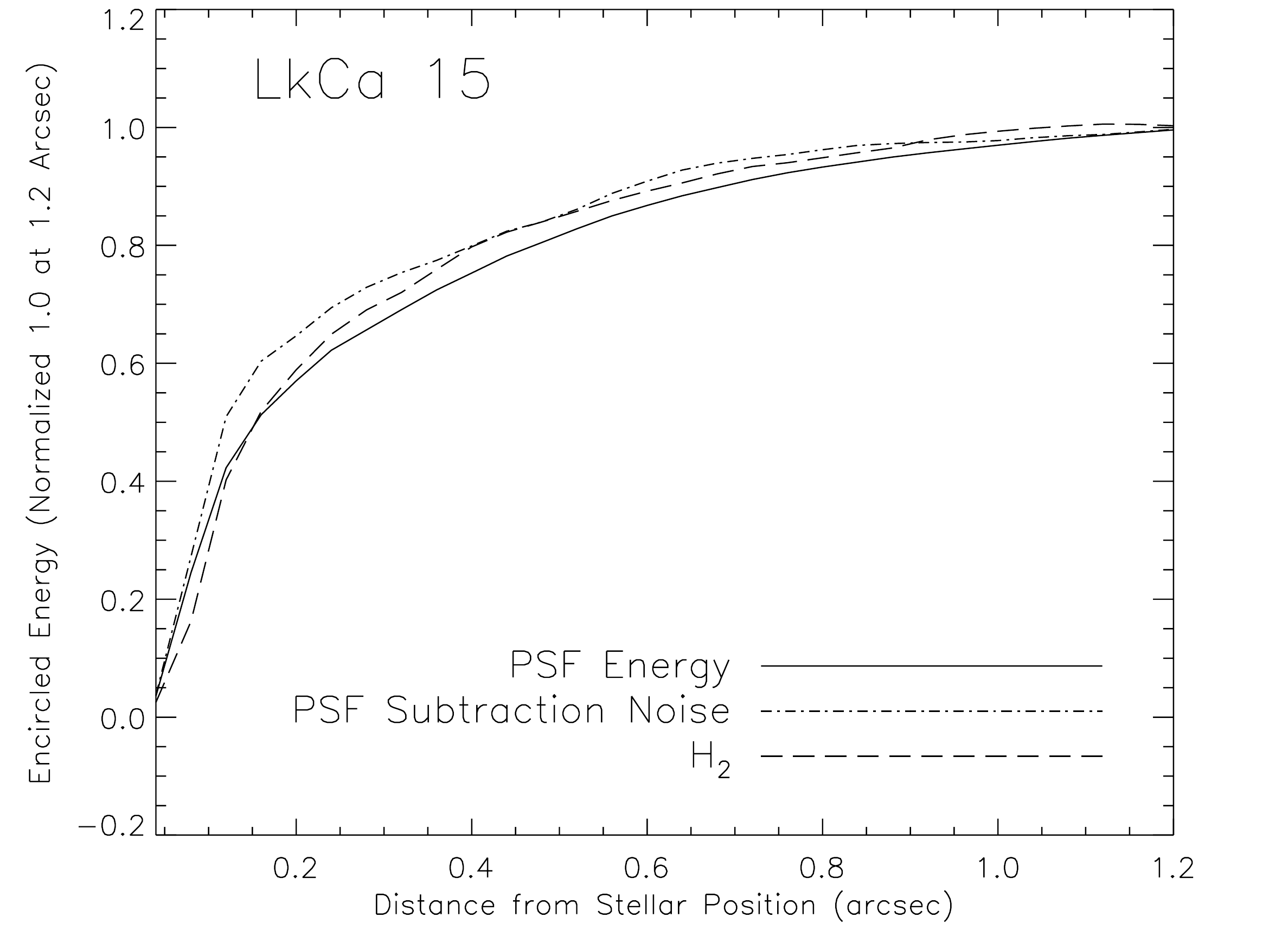}
\caption{ The encircled energy for LkCa~15 plotted versus position for increasing distances from the central star.  Presented are the PSF encircled energy, the continuum subtracted H$_2$, and a measure of the PSF subtraction noise.  The encircled energies are normalized to a value of 1.0 at a distance of 1.$\arcsec$2 from the central star.  For LkCa~15, the continuum subtracted H$_2$ residual emission energy profile is indistinguishable from the PSF and noise spatial profiles, confirming that extended H$_2$ emission is not detected in this source. \label{fig:lkha}}
\end{figure}

\subsection{A Closer Look at AB Aur}
 
Figures~7~and~8 presents further investigation to detect extended near infrared molecular hydrogen in the environment of AB Aur.  The initial analysis shown in Figure~4 did not find appreciable H$_2$ emission greater than $\sim$0.5\% of the integrated peak continuum flux for AB Aur.   Within 0.$\arcsec$1 from the star, the occulting spot used for the observations negatively affects the sensitivity to extended emission.  For this analysis, we masked out this inner region.  Figure~7 shows a zoomed view of the 2.115-2.130$\mu$m spectral region of a continuum subtracted small aperture with 0.\arcsec12$\times$0.\arcsec12 spatial extent, extracted to the north-west of AB Aur at a $\sim$0.\arcsec24 distance.  Careful subtraction of continuum emission measured nearby in wavelength was necessary to measure the H$_2$ from AB Aur because the peak H$_2$ line emission is just 1.5\% above the continuum flux measured at this single spatial position from the star.  The H$_2$ flux measured in this small spatial aperture is included for AB Aur in Table~5.   Integrated over the full IFU field of view, the very low level H$_2$ in AB Aur was swamped by the bright stellar flux at spatial locations that have no H$_2$ and by uncertainties in the continuum subtraction analysis used for Figure~4. 

Figure~8a shows the point source continuum subtracted and integrated emission for AB Aur at the 2.12$\mu$m wavelength corresponding to H$_2$ emission.  This image was created by 1) identifying two nearby 3-dimensional spatial-spatial-wavelength continuum flux regions that were modulated by the occulting disk substrate in a similar manner as the spectral region surrounding the H$_2$ line, 2) averaging these two regions to create a  3-D continuum model,  3) scaling and subtracting this average continuum 3-D datacube from the data cube centered on the H$_2$ emission, and 4) summing the central 4 wavelength channels of the continuum subtracted H$_2$ data cube into a 2-D image.  For AB Aur, this 3-D image subtraction analysis was significantly more accurate at detecting high contrast line flux than the 1-D spectral fitting method presented in Figure~4.  This analysis was also more efficient at removing the slight residual structure from spectral modulation caused by using the occulting disk substrate material.  Figure~8b shows the corresponding average uncertainty in the process used to construct Figure~8a.  Here, instead of subtracting the model data cube from the cube centered on the H$_2$ emission, the continuum model data cube was subtracted from another nearby continuum flux region with the same modulation, and the result was summed into this image.  This image serves as a measure of the average integrated flux level from the noise from the continuum subtraction process.  Panel c) shows the standard deviation of the subtraction flux results shown in panel b).  The images in Figures~8a, b and c share the same flux scaling from 0.0 to 1.0$\times$10$^{-15}$erg/cm$^2$/s.  Panel d) presents panel a) divided by panel c), which is the signal-to-noise of extended H$_2$ line emission in the environment of AB Aur.   There is low level H$_2$ emission surrounding the AB Aur system at a statistically significant level of detection.  The integrated line flux from this analysis is shown in Table~4.  Note that no correction was made for the effects of flux attenuation from the occulting spot used for the AB Aur observation, so this measured line flux level is a lower limit to the true integrated H$_2$ emission.  
 
\begin{figure}[ht!]
\plotone{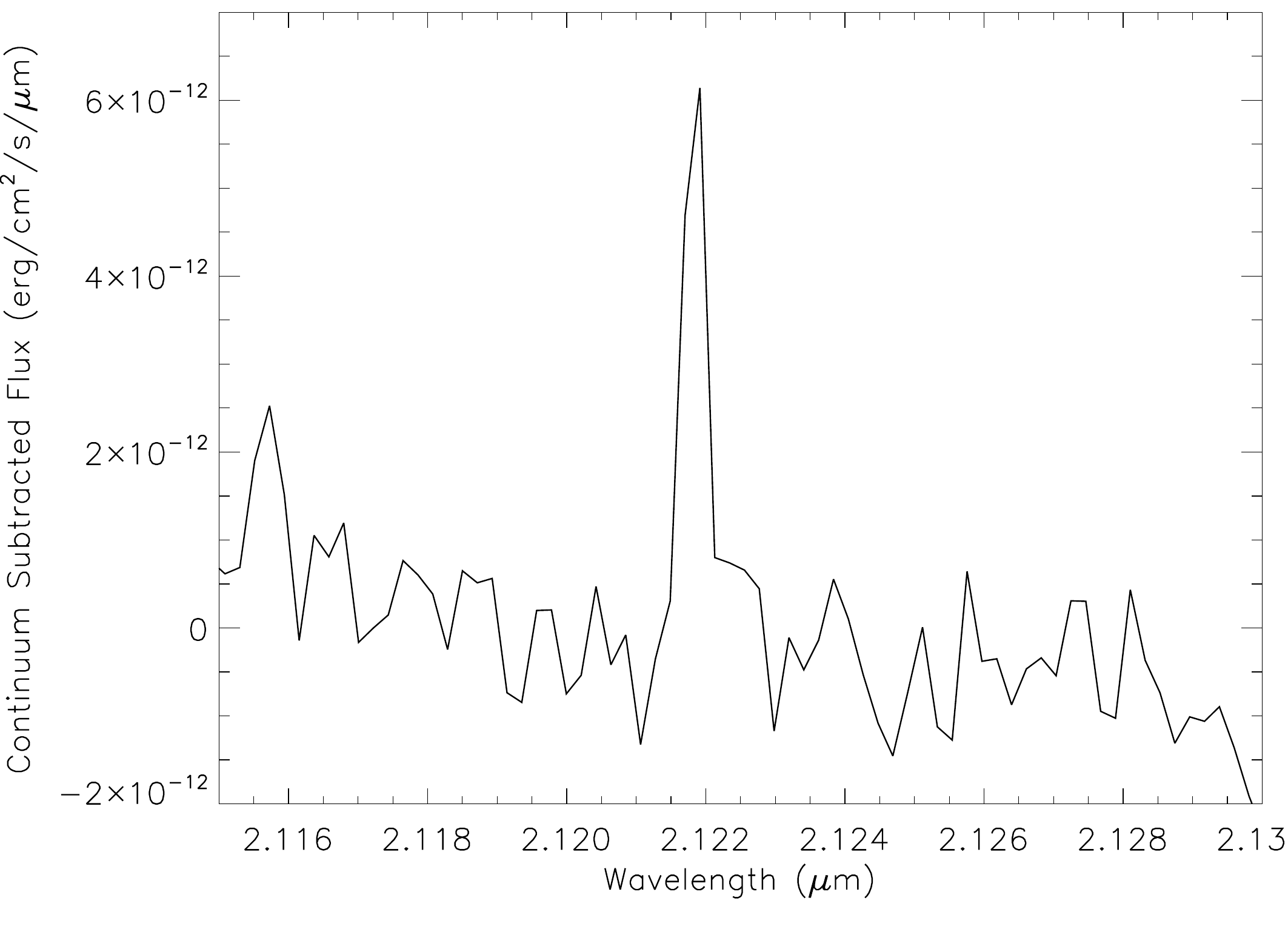}
\caption{The continuum subtracted H$_2$ emission extracted from a small aperture with a 0$\farcs$12 radius, located 0$\farcs$24 to the north of AB~Aur. \label{fig:abaur1}}
\end{figure}

\begin{figure}[ht!]
\plotone{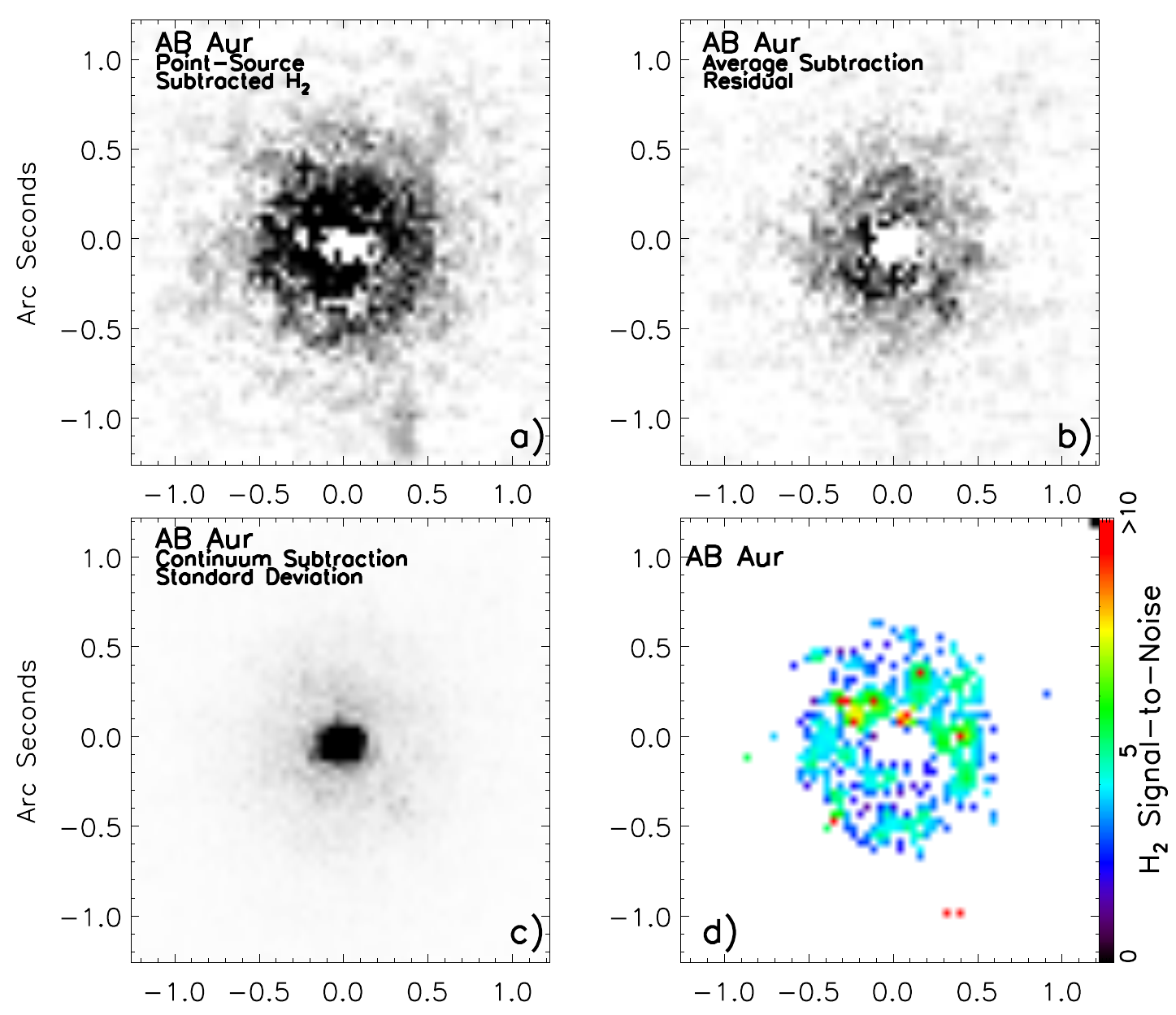}
\caption{Images showing the high contrast near infrared molecular hydrogen in the environment of AB Aur.  The region within 0.$\arcsec$2 from the star has been masked out for this analysis.  a) shows the continuum subtracted and integrated emission at the 2.12$\mu$m wavelength corresponding to H$_2$ emission.   Panel b) shows the continuum subtraction residual flux level (continuum minus continuum), which presents a measure of the flux structure that results from the continuum subtraction process.  Panel c) shows the standard deviation of the flux in the subtraction showed in b).   Panel d) presents panel a) divided by panel c), which is an estimate of the the signal-to-noise on extended H$_2$ line emission in the environment of AB Aur. \label{fig:abaur2}}
\end{figure}

\subsection{1-D Extracted Spectra and Analysis of S and Q-branch H$_2$ Transitions}

The $\lambda$-dimension of the NIFS IFU data allows for extraction of moderate resolution spectra, analogous to traditional long-slit spectroscopy, but located at the position of each 0$\farcs$1~$\times$~0$\farcs$04 pixel.  Figure~9 presents a one dimensional spectrum extracted in a circular aperture that spans the full IFU field for each of the seven stars that have strong H$_2$ emission detections (see Figure~4). The extraction aperture radii and corresponding H$_2$ 2.12~$\mu$m line fluxes are presented in Table~4.  For AA~Tau, DoAr~21 and V773~Tau, the weak H$_2$ emission lines are not easily detected over the bright integrated fluxes of the full field spectra. In all other stars, H$_2$ emission is clearly discernible from the continuum.  All spectra show 2.16~$\mu$m Br$\gamma$ emission from stellar mass accretion and absorption features from species including Al, Mg, Fe, Na, Ca, CO, which are typical photospheric lines observed in young stars with K-M spectral types.  AB~Aur is not presented in Figure~9, because the use of the NIFS occulting spot was necessary to observe this very bright star, data analysis required a different technique as described in \S3.4.  The line flux for AB~Aur presented in Table~4 was derived by spatially integrating the continuum subtracted H$_2$ line emission from Figure~8a.  This is a lower limit to the true line flux for the AB~Aur system because no correction for the occulting spot was applied.  

\begin{deluxetable}{lcc}
\tabletypesize{\scriptsize}
\tablecaption{H$_2$ Line Fluxes from NIFS IFU Data\label{tbl-4}}
\tablewidth{0pt}
\tablehead{
\colhead{Star} & \colhead{Extraction}  & \colhead {1-0 S(1) Line Flux}  \\
\colhead{   } & \colhead{Aperture Radius}  & \colhead{2.12~$\mu$m (\ergcms)}  }
\startdata
AA Tau & 1$\farcs$24 &  3.2$\pm$0.1$\times$10$^{-15}$   \\
AB Aur$^1$ & 1$\farcs$12 &  $>$6.6$\pm$0.7$\times$10$^{-14}$   \\
DoAr 21 & 1$\farcs$12 &  1.3$\pm$0.6$\times$10$^{-15}$   \\
GG Tau A & 1$\farcs$12 & 6.0$\pm$0.3$\times$10$^{-15}$   \\
GM Aur & 1$\farcs$24 & 4.9$\pm$0.1$\times$10$^{-16}$   \\
LkCa 15 & 1$\farcs$12 & $<$1.2$\times$10$^{-15}$   \\
LkH$\alpha$ 264 & 1$\farcs$24 & 4.5$\pm$0.1$\times$10$^{-15}$ \\
UY Aur & 0$\farcs$92 & 9.6$\pm$0.4$\times$10$^{-15}$   \\
V773 Tau & 1$\farcs$24 & 2.6$\pm$0.1$\times$10$^{-15}$   \\
\enddata
\tablenotetext{1}{The integrated {\it v}~=~1-0 S(1) line flux for AB~Aur is derived by summing the flux in a 1$\farcs$12 aperture from the H$_2$ image presented in Figure~8a.} 
\end{deluxetable}

In Figure~9, two spectra are plotted for each of the seven sources where H$_2$ was detected in Figure~4. (AB Aur is discussed in more detail in \S3.4). The top spectrum of each pair was produced by integrating over the entire NIFS field of view and is labeled ``Full Field."  The bottom spectrum of each pair was produced by extracting the spectrum from a 3-pixel (0$\farcs$12) radius aperture centered on the peak of the H$_2$ emission flux found in the continuum subtracted images (see Figure~4) and is labeled ``peak H$_2$." For GM~Aur and LkH$\alpha$~264, the peak flux was centered at the stellar continuum peak, so the smaller aperture ``peak H$_2$" spectrum presented here actually decreases the sensitivity to the H$_2$ line emission compared to the ``Full Field" spectrum.  For all other systems, the 2.12~$\mu$m H$_2$ emission is extended, originating far enough from the bright stellar continuum such that the line emission is significantly stronger relative to the continuum flux than the emission observed in the "Full Field" spectra.

The peak H$_2$ flux typically has between 5 to 10\% of the integrated H$_2$ emission seen in the whole field (Table~5).  In addition to the {\it v}~=~1-0 S(1) feature, several other transitions of H$_2$ are detected in these spectra. Table~5 presents a summary of the extraction aperture size, the {\it v}~=~1-0 S(1) flux level in the small extraction aperture, and the emission line ratios of the other transitions compared to the measured {\it v}~=~1-0 S(1) line.  Three-sigma detection limits are included as appropriate for sources that lacked detection of certain transitions.   The uncertainty in the fluxes for the Q-branch features from 2.40 - 2.42~$\mu$m are large because the long on source exposure times for the observed sources resulted in imperfect correction of narrow telluric absorption features in this spectral range. The uncertainties in the Q-branch emission fluxes were estimated by investigating the residual telluric features in the corrected spectra. The uncertainties are assumed to be lower limits. The H$_2$ flux for AB~Aur is also included in Table~5, the extraction aperture was centered on the discrete knot of emission to the northwest of the star.  No additional H$_2$ features were identifited in the spectrum of AB~Aur.  GM~Aur and V773~Tau had significant telluric correction noise in the spectral region longward of 2.40~$\mu$m, which made uncertainties too large to permit reliable detections of Q-branch line emission.

\begin{figure}[ht!]
\plotone{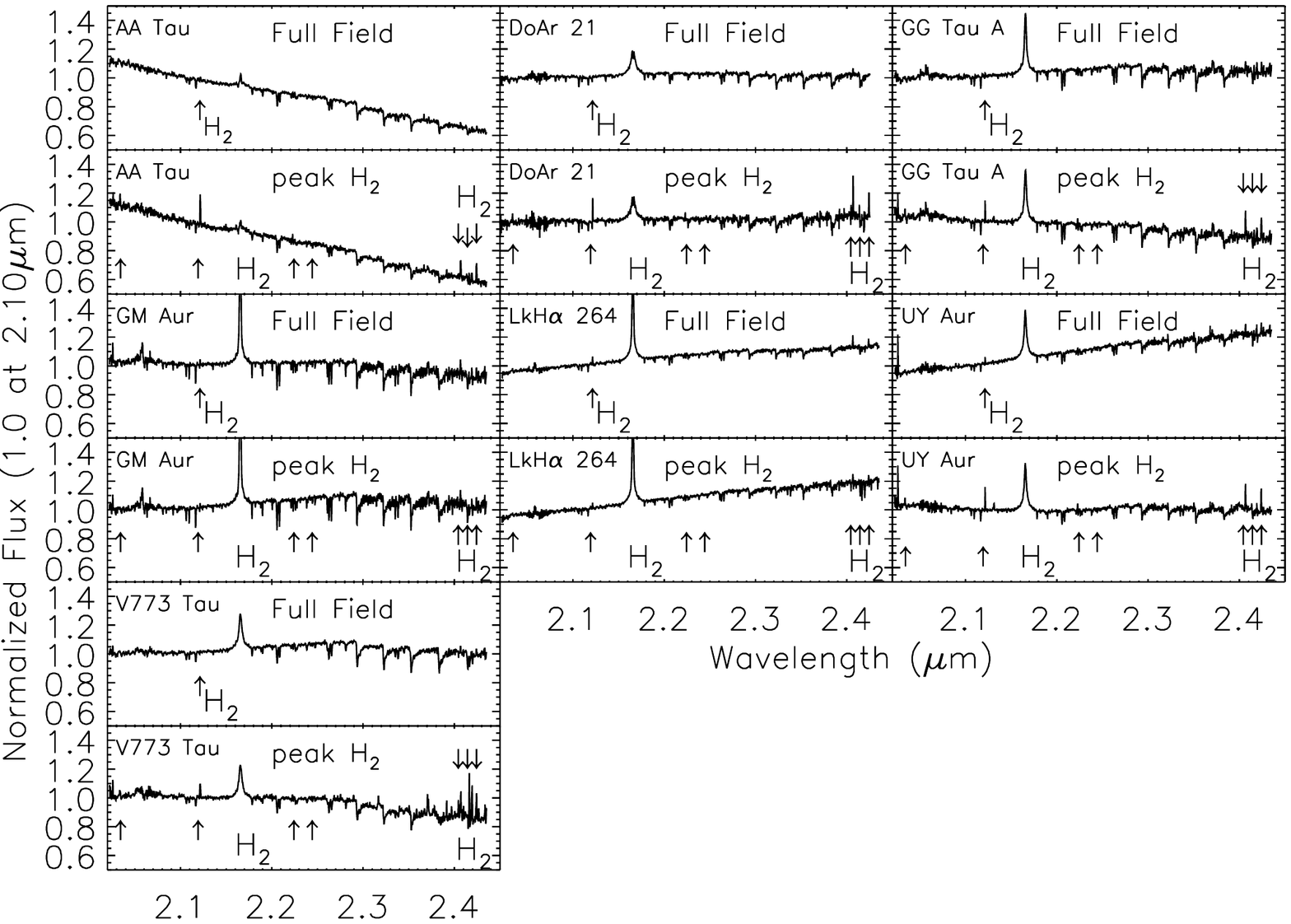}
\caption{The spectra of the seven stars with resolved {\it v}~=~1-0 S(1) H$_2$ emission measured in the analysis carried out for Figure~4.  For each star, the upper spectral plot shows the integrated flux vs. wavelength in a $\sim$1$\farcs$5 radius extraction aperture, approximating a "long-slit" K-band spectrum (called "Full Field").   The {\it v}~=~1-0~S(1) emission at 2.12~$\mu$m is identified by an arrow in all plots, and the S and Q-branch transitions also have arrows if they are detected.  The lower spectral plot for each star presents an 0$\farcs$12 (3 pixel) radius extraction aperture centered on the peak flux of the H$_2$ emission.  Both the ``Full Field" and ``Peak H$_2$" spectra are normalized to a value of 1.0 at 2.10~$\mu$m, so that a direct comparison between the two is possible.  The 2.16$\mu$m HI Br$\gamma$ emission is seen in all systems, as are multiple photospheric absorption features typical of these K and M type stars. \label{fig:kspec}}
\end{figure}

In past studies, the measured {\it v}~=~2-1~S(1) / {\it v}~=~1-0~S(1) line ratio has been proposed or adopted to constrain H$_2$ emission excitation mechanisms \citep{blac87,burt89a,ster89,gred94,gred95,malo96,tine97,taka07,beck08,gree10}. This diagnostic can work well to distinguish between non-thermal H$_2$ excitation processes such as UV pumping and fluorescence \citep[e.g.,][]{blac87,herc02}, which result in large electron populations of the higher vibration levels, from collisional excitation mechanisms such as shock heating in winds or jets. Typically the non-thermally excited H$_2$ emission regions have  {\it v}~=~2-1~S(1)  / {\it v}~=~1-0 \~S(1) line ratios of $\sim$0.25 or higher \citep[UV excitation;][]{blac87}.  In environments where densities are greater than or equal to a few times 10$^4$~molecules~cm$^{-3}$, the gas quickly thermalizes and results in line ratios consistent with the excitation temperature in the environment.  These regions may have H$_2$ emission stimulated by high energy flux from the star or by shock excitation. \cite{beck08} used this line ratio diagnostic to create spatially resolved gas excitation maps in the environment of the T Tau triple system, and \cite{gust10} furthered the analysis on T Tau.

The H$_2$ line ratios measured across the sample of stars in this study have a narrow range from 0.11 seen in AA~Tau to 0.06 measured in GG~Tau A.  These values are more consistent with denser thermalized gas than a pure low density excitation environment where the population of the higher vibration levels are enhanced.  Assuming LTE, our measured line ratios correspond to excitation temperatures in the range of 1700K - 2100K for the four systems that exhibit appreciable {\it v}~$=$~2-1~S(1) emission.   Table~6 presents line ratios and the associated excitation temperatures for LTE gas in these four systems, plus a lower detection limit on DoAr~21. 

\begin{deluxetable}{lcccccccc}
\tabletypesize{\scriptsize}
\rotate
\tablecaption{Peak (Narrow Aperture) H$_2$ Line Fluxes and 3$\sigma$ Detection Limits$^1$\label{tbl-5}}
\tablewidth{0pt}
\tablehead{
\colhead{Star} & \colhead{Extraction}  & \colhead {1-0 S(1) Line Flux}  & \colhead{1-0 S(2)}& \colhead{1-0 S(0)}& \colhead{2-1 S(1)}& \colhead{1-0 Q(1)$^2$} & \colhead{1-0 Q(2)$^2$}& \colhead{1-0 Q(3)$^2$} \\
\colhead{   } & \colhead{Aperture Radius}  & \colhead{2.12~$\mu$m (\ergcms)}  & \colhead{2.03~$\mu$m} &  \colhead{2.22~$\mu$m}& \colhead{2.24~$\mu$m}& \colhead{2.40~$\mu$m}& \colhead{2.41~$\mu$m}& \colhead{2.42~$\mu$m} }
\startdata
AA Tau & 0$\farcs$12 &  2.6$\pm$0.1$\times$10$^{-16}$ & 0.30$\pm$0.04  & 0.16$\pm$0.03 & 0.11$\pm$0.03 & 0.76$\pm$0.18 & 0.2$^{+0.20}_{-0.12}$ & 0.75$\pm$0.15  \\
AB Aur & 0$\farcs$12 &  9.5$\pm$0.8$\times$10$^{-15}$ & -- & -- & -- & -- & -- & --  \\
DoAr 21 & 0$\farcs$12 &  2.5$\pm$0.1$\times$10$^{-16}$ & 0.33$\pm$0.04  & 0.28$\pm$0.03 & $<$0.09 (3$\sigma$) & 1.1$\pm$0.3 & 0.4$\pm$0.2 & 0.9$\pm$0.3  \\
GG Tau A & 0$\farcs$12 & 4.8$\pm$0.3$\times$10$^{-16}$ & 0.37$\pm$0.04 & 0.29$\pm$0.02 & 0.06$\pm$0.02 & 1.1$\pm$0.2 & 0.3$\pm$0.2 & 1.0$\pm$0.2  \\
GM Aur & 0$\farcs$12 & 8.5$\pm$0.4$\times$10$^{-16}$ & $<0.20$ & $<0.24$ & $<$0.18 & --$^3$  & --  & --  \\
LkH$\alpha$ 264 & 0$\farcs$12 & 1.7$\pm$0.2$\times$10$^{-16}$ & 0.27$\pm$0.06 & 0.19$\pm$0.08 & $<$0.12 & 0.8$\pm$0.1  & 0.3$\pm$0.1  & 0.8$\pm$0.2 \\
UY Aur& 0$\farcs$12 & 2.7$\pm$0.1$\times$10$^{-16}$ & 0.29$\pm$0.03 & 0.28$\pm$0.04 & 0.07$\pm$0.02  & 0.9$\pm$0.1  & 0.3$\pm$0.1  & 0.8$\pm$0.2 \\
V773 Tau& 0$\farcs$12 & 3.4$\pm$0.1$\times$10$^{-16}$ & 0.26$\pm$0.03 & 0.26$\pm$0.02 & 0.08$\pm$0.03 & --$^3$  & -- & -- \\
\enddata
\tablenotetext{1}{The integrated {\it v}~$=$~1-0 S(1) line flux is presented in the third column.  All other H$_2$ emission line fluxes are presented as ratios with respect to the measured {\it v}~$=$~1-0 S(1) line flux.}
\tablenotetext{2}{Q-branch flux uncertainties are very high because of imperfect correction of narrow telluric features in the 2.40-2.44~$\mu$m region.  Uncertainties in Q-branch emission fluxes were estimated by investigating the residual telluric features in the corrected spectra. This is assumed to be lower limit on the uncertainties.}
\tablenotetext{3}{Poor telluric correction made it impossible to define the continuum flux level and unambiguously identify Q-branch features.  Accurate flux ratio limits could not be derived.}
\end{deluxetable}

\begin{deluxetable}{lccc}
\tabletypesize{\scriptsize}
\tablecaption{H$_2$ Excitation Temperature from {\it v}~$=$~2-1 S(1)/{\it v}~$=$~1-0 S(1) Ratio \label{tbl-6}}
\tablewidth{0pt}
\tablehead{
\colhead{Star} & \colhead{Extraction}  &  \colhead{2-1 S(1) / 1-0 S(1)}  & \colhead {Excitation }  \\
\colhead{} & \colhead{Location}  &  \colhead{Line Ratio}  & \colhead { Temperature (K)}   }
\startdata
AA Tau & Peak H$_2$ (Table~4)  &  0.11$\pm$0.03 & 2220$\pm$240  \\
DoAr 21 & Along H$_2$ Ridge (\S 4.5.3) &  $<$0.03 & $<$1460   \\
GG Tau A & Peak H$_2$ (Table~4) & 0.06$\pm$0.02 & 1790$\pm$200   \\
UY Aur & Peak H$_2$ (Table~4) & 0.07$\pm$0.02 & 1890$\pm$190\\
V773 Tau & Blue-shifted Outflow (Table~4) &  0.08$\pm$0.03 &  1980$\pm$280 \\
\enddata
\end{deluxetable}

The H$_2$ {\it v}~$=$~1-0 S(1) and {\it v}~$=$~1-0 Q(3) transitions arise from the same upper state in the H$_2$ molecule, and the intrinsic Q(3) / S(1) line ratio of 0.7 is unaffected by physical conditions in the gas emission environment.  The ratio of the flux from these transitions can be altered from the intrinsic value by line flux scattering off of dust grains which decreases the ratio, or by attenuation from obscuring material along the line of sight to the emitting region which increases the intrinsic ratio.  Extended dust structures can cause variations in obscuration in the environments of young stars, and past analysis of IFU data suggests that significant spatial variations in dust may be revealed using the H$_2$ emission line ratio \citep{beck08,gust10}.  Figure~10 shows the {\it v}~$=$~1-0 S(1) and  {\it v}~$=$~1-0 Q(3) emission line images and their ratio for AA~Tau and UY~Aur.   The ratio images were created using only the spatial locations that had {\it v} = 1-0 S(1) line flux that was brighter than 7.5\% of the peak H$_2$.  The relative sensitivity to line flux at the central stellar positions is lower than at extended distances because of bright continuum emission, so ratio values at spatial locations where the stellar flux was brighter than 10\% of the continuum peak are also not presented.   

The brightest extended knots of H$_2$ emission in both AA~Tau and UY~Aur have  {\it v}~$=$~1-0 Q(3)  / {\it v}~$=$~1-0 S(1) line ratios that are consistent within the uncertainties to the intrinsic value of 0.7 (purple or blue in the maps), implying no effects from scattering or line of sight extinction.  The H$_2$ emission in AA~Tau that extends to the southeast of the star has a higher line ratio, increasing to levels of $\sim$0.77.  This ratio corresponds to a visual extinction level of $\sim$5 magnitudes.   This could be indicative of higher extinction along the line of sight to this H$_2$ emitting region, perhaps caused by obscuration from extended regions of an outer disk.  The {\it v}~$=$~1-0 Q(3) flux is relatively stronger in the spatial regions around UY~Aur~B compared to A, which may imply greater line-of-sight obscuration toward this known infrared luminous companion.   The extension to the north of UY Aur B traces an increased line ratio at a level of 0.9~-~1.0, which translates to strong levels of extinction of 12~-~18 magnitudes in this H$_2$ emission region.  However, the absolute {\it v}~$=$~1-0 Q(3)  / {\it v}~$=$~1-0 S(1) line ratios typically have uncertainties of $\pm$0.1-0.2, which corresponds to A$_v$ differences of 7~-~12 magnitudes of flux attenuation.  Hence, the absolute magnitude of the A$_v$ variation is less important than the overall relative spatial structure in these line ratio maps.  DoAr~21, GG~Tau, GM~Aur, LkH$\alpha$~264 and V773~Tau produced detectable levels of {\it v}~$=$~1-0 Q(3) emission in the ``Full Field" and/or "Peak H$_2$" spectra in Figure~9, but the extended emission was either too faint or the fluxes were too strongly affected by telluric correction uncertainties to provide a reliable line ratio image at each spatial position.  

\begin{figure}[ht!]
\plotone{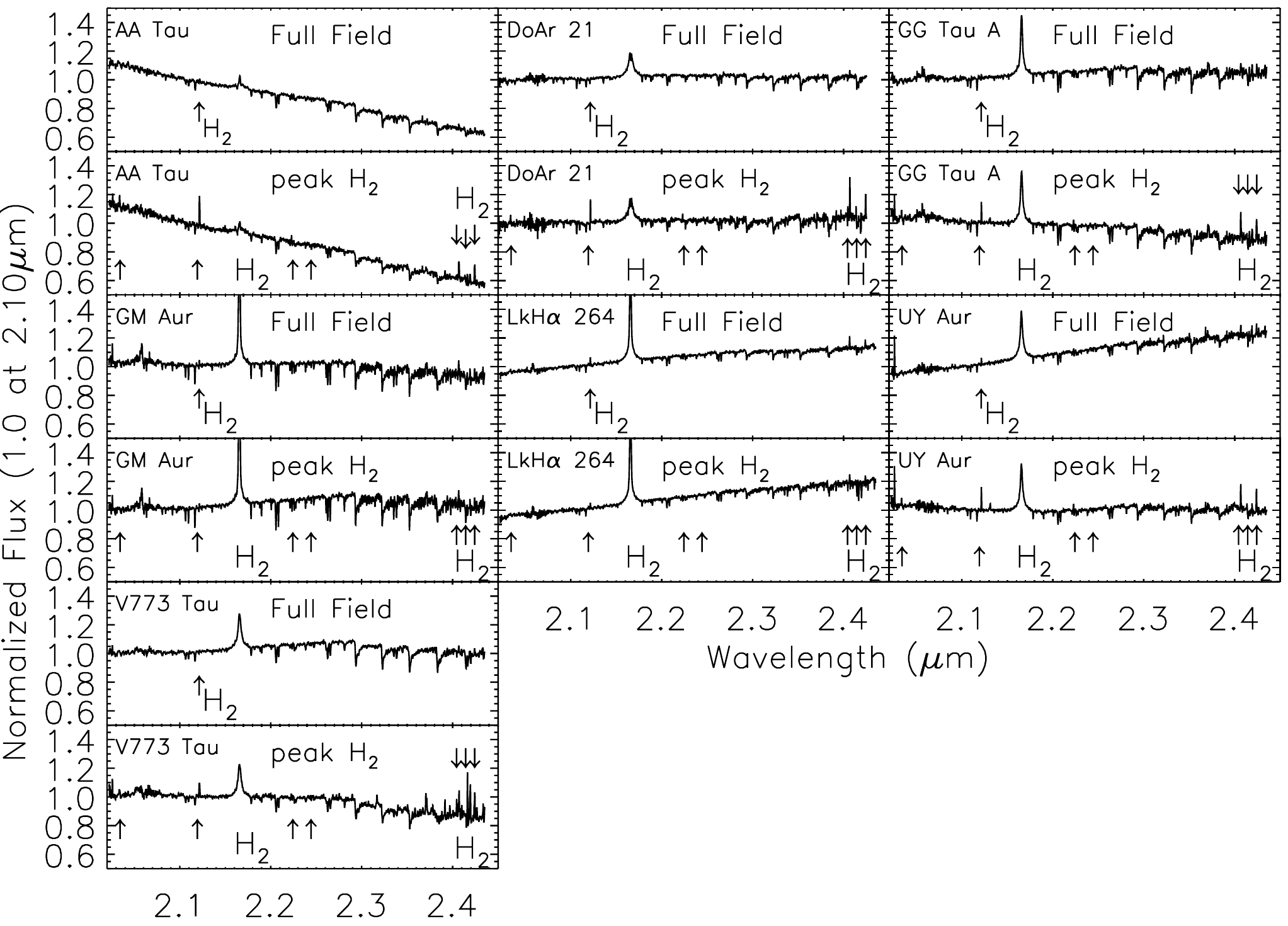}
\caption{The continuum subtracted H$_2$ {\it v}~$=$~1-0 S(1) (2.12~$\mu$m) and {\it v}~$=$~1-0 Q(3) (2.42~$\mu$m)  images for AA~Tau and UY~Aur are shown in the left and center panels, respectively.  The ratio of the {\it v}~$=$~1-0 Q(3)  to {\it v}~$=$~ 1-0 S(1) emission is shown to the right for each star.    \label{fig:h2ratios}}
\end{figure}

\subsection{Kinematics of the H$_2$ Emission}

The IFU data presented in this study possess inherent uncertainties in absolute velocity calibration due to coadding exposures collected over several nights, as described in \S2.  With the exception of UY~Aur, the data has a velocity resolution accuracy of no better than 30~km/s.  For UY~Aur, where all of the IFU data was acquired on a single night (Table~2), the velocity accuracy is estimated to be 12~km/s. In all other systems, absolute or relative velocity structures that are less than the accuracy of $\sim$30~km/s are unresolved. Clear detections of H$_2$ kinematics are seen in UY~Aur and V773~Tau, which both exhibit velocity structure at greater than 30~km/s.  Figures~11 and 12 show the kinematics of the H$_2$ emission seen in these two systems.  In both cases a) shows the barycenter velocity (flux profile weighted average velocity) of the H$_2$ emission for each spatial location that had a line flux of greater than 8$\sigma$ over the noise.  A more detailed description of the velocity barycenter analysis methodology for IFU data is presented in \cite{beck08}.  


\begin{figure}[ht!]
\plotone{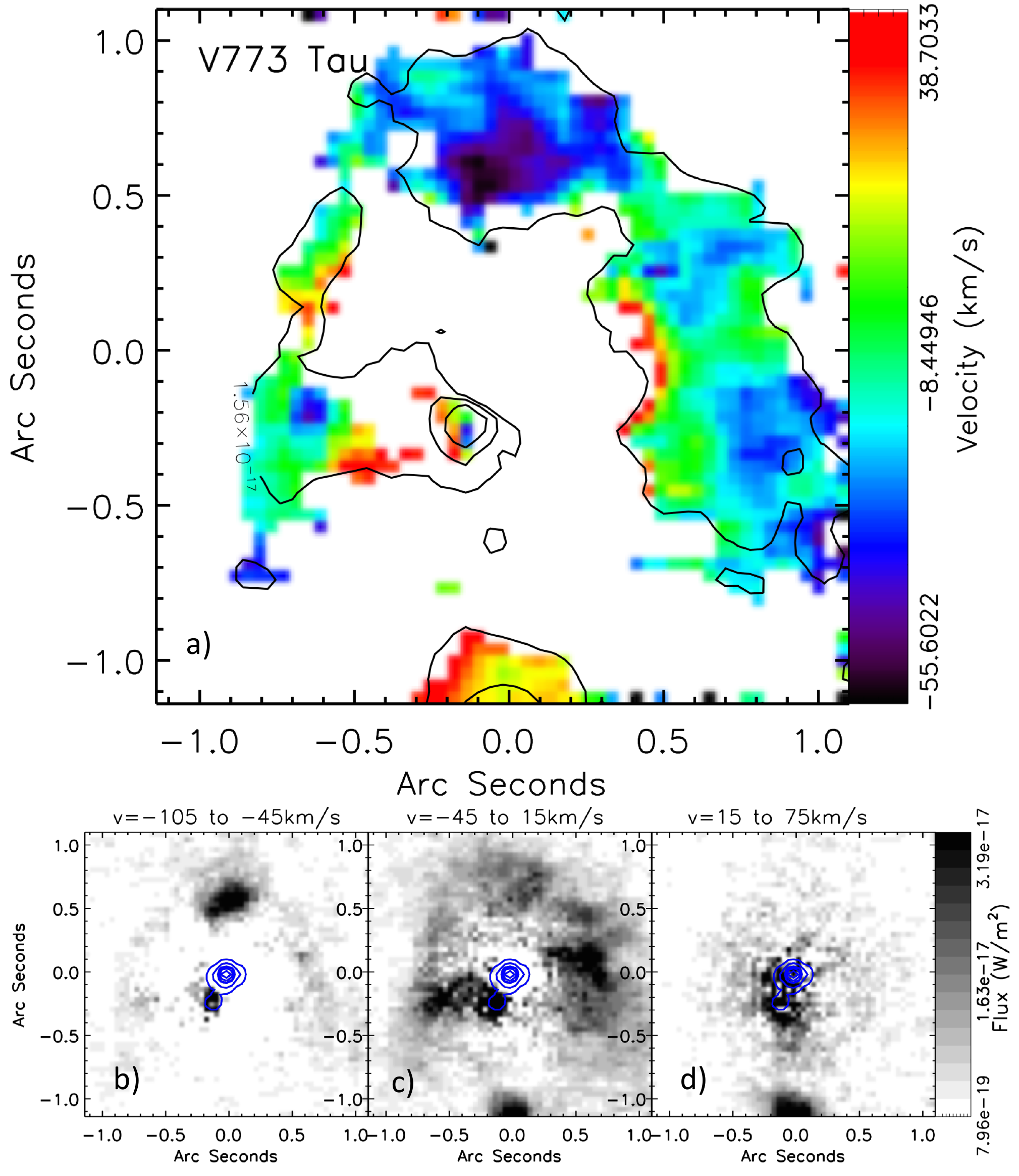}
\caption{The velocity structure seen in the H$_2$ emission in the environment of the young sub-arcsecond multiple system, V773 Tau.  In (a) the flux weighted velocity barycenter is presented in color.  Contours of the {\it v}~$=$~1-0 S(1) emission at the 10\%, 35\%, 50\%, and 70\% level compared to the peak H$_2$ flux are over plotted. Over-plotted in blue are contours of the continuum flux at 2.12~$\mu$m, highlighting the location of the stars that were spatially resolved in these observations.  Panels (b) - (d) shows three H$_2$ velocity channel maps from flux summed through the specified velocity ranges. \label{fig:v773vel}}
\end{figure}

\begin{figure}[ht!]
\plotone{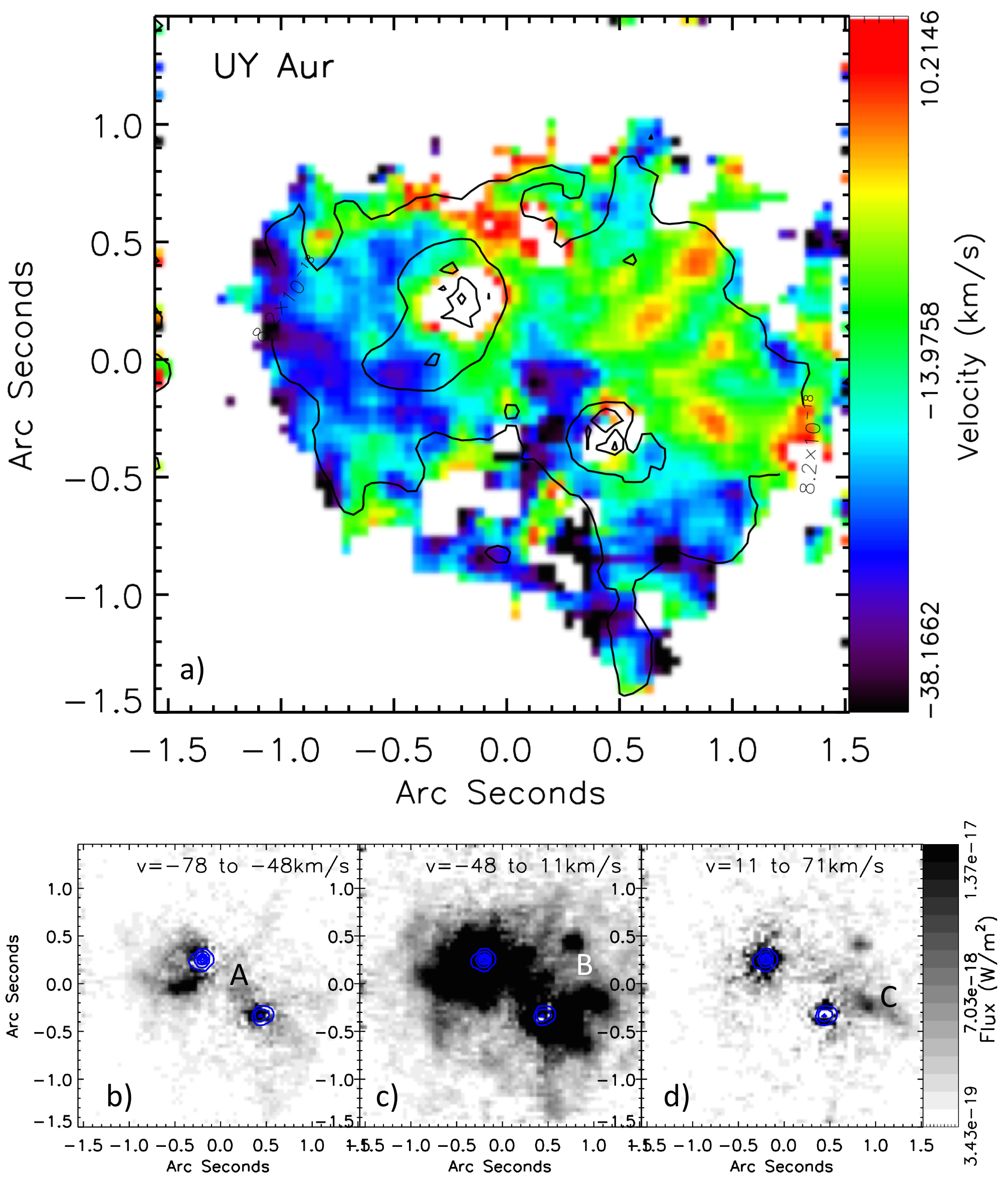}
\caption{Similar to Figure~10, but for UY~Aur. \label{fig:uyaurvel}}
\end{figure}

Figure~11b, the blue-shifted emission is seen to exhibit a strong knot of H$_2$ to the north of V773~Tau.   In Figure~11c, the low velocity channel map of the H$_2$ emission shows the gas distributed in knot-like structures all around V773~Tau with more diffuse emission coming from gas extending to distances of $\sim$200~AU or greater.  The position of the peak in H$_2$ flux is coincident with the location of the companion, 0$\farcs$2 from the primary source.  In Figure~11d, a bright knot of red-shifted H$_2$ emission is seen to the south of the V773~Tau system.  High velocity blue and red-shifted knots of emission around V773~Tau are obvious in Figure~11, presenting evidence of a bi-polar outflow from this complex young multiple system.  To our knowledge, this represents the first detection of spatially extended lobes of outflow emission from the V773 Tau system.

UY Aur shows blue-shifted emission to the south-east of the primary in the system (Figure~12b).  UY~Aur also shows knot-like emission and bright arcs that extend away from both stars.  The brightest emission region is to the south-east of the primary, UY~Aur~A (Figure~12c), and appreciable redshifted flux is seen to the north-west of UY~Aur~B (Figure~12d).  The kinematic structure in the H$_2$ observed around UY~Aur shows a gradient of $\sim$20~km/s from blue-shifted H$_2$ to the southeast of the field of view, to redshifted emission in the northwest.  Interestingly, this observed velocity gradient in the UY~Aur system is perpendicular to the position angle of the jets identified in \citet{pyo14}.  Moreover, multiple knots of emission from blueshifted velocities of -40km/s (position A, Figure~12b) to redshifted at +10km/s (position C; Figure!12d) correspond to features observed in the [Fe~II] jets from UY Aur as measured in \citet{pyo14}.  Cross referencing the kinematic features with the proposed system geometry from \citet{pyo14} clarifies the nature of the H$_2$ emission from UY Aur.  This is discussed in detail in \S4.1.

\subsection{Mass Accretion Rates and HR Diagram Masses and Ages}

In order to compare H$_2$ emission line luminosity to other characteristics of the observed systems, we use information from the literature and from our K-band IFU spectra to collect and derive stellar parameters.  The distances to our surveyed systems are taken from Gaia DR2 catalog for most stars \citep{bail18}.  An ensemble average for the Taurus dark cloud of 140pc is used for GG Tau A because the DR2 distance seems affected by the 0.$\arcsec$25 separation multiple system.  Integrated H$_2$ emission line luminosities and Br$\gamma$ fluxes are from this study.  X-ray luminosities are from the XEST survey for all available systems \citep{gude07, gude10}. X-Ray measurements of AB Aur and GG Tau A  came from \cite{tell07} and P. C. Schneider (private communication), respectively.  Visual extinctions, temperatures and luminosities are taken from \cite{herc14}, who compiled a consistent dataset for all of the systems in our survey.  The luminosities for the young multiples are adjusted based on K-band flux ratios for the spatially unresolved systems from \cite{herc14}, and presented values are for the primary stars in the systems.  DoAr 21 was analyzed assuming the source is an equal mass double star \cite{loin08}.  Stellar masses and ages are derived here, using these data from \cite{herc14} and the evolutionary tracks of \cite{bara15}. The more massive DoAr 21 and AB Aur systems required use of information from \cite{feid16}.  The results for masses and ages are very consistent with published values for each of these systems, but we have re-derived the information to ensure a consistent comparison.  The mass accretion rates are determined using our measured HI Br$\gamma$ line flux and converting to an accretion luminosity using the methodology of \cite{gull98}, and using the existing scaling relations to equate that to mass accretion \citep{mana13,muze98}.  All results are presented in Table~7 and used for comparison and discussion in \S4.

\begin{deluxetable}{lcccccccccc}
\tabletypesize{\scriptsize}
\tablecaption{Stellar Data for H$_2$ Comparison}
\tablewidth{0pt}
\tablehead{
\colhead{Star} & \colhead{Distance}   &  \colhead{A$_v$}  &  \colhead{H$_2$ 1-0 S(1) Line}  & \colhead {H {\scshape i} Br $\gamma$ }  & \colhead {X-Ray } & \colhead {Temp} &  \colhead {Stellar} &  \colhead {Mass } &  \colhead {Mass Accretion} &  \colhead {Age}  \\
\colhead{} & \colhead{}   &  \colhead{ }  &  \colhead{Luminosity}  & \colhead {Line Flux}  & \colhead {Luminosity} & \colhead {} &  \colhead {Luminosity} &  \colhead {} &  \colhead {Rate (Log)} &  \colhead {(Log)}  \\
\colhead{} & \colhead{(pc)}  &  \colhead{(mag)}  & \colhead{(Log L)}  & \colhead {(\ergcms)} & \colhead {(Log L)} & \colhead {(K)} &  \colhead {(Log L/L$_{\odot}$)} &  \colhead {(M$_{\odot}$)} &  \colhead {(Log M$_{\odot}$ yr$^{-1}$)} &  \colhead {(Myr)}  \\
\colhead{ } & \colhead{(1)}   &  \colhead{(2) }  &  \colhead{(3) }  & \colhead { (3) }  & \colhead {(4) } & \colhead {(4)} &  \colhead {(3)} &  \colhead {(5)} &  \colhead {(5)} &  \colhead {(5)} }
\startdata
AA Tau & 136.7 & 0.40 & 27.85 & 2.40e-17 & 29.66 & 3800 & -0.35 & 0.53 & -8.6 & 2.3 \\
AB Aur &162.1 & 0.55 & $>$29.31 & $>$8.83e-15 & 27.86 & 9910 & 1.39 & 2.4 & $>$-6.0 & 4.0 \\
DoAr 21 &133.8 & 7.10 & 27.44 & 2.48e-15 & 31.80 & 5760 & 0.92 & 1.88 & -6.3 & 6.3 \\
GG Tau A & 140 & 0.60 & 28.15 & 1.4e-16 & 29.30 & 3960 & -0.07 & 0.58 & -7.6 & 2.3 \\
GM Aur & 158.9 & 0.30 & 27.17 & 3.17e-16 & 30.08 & 4150 & -0.31 & 0.87 & -7.5 & 7.0 \\
LkCa 15 &	158.1 & 0.30 & $<$27.55 & 1.89e-17 & 30.40 & 4185 & -0.09 & 0.85 & -8.9 &  1.8 \\
LkH$\alpha$ 264  &  246.4 & 0.0  &  28.51 & 1.64e-16 & 28.38  & 4185 & -0.36 & 0.90 & -7.9 & 7.8 \\
UY Aur  &154.9 & 1.00 & 28.43 & 4.02e-16 & 29.41 & 4020 & -0.17 & 0.69 & -7.1 & 1.5 \\
V773 Tau  & 127.7 & 0.95 & 27.70 & 9.41e-16 & 31.0 & 4045 & 0.002 & 0.67 & -6.6 & 1.1 \\
\enddata
\tablecomments{References - (1): \cite{bail18}, except GG Tau A, which is an adopted average distance for Taurus-Aurigae. (2): \cite{herc14}   (3):  This study, (4): \cite{gude07,gude10,tell07} and P. C. Schneider (private communication) for GG Tau A (5): New derivation from this study.}
\end{deluxetable}

\section{Discussion: Ro-vibrational H$_2$ In the Environments of Young Stars}

The availability of high sensitivity adaptive optics fed integral field spectroscopy makes it possible to achieve the contrasts necessary to study the spatial morphology and excitation of ro-vibrational H$_2$ in young star environs.  In this project, we searched for spatially extended 2.12$\mu$m H$_2$ emission in a sample of nine sources.   The stars were selected because they had H$_2$ gas emission features that suggested the ro-vibrational emission might be spatially resolvable from the inner disks.  We detected and spatially resolved the H$_2$ in eight targets.  As summarized in Table~1, the 2.12$\mu$m ro-vibrational H$_2$ transitions can be stimulated by shock excitation in outflows, UV stellar Ly$\alpha$ pumping (as dominates for the UV electronic transitions), stellar heating of ambient gas, and by X-ray ionization and heating to temperatures that stimulate H$_2$ transitions \citep{fran12,nomu07,malo96}.  Of the eight systems where we detect H$_2$, two show clear emission from outflow components, three are young spatially resolved multiple star systems, and two have extremely strong X-ray flares reported in the literature.  As found by \cite{bary03, carm08b, itoh03}, the ro-vibrational H$_2$ emission for all eight systems amounts to several Lunar masses worth of emitting gas.

In \S 3.7 we collected, derived and presented stellar data for the nine systems in this sample to aid in our investigation of the origin of the ro-vibrational H$_2$ emission.  If X-ray ionization and heating provides a strong excitation mechanism for ro-vibrational H$_2$, we might expect a correlation between emission H$_2$ and X-ray luminosity.  Also H$_2$ can be stimulated by shocks in outflows and wide-angle winds, and hence we might expect a correlation between H$_2$ line luminosity and mass accretion rate, since mass outflow correlates with mass accretion \citep{hart95}.  As disk gas is accreted onto the central star and material in the disk coalesces to form planets, we might also expect to see a relation between H$_2$ line luminosity and stellar age.  Figure~13a - 13d plot several of the collected and derived parameters presented in Table~7 versus the {\it v}=1-0 S(1) H$_2$ emission line luminosity for the nine sources in this survey.  H$_2$ data for LkCa 15 shows the detection limit and AB Aur traces the lower luminosity bound because of the flux occulting spot used for the observations.  Interestingly, Figure~13a shows an apparent anti-correlation between X-ray and H$_2$ line luminosity.  This makes sense given that the H$_2$ disk gas and shocked outflow emission will decrease as a CTTS evolves into the more X-ray luminous weak-lined T Tauri Star (WTTS) phase \citep{neuh98}.  However, this anti-correlation also implies that X-ray ionization and heating is not a dominant excitation mechanism for the measured ro-vibrational H$_2$ excitation in CTTSs.   This is consistent with the analysis and results of \cite{bary03} and \cite{itoh03} who used the equations and analysis of \cite{malo96} with simple CTTS disk models and found that the H$_2$ stimulated by the X-ray luminosity is as much as two orders of magnitude lower than the line flux observed in the CTTS systems.  Additionally, \cite{espa19} also found no relationship between the FUV H$_2$ emission bump and X-ray luminosity, adding to the growing evidence that the X-Ray ionization does not significantly affect the H$_2$ gas in the inner disks of CTTSs.  In our admittedly small 9 source sample, no obvious correlations are seen between the H$_2$ line luminosity and other system characteristics such as age, mass accretion rate or stellar mass (primary mass for multiple systems).  Interestingly, \cite{espa19} do find a correlation between the FUV H$_2$ emission and mass accretion, which we do not see.  This implies that processes that dominate the hot gas of the inner disk do not extend to the larger disk radii and denser layers of material where the bulk of the ro-vibrational H$_2$ emission originates.  

We also investigated the near-IR H$_2$ S(1) line luminosity plotted versus UV H$_2$ Luminosity for four systems that overlap between this study and published measurements (AA Tau, GM Aur, LkCa 15; including RW Aur from \cite{beck08};\cite{fran12}, and versus mid-IR line luminosity for AB Aur, RW Aur and HL Tau \citep{beck08,bitn08}.  We found potential weak correlations, but the number of overlapping sources in the surveys made for unconvincing plots that we chose not to include here.


\begin{figure*}
\gridline{\fig{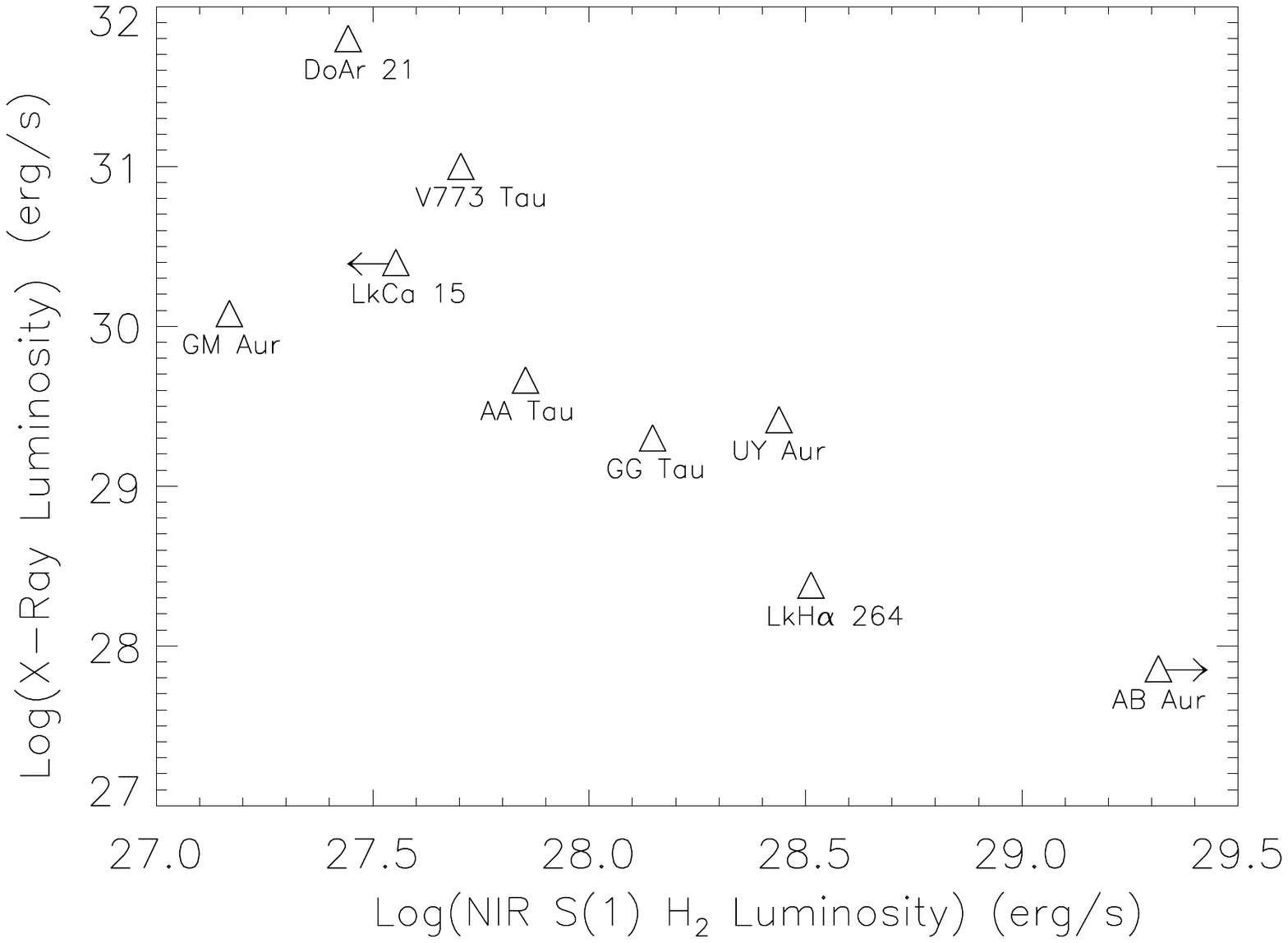}{0.48\textwidth}{}
          \fig{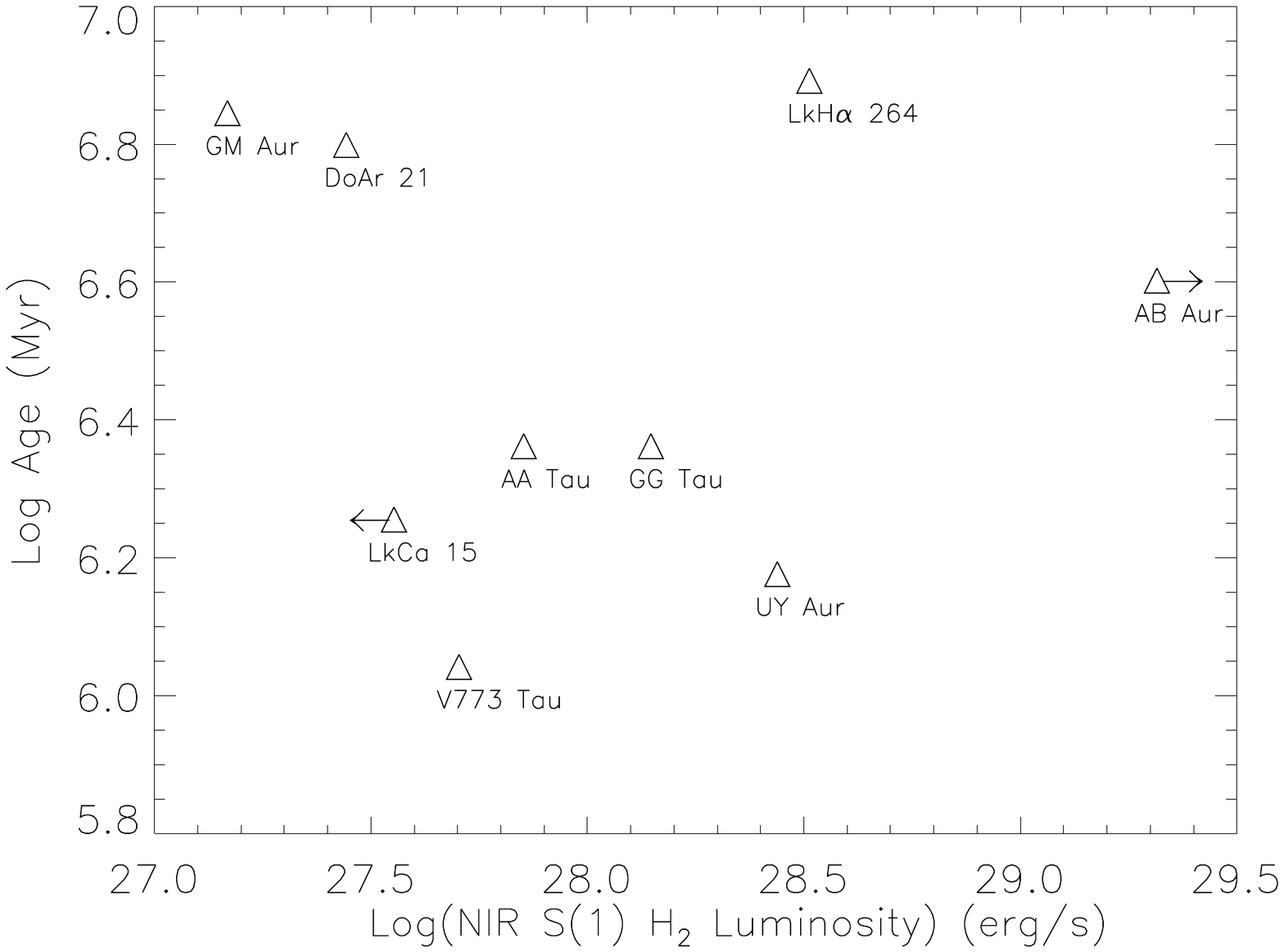}{0.48\textwidth}{}
          }
\gridline{\fig{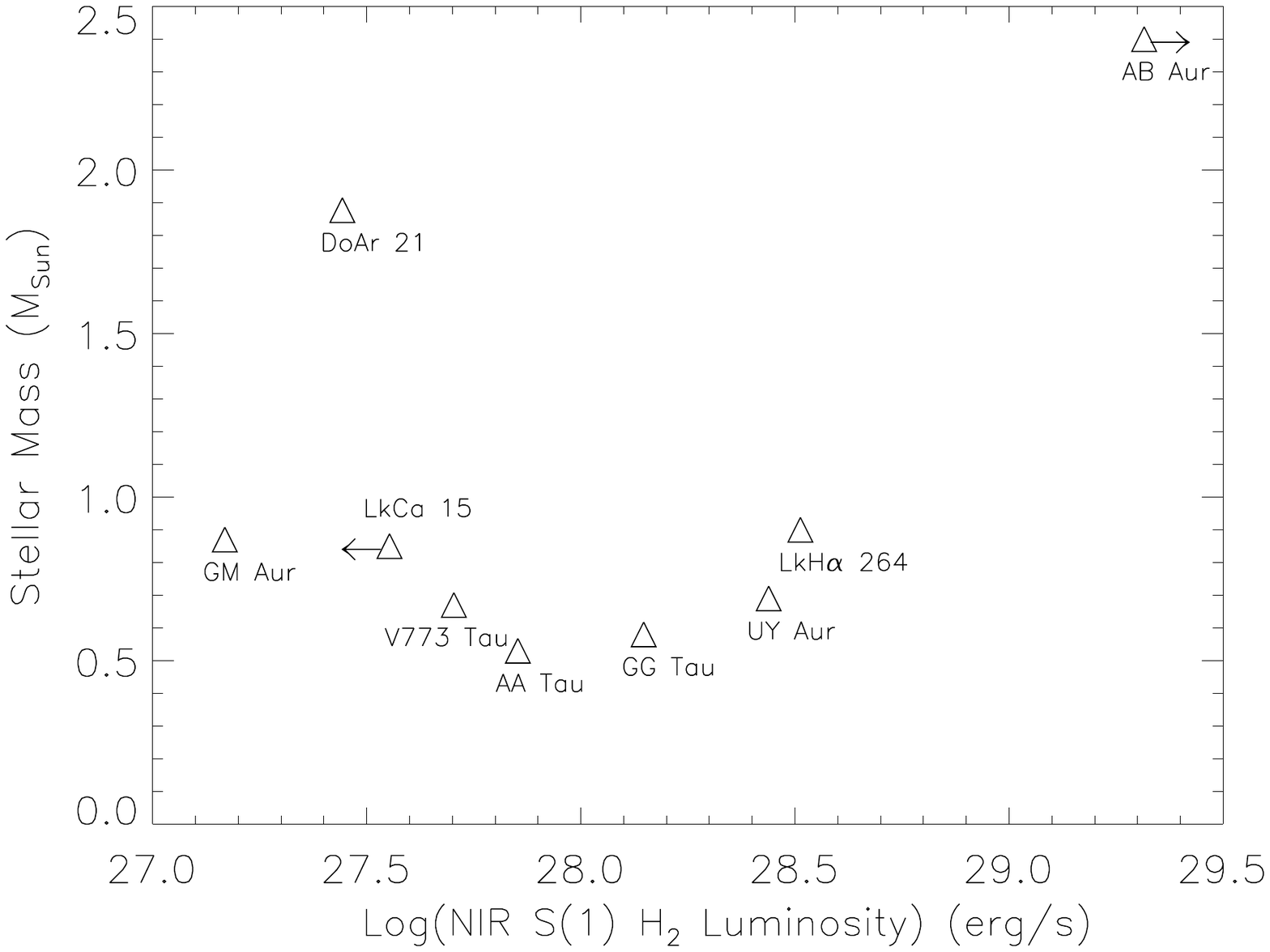}{0.48\textwidth}{}
          \fig{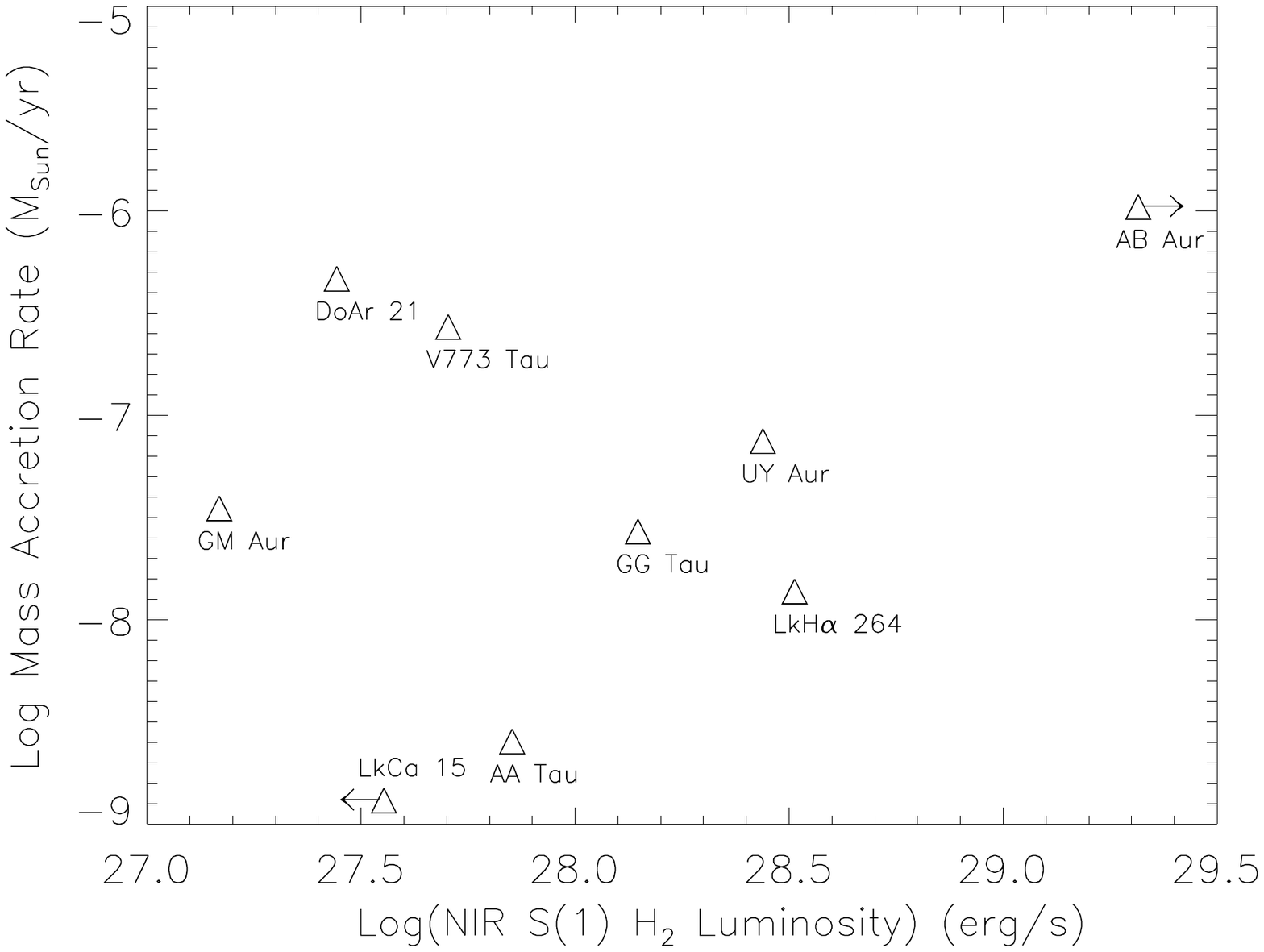}{0.48\textwidth}{}
          }
\caption{The (Log) X-Ray Luminosity (a), stellar ages (b), stellar masses (c) and (Log) mass accretion rates (d) plotted versus the (Log) 2.12$\mu$m ro-vibrational H$_2$ line luminosity.   Measurement of the H$_2$ luminosity in AB Aur is a lower limit because of the occulting spot used for the observation.  The H$_2$ luminosity plotted for LkCa 15 is the 3$\sigma$ detection limit.  \label{fig:pyramid}}
\end{figure*}


\subsection{Ro-vibrational H$_2$ from Outflows} 

In \cite{beck08}, six stars with known Herbig-Haro outflows were surveyed for spatially extended ro-vibrational H$_2$ emission.  Of those six stars, all were found to have appreciable H$_2$ emission that was predominantly excited by shocks in the jets or winds.  In this project, we specifically targeted sources that had evidence suggestive of a more disk-like origin for the H$_2$ emission.  In the eight systems where we measure the ro-vibrational H$_2$, strong centralized gas emission from the inner disks likely exists in all stars except for DoAr 21.    AA Tau, UY Aur and V773 Tau show shock-excited H$_2$ emission from known and newly revealed inner outflows.

AA Tau exhibits a slight blue-shifted line H$_2$ line profile in its high resolution spectrum (Table~2), and the brightest knot of emission to the southwest at 0.\arcsec7 (100AU) distance from AA Tau lies along the axis of the known jet \citep{cox13}.  The 0.\arcsec2 aperture "peak H$_2$" spectrum presented in Table~5 and Figure~9 was extracted at this location.  The measured {\it v} = 1-0 Q(3)  / {\it v} = 1-0 S(1) line ratio at the knot position reveals a very low level of line-of-sight obscuration toward the extended region of the jet (Figure~10).  This bright knot of jet emission also shows measurable {\it v}=2-1 S(1) flux with a ratio of 11\% compared to the 1-0 S(1) transition.  For gas in LTE, this emission corresponds to an excitation temperature of $\sim$2200K (Table~6).  This emission at 100AU distance along the jet channel with a temperature of over 2000K clearly points to shocks in the outflow as the responsible excitation mechanism for the H$_2$.  

Figure~11 shows the average barycenter velocity for the H$_2$ emission from V773 Tau, revealing north - south bi-polar knots.  The knots are positioned asymetrically around the system; the blueshifted knot is $\sim$0$\arcsec$6 to the north of V773 Tau A, and the southern redshifted knot is $\sim$1$\arcsec$1 to the south and half off of the field of view (Figure~11).   It is not clear which star in this sub-arcsecond quintuple system might be exciting this outflow.  V773 Tau shows the strongest blue and red-shifited H$_2$ of any of the stars in the current nine source sample.  In fact, of the fifteen stars now surveyed for H$_2$ emission \citep{beck08}, V773 Tau shows the second strongest kinematic structure in the H$_2$ emission profile, behind only the anomalously strong redshifted jet seen in RW Aur.  

UY Aur has a known Herbig-Haro flow (HH~386; \cite{hirt97}), with spatially resolved inner jets from both the A and B stars and a wide angle wind from UY Aur A \citep{pyo14}.  The collimated blueshifted jet from UY Aur B extends to the northeast toward UY Aur A, and the red-shifted flow from UY Aur A extends to the southwest toward UY Aur B.  The apparent outflows nearly align with the binary orientation.  We clearly detect strong H$_2$ emission in several knots that can be traced to these known [Fe~II] jets \citep{pyo14}.   In Figures~12b-d and Figure~14, three positions A, B and C are designated.  Position A is the location of the strongest measured blueshifted H$_2$ in the environment of UY Aur.  Comparison with the [Fe~II] maps of \cite{pyo14} shows that this emission traces the inner blueshifted jet from UY Aur B.  Position B traces a strong knot of slightly redshifted outflow emission and Position C shown in Figures 12 and 14 traces strongly redshifted emission from the southwestern jet from UY Aur A.  The bright arc of H$_2$ emission that we detect to the southeast of UY Aur A has a counterpart in the low velocity [Fe~II] blueshifted maps presented by \citet{pyo14}.      

The UY Aur A+B binary exhibits extensive H$_2$ emission that fully encompasses this young system (Figure~14a).  Three arcs surrounding UY Aur A and B are highlighted in green, yellow and cyan in Figure~14a.  The green arc spans a range of nearly 180$\deg$, opens toward the north-northeast, and fully encompass the south side of UY Aur A.   \cite{pyo14} also saw the brighter inner regions of the green arc in low velocity [Fe~II] emission and interpreted this structure as arising from a blueshifted wind with a very wide opening angle (90$\deg$ from \cite{pyo14}) and a special viewing geometry tilted toward the observer.    The yellow line overplotted in Figure~14a is another arc of emission around UY Aur A that in the H$_2$ map.  This arc starts at the northern side of UY Aur A and wraps around the east to the south-southeast of the star.  This particular arc does not seem to have a direct counterpart in [Fe~II] emission measured by \cite{pyo14}, but the overall morphology is similar to the geometry that they propose for the corresponding redshfited wide angle outflow from UY Aur A.  However, inner redshfited outflows are usually very difficult to measure because the optically thick central dust disk blocks clear detection of the emission on the far side of the disk.  UY Aur A has a dense inner dust disk measured by ALMA \citep{tang14}.   If the line traced in yellow is an arc from the redshifted wind, the fact that we see this arc initiate within $\le$0.$\arcsec$15 from the north of UY Aur A and wrap around to the east-southeast necessitates a special viewing geometry and a sparse amount of obscuring dust in the inner $\sim$20~AU of the circumstellar disk.   The third line overplotted in cyan in Figure~14a traces a similar arc of emission from UY Aur B.

The UY Aur system geometries presented in Figure~5 of \cite{pyo14} are updated and shown here in panels Figure~14b and 14c.   Figure~14b presents our observed view, and Figure~14c shows an expanded view of the geometry as viewed from the right side (consistent with \cite{pyo14}).  The location of the measured collimated jets, wide angle wind components and the regions with strong H$_2$ emission are highlighted.   The strongly redshifted knot C is associated with the outskirts of the jet from UY Aur A \citep{pyo14}.   These schematic geometrical descriptions of the UY Aur system also reveal that the bright H$_2$ knot designated B in Figures~12 and 14 is likely associated with extended regions in the wide angle redshifted flow from UY Aur A.   We hypothesize that the extended arc-like H$_2$ structures seen in the environment of UY Aur A are excited in part by shocks at the outskirts of the wide angle outflows, possibly through interaction of the multiple wide-angle flow components that are slightly inclined with respect to each other \cite{pyo14}.   For example, the collimated blueshifted jet from UY Aur B seems like it could be interacting with the wide angle wind from UY Aur A, producing the extended arc of emission highlighted in green in Figure~14a.  The great extent of this H$_2$ arc implies that there might be a wide angle wind component from UY Aur B that is also colliding and shocking the H$_2$.   Hence, we postulate that young star multiples have outflow-outflow interactions as an additional H$_2$ shock excitation mechanism that does not exist in single star systems.


\begin{figure}[ht!]
\plotone{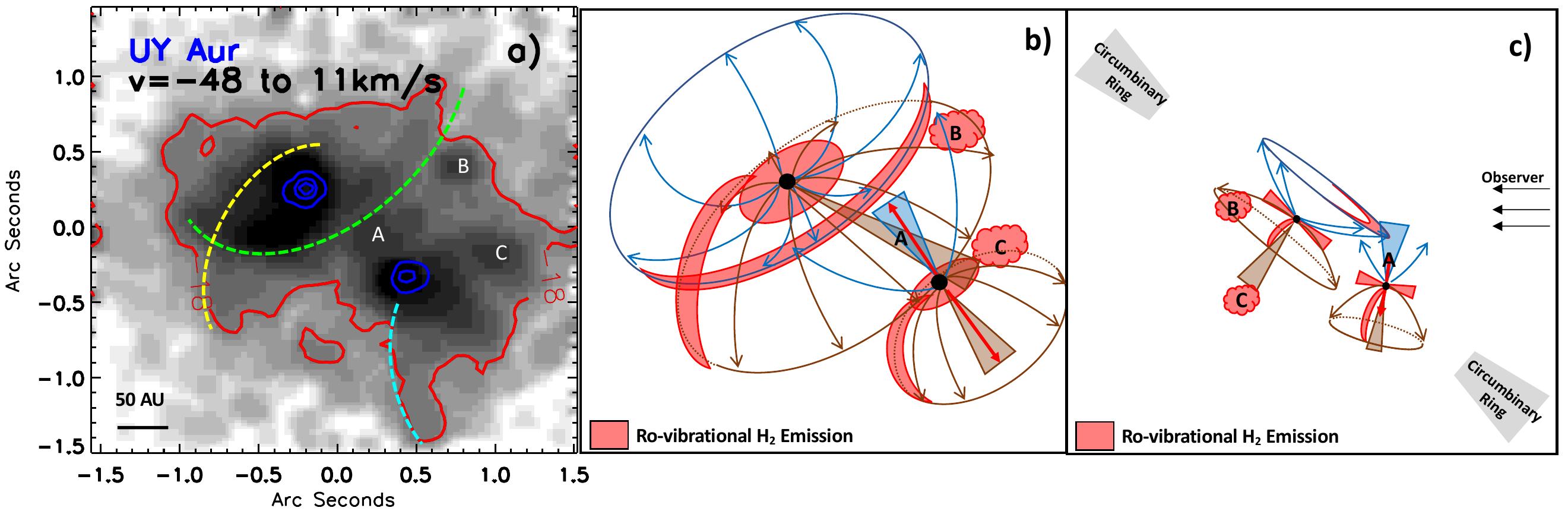}
\caption{The integrated 2.12$\mu$m ro-vibrational H$_2$ line emission image from velocity channels -45~km/s to +15~km/s in the environment of the young multiple system UY Aur (a).  The positions of UY Aur A and B are highlighted in (a) by blue contours of 10\%, 40\% and 70\% of the peak 2$\mu$m continuum flux.  The display is scaled logarithmically from 1\% to 75\% of the peak H$_2$ line emission and shows one red contour that presents the 10\% H$_2$ flux level with respect to the peak line emission.  The green, yellow and cyan lines trace the positions of arc-like structures seen in the H$_2$ morphologies.  (b) and (c) Show updated geometrical descriptions of UY Aur and are based on Figure~5 from \cite{pyo14}.  The knots and arcs shown in red in (b) and (c) highlight the interpreted positions of strongest H$_2$ emission structures from the disks, winds and collimated outflows. H$_2$ emission regions labeled A, B, and C are shown in all panels. \label{fig:H2imguyaur}}
\end{figure}

Combined with the results of \cite{beck08}, of the fifteen sources surveyed for extended H$_2$ emission, eight systems show strong evidence for H$_2$ in extended outflows.  Four of these exhibit resolved kinematic structure beyond the 30km/s instrumental limits: RW Aur, T Tau, UY Aur and V773 Tau.  We plotted mass accretion rate, age and stellar mass versus H$_2$ emission luminosity for the \cite{beck08} sample of six stars driving HH outflows, as in Figure~13.  We find no apparent correlations.  The observed anti-correlation between X-ray luminosity and H$_2$ emission luminosity shown in Figure~13a is not as strong if the six targets from \cite{beck08} are included in the analysis.  This is perhaps because several of the targets in that survey were very young ($<\sim$1 Myr) and/or highly obscured, which limits measurement of the X-ray luminosity.



\subsection{H$_2$ in the Environments of Young Multiple Stars} 

Stars in the GG Tau A, UY Aur, and V773 Tau young star multiple systems with separations greater than 0.$\arcsec$2 were spatially resolved by our study (Figure~3).  DoAr 21 is a tight binary system; it was resolved into two stars in 3.6 cm radio continuum emission with a separation of 5mas (0.6AU; \cite{loin08}).   For the case of the sub-AU separation for DoAr 21, heating from two stars in a central binary should increase the detected H$_2$ disk emission luminosity and extent compared to a single central star of equivalent spectral type.  Interestingly, DoAr 21 exhibits no compact and central H$_2$ emission.  It is the only system in our sample with appreciable H$_2$, but no evidence for an inner flux distribution suggesting emission from a central disk component.  In fact, there appears to be little to no H$_2$ at any location within a $\sim$60-70AU distance from DoAr21.  All of the H$_2$ emission arises from the extended distribution spanning a 120$\deg$ arc from the north to southwest of the central DoAr 21 binary system (Figure~4).

GG Tau A and UY Aur show H$_2$ morphologies with arcs of emission that extend away from inner stars in the systems (Figure~14a and 15a).  As discussed in the previous section, the arcs encompassing UY Aur are seen in between stars and are likely shock excited by outflows and possibly interacting winds.  We detect two arcs of emission (yellow and cyan in Figure~14a) that are interpreted to arise from the extended redshfited outflow from both UY Aur A and B based on the [Fe~II] kinematics and geometry found by \cite{pyo14}.  The yellow arc is seen to arise from within $<$0.$\arcsec$15 of UY Aur A, at a location where significantly redshifted outflow emission should be obscured by the optically thick inner dust disk.  \cite{tang14} finds highly complex dynamics of the CO gas surrounding the UY Aur binary within the circumbinary ring cavity from interaction of material in Keplerian rotation with gas tracing both the mass accretion infall and outflow.  Their study does not clearly resolve the gas in the inner $\sim$1$\arcsec$ of UY Aur A.  While we find that shock excitation in the wide-angle redshifted outflows is the most likely explanation for these two arcs of emission, alternative excitation of gas in mass accretion infall shocks cannot be firmly excluded based only on the morphology of the H$_2$.  Streamers of infalling material would also appear to be slightly redshifted at a $\sim$20-30~km/s level \citep{pyo14}, and might be observable at $<$0.$\arcsec$15 distances from UY Aur A in the foreground of the circumstellar dust disk.

In the case of the complex sub-arcsecond triple system GG Tau A \cite{difo14}, the H$_2$ emission arcs encompass the stars in this young system (Figure~15a).  The brightest region of H$_2$ emission in the environment of GG Tau A is $\sim$40AU to the northeast of the stars.  This H$_2$ emission has an estimated LTE excitation temperature of $\sim$1700K, estimated from the detected level of {\it v}=2-1 S(1) line emission.  Based on models for H$_2$ excitation in circumstellar disks (Nomura et al 2007), at a distance of $\sim$40AU from the stars, this temperature is on the high side to be excited only by flux from the central stars.  However, this region of strongest H$_2$ gas emission is not in a disk, it is located in a dynamically unstable region that must be continuously replenished or it will be cleared on $\sim$100~year timescales.  Streamers of infalling material that transfer mass from a circum-system disk to the inner stars are predicted by hydrodynamical theories of binary star evolution \citep{arty96}.  Streams of CO gas that connect the outer GG Tau A circum-system ring to the inner H$_2$ gas distribution have been measured \citep{dutr14}.   GG Tau A is a complex triple system, and \cite{beck12} postulated that this extended distribution of H$_2$ emission might also be excited in part by low velocity shocks from accretion infall as material streams inward from the massive circumsystem ring that surrounds the stars.   The accretion infall shock velocity would result from a combination of the gas free-fall velocity onto the system ($\sim$~15~km/s ) with a component from the Keplerian motion of material in the inner system.  The presence of the sub-40~AU triple system would further complicate gas kinematics of the regions surrounding the stars.  Shock velocities in the lower 20 to 30~km/s range of models \citep{lebo02} could explain the H$_2$ in the GG Tau A system \cite{beck12}, but these velocities are quite large compared to infall+Keplerian kinematic measurements of cooler denser circumbinary disk gas (e.g., \cite{dutr14,dutr16,tang14}).   Figure~15a highlights in green the location of two of the obvious extended arcs of emission that encompass the GG Tau A triple star, the eastern arc traces the inner edge of the CO streamer \cite{dutr16}.

\begin{figure}[ht!]
\plotone{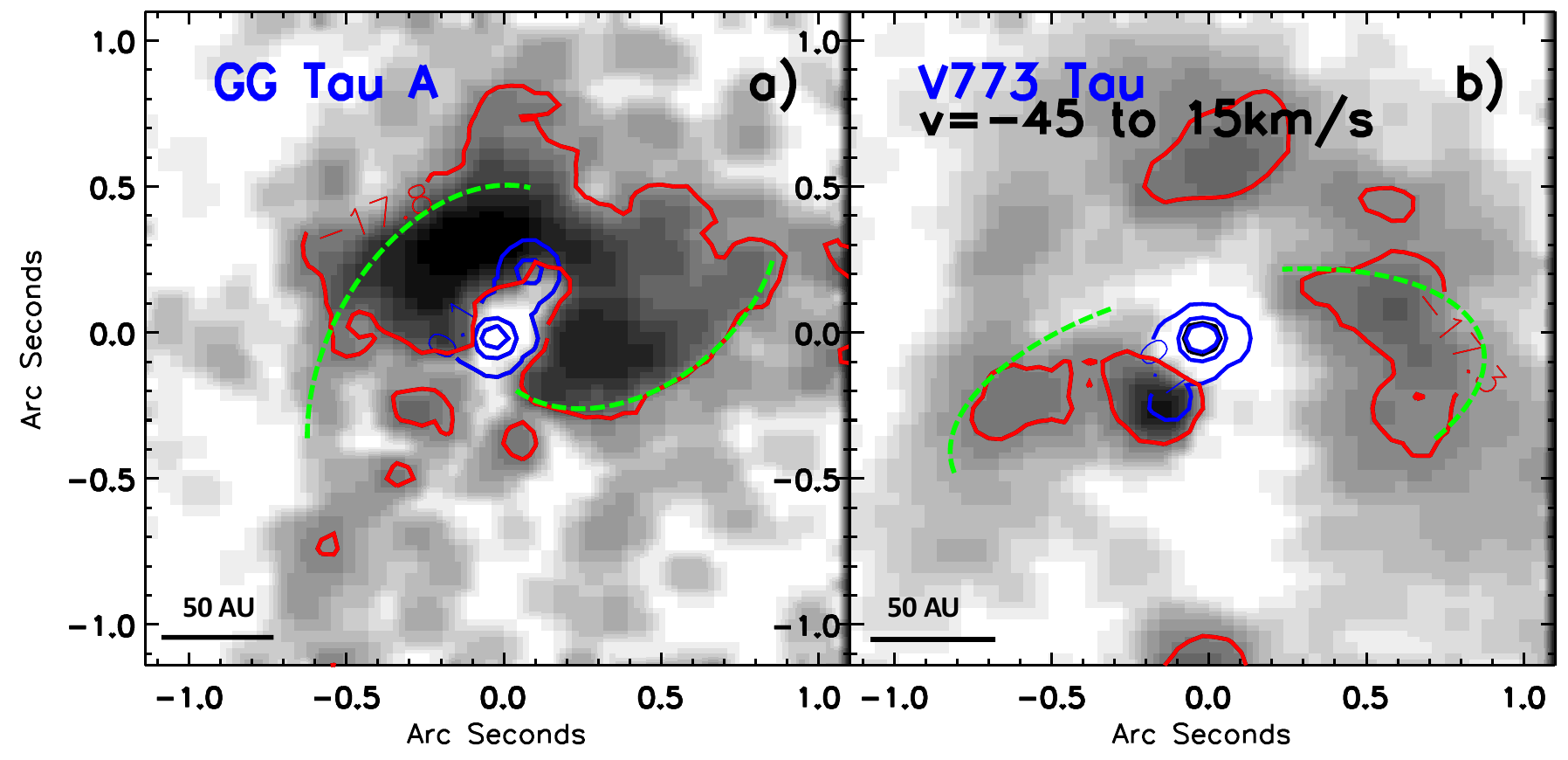}
\caption{The integrated 2.12$\mu$m ro-vibrational H$_2$ line emission from the young multiple systems GG Tau A (a) and V773 Tau (b).  Stellar positions are highlighted in blue contours that trace the 10\%, 40\% and 70\% flux levels with respect to the peak continuum emission.  Both displays are scaled logarithmically from 1\% to 75\% of the peak H$_2$, and each image shows one red contour that presents the 10\% H$_2$ flux level with respect to the peak line emission.  The green lines trace the positions of arc-like structures seen in the H$_2$ morphologies that encompass the stars in these young multiple systems.    \label{fig:H2imgmult}  }
\end{figure}

V773 Tau is a putative young sub-arcsecond quintuple \cite{ghez93, lein93, duch03, bode12}.  V773 Tau A was found to be a double-lined optical spectroscopic binary (V773 Tau Aa+Ab SB2; \cite{welt95}).  Orbital monitoring and dynamical mass fits of nearby V773 Tau B suggest that it is also an $\sim$AU-scale binary (V773 Tau Ba+Bb; \cite{bode12}).  In our continuum image of V773 Tau in Figure~3, The V773 Tau A+B quadruple system is the spatially unresolved brighter point-like component.  \cite{bode12} showed that the A+B separation at the time of our 2009 observation was $\sim$50milli-arcseconds, and so it was below our spatial resolution limit.  The 0.$\arcsec$25 separation companion that we do spatially resolve in Figure~3 is V773 Tau C (designated "D" in the discovery paper by \cite{duch03} but later renamed by \cite{bode12}).  V773 Tau C is an infrared luminous companion, it is the brightest star in the system at wavelengths long-ward of 4$\mu$m \cite{duch03}.  Figure~15b shows the image of the low velocity (v=-45km/s - 15km/s) H$_2$ emission from V773 Tau.   The brightest region of H$_2$ emission is seen at the position of V773 Tau C.  The blue and redshifted knots in the bi-polar outflow are seen to the north and south of the stars.  The two green lines in Figure~15b trace distributions of brighter H$_2$ emission arcs that are near the stellar rest velocity.  These arcs seem to partially encompass the stars in the V773 Tau system in a direction that is perpendicular to the outflow knots. The morphology of the two arcs resembles a circumsystem distribution of material that encircles the stars from the east to the west.  However, it is unfortunately not possible to discern from the image if these two distributions of H$_2$ emission are related or independent, or what the excitation mechanism might be. 

When combined with the six sources from \cite{beck08}, fifteen young stars have been surveyed with Gemini+NIFS for spatially resolved ro-vibrational molecular hydrogen.  Of these fifteen, seven are young single star systems (AA Tau, AB Aur, DG Tau, HL Tau GM Aur, LkCa 15 and LkH$\alpha$ 264) and eight are young multiple stars (DoAr 21, GG Tau A, HV Tau C, RW Aur, T Tau, UY Aur, XZ Tau and V773 Tau).  Figure~16 investigates the H$_2$ extent and brightness across the full sample of stars.    This shows the H$_2$ line luminosity versus this H$_2$ emission area measure for single stars (triangles), resolved multiple stars (diamonds), sub-AU multiples (asterisks), and the two high A$_v$ sources are also encompassed by a square.  The H$_2$ sky emission area used in Figure~16 shows the region of the spatial field (in AU$^2$) that exhibits greater than 5\% of the H$_2$ flux compared to peak line emission pixel position.  This area measure is calibrated for distance in the stellar frame (Table~7), and is not affected by the signal-to-noise of the observation because all measurements were sensitive to less than 5\% of the peak flux.   Overplotted as dash-dotted lines are the averages and ranges for single stars and multiples; the lines cross at the average position and the range is defined by the span of the triangle and diamond points, respectively.  AB Aur, HL Tau and HV Tau C were omitted for the average and ranges because of the occulting spot used for the observations (AB Aur, HL Tau) and the high source extinction (HL Tau, HV Tau C) preferentially increases sensitivity to extended H$_2$ emission.  Figure~16 shows that the average H$_2$ emission area is $\sim$3 times greater in the young multiples versus single stars, and the average  {\it v}=1-0 line luminosity is nearly a factor of ten higher.  DoAr 21 has an H$_2$ line luminosity and spatial extent that is similar to the measured singles rather than the multiple systems.  

The young star multiple systems surveyed thus far have stronger H$_2$ emission line luminosity over a greater overall spatial area than the single star systems.  This is naturally expected for spatially resolved binaries or higher order multiples because the two (or more) disk+outflows have greater combined flux and span a larger overall area on the sky compared to single star systems (e.g., Figure~1).  Interpretation of results for GG Tau A and UY Aur suggest that at least two additional H$_2$ gas excitation mechanisms may exist solely in young multiple systems:  shock excitation from system dynamics and mass accretion infall (GG Tau A), and shocked arcs from interacting jets and wide-angle outflows in slightly mis-aligned systems (UY Aur).

\begin{figure}[ht!]
\plotone{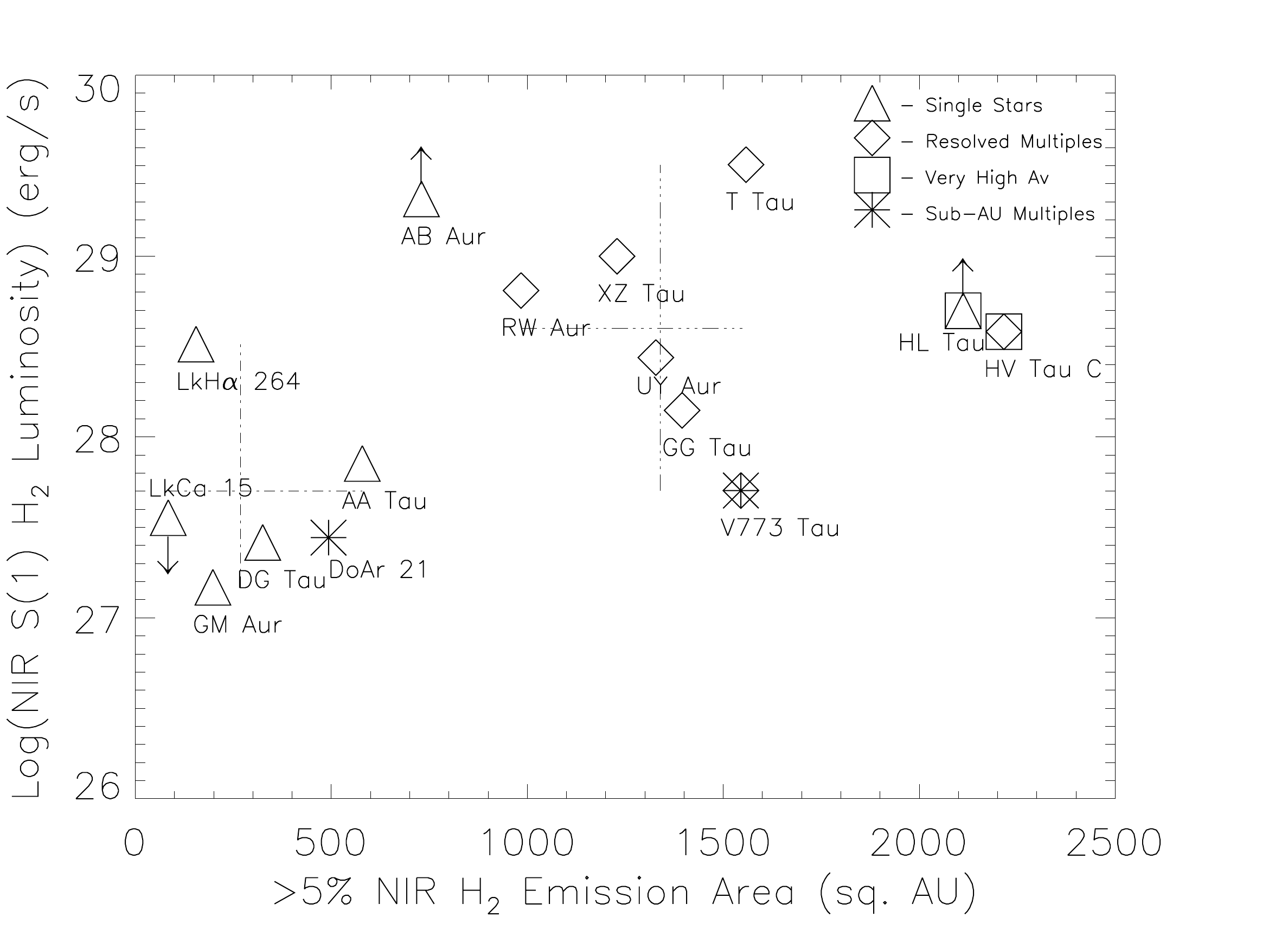}
\caption{The integrated (log) 2.12$\mu$m ro-vibrational H$_2$ line luminosity plotted versus the area on the sky in AU$^2$ that exhibits H$_2$ flux at a level greater than 5\% of the measured peak.  Single stars are plotted as triangles, Multiple star systems are diamonds, and the two very high visual extinction systems are encompassed by squares.  Measurement of the H$_2$ luminosity in AB Aur and HL Tau are lower limits because of the occulting spot used for these observations.  The H$_2$ luminosity plotted for LkCa 15 is the 3$\sigma$ detection limit.  \label{fig:xrayH2}}
\end{figure}


\subsection{Ro-vibrational H$_2$ from the Disks of Young Stars} 

Of the nine stars surveyed for this project, we find that LkCa 15 and DoAr 21 exhibit no central H$_2$ emission associated with the stellar point source constraining molecular gas in the inner disks.  All other systems have appreciable central H$_2$ flux that likely has some contribution from disk gas excited by central emission from the star(s).   AA Tau, GG Tau A, UY Aur and V773 Tau also have shocks excitation mechanisms stimulating spatially extended H$_2$, which makes line ratio and spatial morphology analysis of the central disk emission difficult.  However, AB Aur, GM Aur and LkH$\alpha$ 264 exhibit near-IR H$_2$ that is most consistent with a pure disk origin.  Emission in these systems is characterized by spatially compact and centrally dominated line flux with no measurable kinematic structure, and no evidence for extended discrete knots or high LTE gas temperatures suggesting shock excitation or an outflow origin.  

Figure~17 shows the log-scaled continuum subtracted H$_2$ images of GM Aur (a), AB Aur (b) and LkH$\alpha$ 264 (c).  The cyan ellipse in the GM Aur view (a) presents the orientation of circumstellar disk material sampled by the 0.9mm dust continuum emission, shown here is the location of the 50$\times$ rms dust contour from Figure~1a of \cite{maci18}.     The near-IR H$_2$ emission from GM Aur is spatially compact, the majority of the flux arises from within 0.$\arcsec$2 ($\sim$32~AU) of the central star.  GM Aur has an inner cavity in the dust disk within 35~AU \citep{hugh09,maci18}, and the H$_2$ gas fills this region.  The low level H$_2$ from GM Aur shows non-axially symmetric structures with a slight overall north~-~south extension.  

Two green lines overplotted on the log-scaled H$_2$ image of AB Aur (Figure~17b) highlight the position of the CO gas spirals measured by \cite{tang17}.  These spirals exist within a dust cleared ring that encircles AB Aur at a distance of $\sim$120~AU.  The $^{12}$CO 2-1 emission spirals trace optically thick cool disk material at a temperature of $\sim$140K \citep{tang17}.  We see spatially coincident H$_2$ emission that traces warmer gas in the lower density upper layers of the spirals.  Hence, the spiral-shaped density enhancements seen around AB Aur exist in multiple gas species that trace a range of scale heights in the disk.  Moreover, the low level H$_2$ flux seen around AB Aur also shows arc-like spiral extensions (e.g., to the north in Figure~17b), tracing sculpted gas at greater distances than revealed in the CO.  \cite{tang17} postulate that the spiral arms in the dust-cleared AB Aur disk cavity might be shaped by tidal disturbances in the gas imparted by two planets at separations of $\sim$30~AU and $\sim$60-80~AU.

The disk emission from LkH$\alpha$ 264 is spatially compact, and like GM Aur the majority of the flux arises from within 0.$\arcsec$2 ($\sim$32~AU) of the central star.  The overall emission is slightly asymmetric toward the west, two linear extensions of H$_2$ emission are seen to $\sim$0.$\arcsec$6 (150~AU) the west of the star.  The origin of these extensions is not clear, the H$_2$ shows no obvious kinematic characteristics of shock-excited gas \citep{itoh03, carm08b} and LkH$\alpha$ 264 is not known to have an extended inner jet.

\begin{figure}[ht!]
\plotone{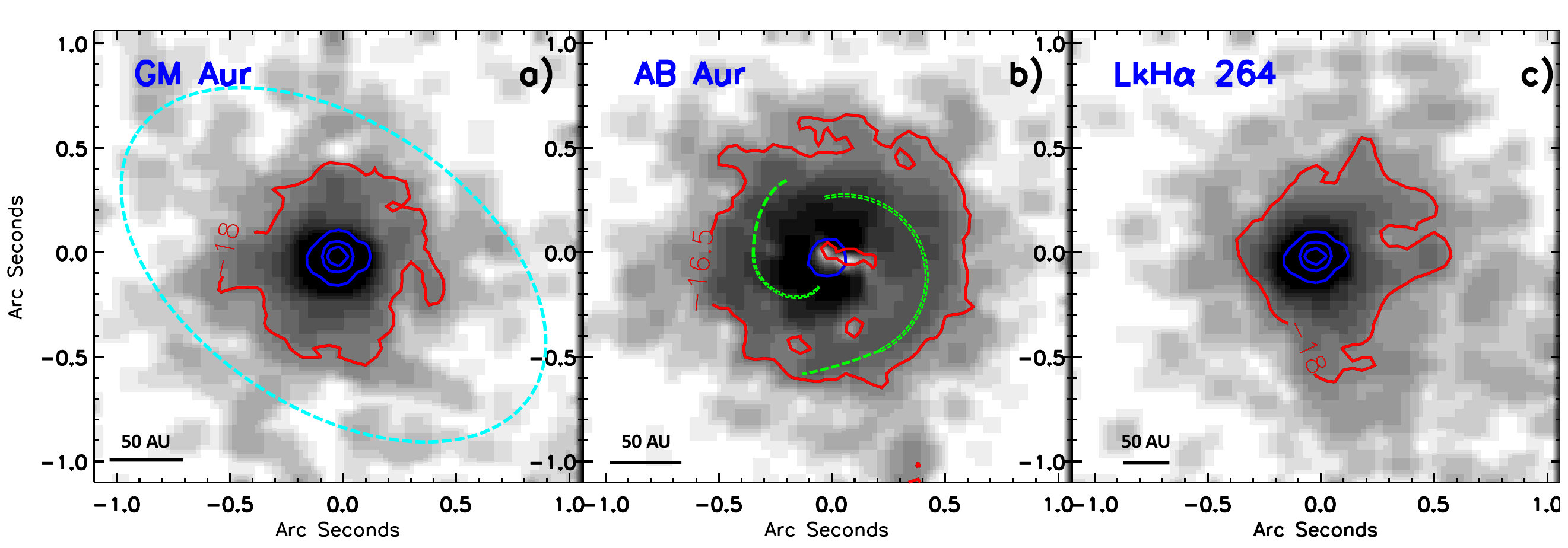}
\caption{The continuum subtracted H$_2$ emission from the disks of GM Aur (a), AB Aur (b) and LkH$\alpha$ 264 (c).   The image displays are scaled logarithmically from 1\% to 75\% of the peak H$_2$ line emission.  GM Aur and LkH$\alpha$ 264 have three blue contours over-plotted that designate the 10\%, 40\% and 70\% continuum flux levels, and one blue contour is included for AB Aur show the position of the coronagraphic occulting spot used for the observation.  Each image shows one red contour that presents the 10\% H$_2$ flux level with respect to the peak line emission.   Overplotted in cyan in (a) is an ellipse that traces the position and orientation of the 50$\times$ rms contour of 0.9mm dust disk emission from Figure~1a of \cite{maci18}.   The green arcs on the image of AB Aur (b) show the approximate position of the two inner disk spiral arms traced in CO emission by \cite{tang17}, which are also seen here in the H$_2$ emission map (and in Figure~7).  \label{fig:diskH2l}}
\end{figure}



\begin{figure}[ht!]
\plotone{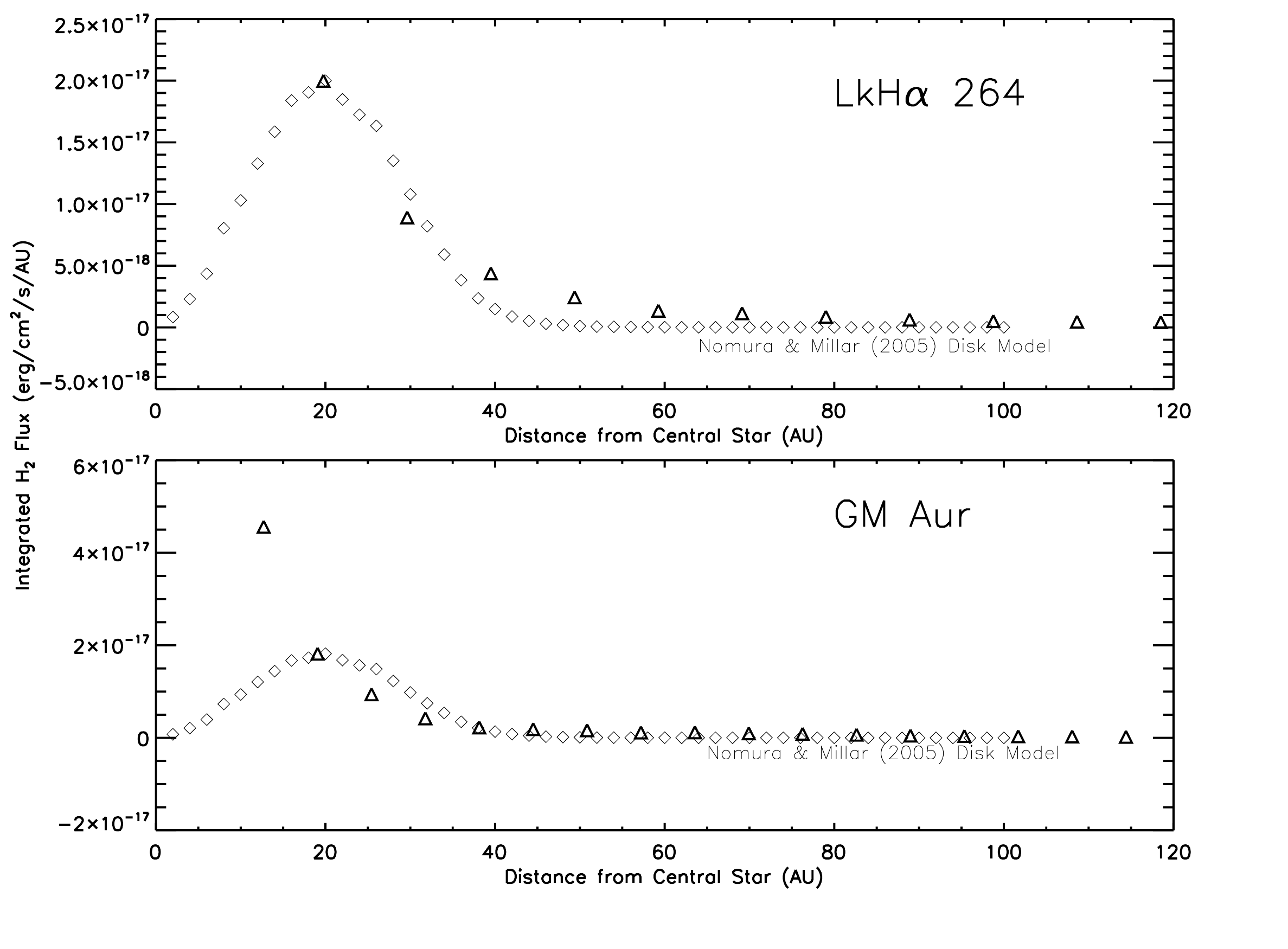}
\caption{The ro-vibrational {\it v} = 1-0 S(1) flux plotted versus distance from the central star for GM Aur and LkH$\alpha$ 264.  Overplotted is the disk model from \cite{nomu05}, scaled to the observed system flux at 20AU distances from the stars. \label{fig:h2models}}
\end{figure}

\begin{figure}[ht!]
\plotone{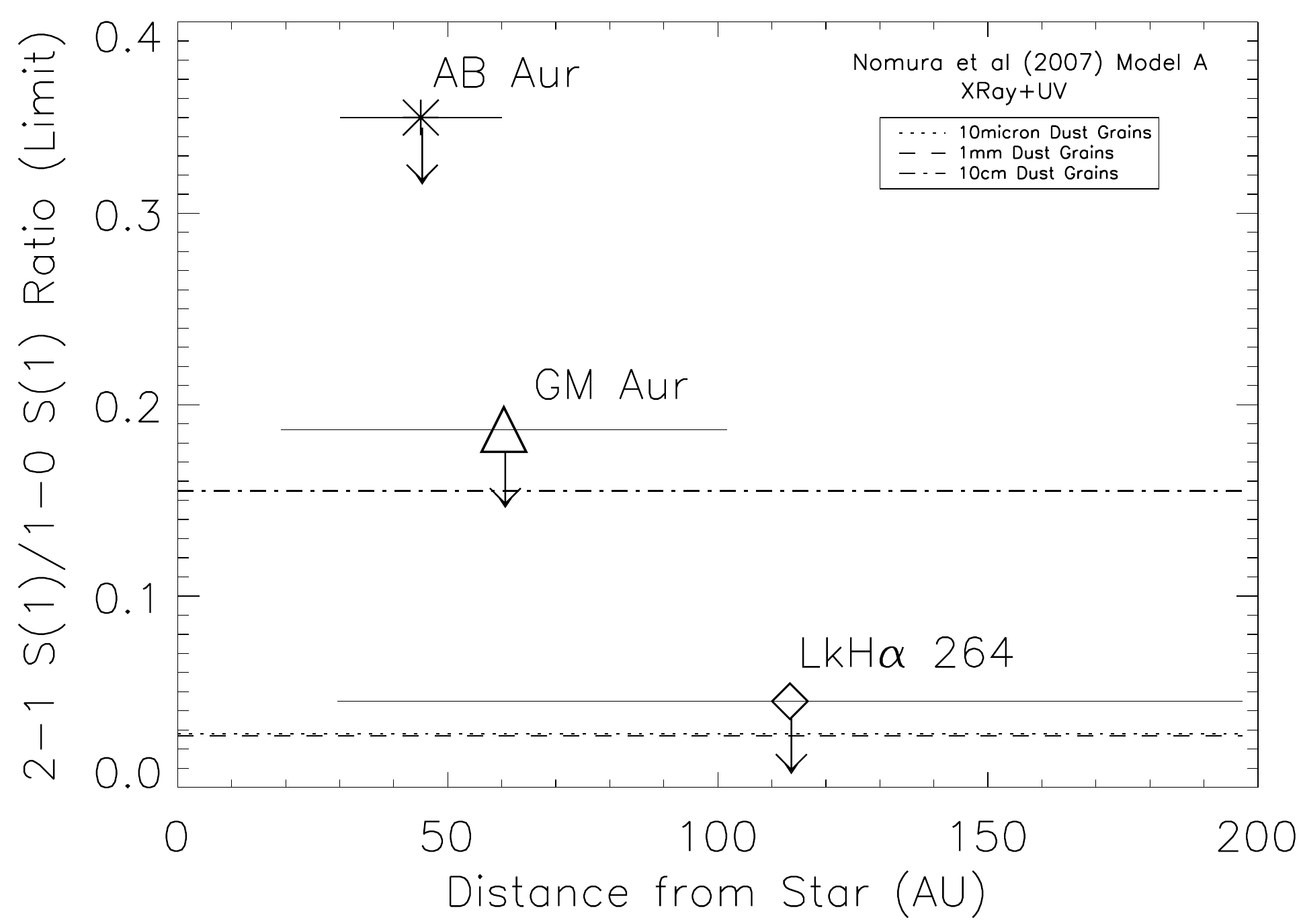}
\caption{The line ratio limits of ro-vibrational {\it v} = 2-1 S(1) / {\it v} = 1-0 S(1) flux plotted versus distance from the central star for GM Aur, AB Aur and LkH$\alpha$ 264.  Overplotted are the three X-ray+UV "Model A" disk models for 10$\mu$m, 1mm and 10cm sized dust grains from \cite{nomu07}, showing how dust evolution and grain growth may alter the line ratios measured in the disks.  \label{fig:h2dustevolution}}
\end{figure}

\cite{nomu05} constructed a self-consistent density and temperature model that accurately reproduces molecular hydrogen emission in a young star disk that has strong UV excess radiation.  The model was constructed for a generic T Tauri star of 0.5M$_{\odot}$, a radius of 2.0 R$_{\odot}$ and a temperature of 4000K.   Their Figure~10 presents the {\it v} = 1-0 S(1) flux model calculated as a function of distance from 0.1 to 100~AU from the central star.  Figure~18 plots the average measured flux as a function of distance from GM Aur and LkH$\alpha$ 264, with the \cite{nomu05} model scaled to the emission level at 20~AU and overplotted.  \cite{nomu05} find that the physical and chemical structure of protoplanetary disks is affected by the dust, and that H$_2$ emission in the central region peaks at a $\sim$20~AU distance from the star.  We do not see an inner decline in H$_2$ emission within 20~AU in the disk of GM Aur, and LkH$\alpha$ 264 is too distant to resolve the inner 20~AU region.  Beyond 20~AU, the average radial H$_2$ flux structure in GM Aur and LkH$\alpha$ 264 closely resembles the \cite{nomu05} model.  It should be noted that the inner H$_2$ emission from GM Aur arises from within the dust cleared inner disk, and the \cite{nomu05} model may not be directly applicable inside the cavity.  \cite{hoad15} found that UV H$_2$ emission in transition disk systems, such as GM Aur, extended to four times greater radii than non-transition disks, and that all of the warm H$_2$ came from within the inner dust cleared cavity.  We see ro-vibrational H$_2$ that extends beyond the $\sim$40AU radius inner dust cleared ring of GM Aur \cite{maci18}, confirming that the near-IR H$_2$ traces gas to larger radii than the UV transitions.

Expanding upon the work of \cite{nomu05}, \cite{nomu07} updated the H$_2$ disk models to include effects of disk evolution and dust grain growth.  As the dust particles evolve in protoplanetary disks, the gas surface temperature drops because grain photoelectric heating becomes inefficient.  The effect of this grain growth manifests itself in an increase in non-thermal pumping versus the collisionally dominated gas in LTE, and this can be measured as an increase in the {\it v} = 2-1 S(1) / {\it v} = 1-0 S(1) near-IR H$_2$ line ratio.  Particularly, models with significant grain growth (to 10 cm) are traced by near-IR H$_2$ gas line ratios of up to 0.15 \cite{nomu07}.  We do not detect the {\it v} = 2-1 S(1) emission in GM Aur, AB Aur or LkH$\alpha$ 264, so Figure~19 plots the {\it v} = 2-1 S(1) / {\it v} = 1-0 S(1) line ratio limit for these three systems from the integrated disk flux apertures (shown on the x axis).  Overplotted are the line ratio calculations for \cite{nomu07} "Model A" models that incorporate X-Ray plus UV flux for small (10$\mu$m) dust grains, medium (1mm) and large (10cm) dust.  The observe line ratio limits rule out significant grain growth to 10cm grains for the LkH$\alpha$ 264 system.  The detection limits on the {\it v} = 2-1 S(1) line flux for GM Aur and AB Aur are not restrictive enough to measure any grain growth in the inner dust cleared regions of these systems where the effect might be most prominent.  Additionally, an increased {\it v} = 2-1 S(1) / {\it v} = 1-0 S(1) line ratio in the inner regions could also be mis-interpreted as shock excited emission, such as from an inner disk wind, unless the ratio is greater than $\sim$0.25 and more indicative of a non-thermal origin.

The  {\it v} = 2-1 S(1) / {\it v} = 1-0 S(1) line ratio of 0.046 measured in the 30-195~AU aperture for the disk of LkH$\alpha$ 264 places a temperature upper bound of 1650K for the LTE gas.  Prior studies have estimated molecular hydrogen disk masses under the assumption that the bulk of the ro-vibrational H$_2$ emission arises from a collisionally dominated region with an average gas temperature of $\sim$1500K \citep{bary03, itoh03, carm08b, bary08}.  Our results and line ratio limits are consistent with these past assumptions.  However, proper calibration of the spatially resolved molecular hydrogen disk mass structure in GM Aur, AB Aur and LkH$\alpha$ 264 would benefit from treatment with an updated 3-dimensional model of these disks.  Particularly, the H$_2$ emission from the inner disk regions of GM Aur and AB Aur arises from dust cleared cavities, and existing models use dust as a dominant disk opacity source that affects the inner radiation field and characteristics of the emitting gas.  Further modeling of the H$_2$ in these disks a topic for a follow up project.

\subsection{The Individual Systems} 

\subsubsection{AA Tau}  

AA Tau is a single spectral type M0 star at 136.7pc distance in the Taurus-Aurigae association.  It is a prototype for large scale periodic photometric variability caused by irregular warped material at the inner edge of a circumstellar disk \citep{bouv99}; the so-called "dipper" systems.  The inclination of the inner warm material and scattered light disk component has been modeled to have a close to edge-on viewing geometry of 71 - 75$\deg$ \citep{osul05,cox13}.   ALMA observations have revealed three concentric disk rings inclined at 59$^{\circ}$ encompassing the inner disk, with non-axisymmetric characteristics that may originate from gap-crossing accretion streams responsible for the observed optical photometric variability of AA Tau \citep{loom17}.  High contrast HST observations of AA Tau revealed an inner scattered light disk component, and extended emission from an inner jet oriented at $\sim$195$\deg$ east of north \citep{cox13}.

Our high resolution spectrum presented in Figure~2 reveals the {\it v}=1-0 S(1) H$_2$ line emission measured in AA Tau.  The inner hot disk of AA Tau exhibits known UV H$_2$ emission \citep{fran12}, and the cooler outer disk has mid-IR H$_2$ \citep{carr11}.  \cite{fran12} observed the UV H$_2$ from AA Tau using HST + COS and found Gaussian emission profile centered and the systemic velocity and fit by a rotating inner gas disk with radius of 0.69AU.  Our high resolution  detection of H$_2$ from AA Tau reveals an average integrated spectral profile that is blue-shifted (e.g., Figure~2) compared to the systemic radial velocity of +16.9km/s \cite{nguy12}.   The NIFS full-field spectrum (Figure~9) is consistent with this measurement, although the NIFS IFU H$_2$ emission from AA Tau is spectrally unresolved.  Our observations of AA Tau were acquired prior to the large-scale photometric dimming that AA Tau exhibited in the early 2010's \cite{bouv13}.  The morphology of the measured H$_2$ from AA Tau shows two obvious spatial components: 1) an extension and knot toward the southwest of the star at a PA of $\sim$195$\deg$ east of north from the star and 2) a linear extension to the east-southeast at a PA of $\sim$100$\deg$ east of north.  AA Tau has a knot of shock excited ro-vibrational H2 from its known jet, and also inner H$_2$ near the blue-shifted inclined inner disk locations.  The asymmetric eastern morphology of the measured inner H$_2$ emission may mean that some special viewing angle of the inclined outer disk may affect the extended emission character of the inner H$_2$.



\subsubsection{AB Aur}  

AB Aur is a spectral type A1 single star in the Taurus-Aurigae association.  At 144pc distance, AB Aur is one of the closest Herbig Ae stars.  As a result, it has been a prime laboratory for sensitive, high spatial resolution observations of the gas and dust in its protoplanetary disk.  Successive observations with increasing spatial resolution have found an inner dust disk ($\sim$11 AU), a cleared cavity, and an outer dust ring with radius 120AU (\cite{piet05, tang12, tang17}).  Spiral dust structures have been seen in the outer disk to $\sim$500AU in optical and near-IR scattered light \citep{hash11,grad99,fuka04,perr09} and large-scale CO spirals, extending beyond the dust disk ring have also been reported \citep{lin06,tang12}.  High resolution ALMA maps of $^{12}$CO 2-1 emission revealed gas inside the dust cavity at $<$0.\arcsec5 spatial scales, with morphology indicative of two inner spiral arms.  These structures are interpreted by \cite{tang17} as arising from potential proto-planets located inside the dust disk, shaping the inner gas into the observed spiral structures.

Figure~8 presented our optimal analysis to detect the {\it v}=1-0 S(1) line emission from AB Aur.  Figure~13b presents the same H$_2$ image with a logarithmic scaling from 1\% to 70\% of the peak line emission flux to bring out structure in the low level extended line emission.  The red contour shows the 10\% continuum subtracted H$_2$ flux level, and the three blue contours trace the level of 10\%, 40\% and 70\% of the peak measured continuum flux.  Overplotted in green in Figure~17b is the location of the inner CO gas and scattered light dust spirals measured by {tang17} and {hash11}, respectively.  Enhancements in H$_2$ line strength follow the curves traced by these known inner spiral arm structures.  This is also clearly seen in the H$_2$ S/N color image in Figure~8d; the green inner regions of S/N$\sim$5 trace these two roughly spiral patterns.


\subsubsection{DoAr 21}  

DoAr 21 is a young spectral type G1 sub-AU binary star \citep{loin08} at 133.8pc distance in the in the $\rho$ Ophiuchus star forming region.  DoAr 21 possesses a narrow 2.12~$\mu$m H$_2$ emission feature, which Bary et al.\ (2003,2008) model as arising from gas stimulated by high-energy photons and confined to a central circumstellar disk.  \cite{jens09} studied the infrared excess, polycyclic aromatic hydrocarbon (PAH), and X-ray emission from DoAr~21 and found little evidence for emission from dust and gas at radii less than 100~AU.  They measure strong PAH emission from an emission arc at large distances.  Based on the lack of central dust emission and the extended nature of the PAHs, these authors suggest that the previously measured H$_2$ is likewise extended and arises from the same spatial region, rather than from an inner circumstallar disk.  

We find that the the all of H$_2$ emission from DoAr 21 is spatially extended, and arises entirely from a $\sim$130$^{\circ}$ arc that extends from 60-70 AU North to 110-120 AU southwest of the central star.  This measured H$2$ arc is coincident with the mid-infrared and PAH emission measured by \cite{jens09}.  Neither the integrated full field spectrum nor the small aperture peak H$_2$ spectrum (Figure~9) measurements were successful at detecting the higher vibration level line of  {\it v}=2-1 S(1) (2.24$\mu$m) emission, which is often used as a gas excitation diagnostic.  \cite{jens09} postulated that DoAr 21 might have a high 2-1 S(1)/ 1-0 S(1) line ratio with a non-thermally pumped origin because of the strong stellar UV flux, high energy flares, and detected PAH emission.  To increase our sensitivity to the {\it v}=2-1 S(1) (2.24$\mu$m) emission line (beyond the limit presented in Table~5), a 1-D spectrum was extracted using a tailored shaped linear aperture that extended along the spatial arc of H2$_2$ from the north to the west of DoAr 21.  The goal was to sum as many spatial elements with strong {\it v}=1-0 S(1) flux as possible to provide a more stringent limit on the detection of the {\it v}=2-1 S(1) transition.  In this summed spectrum, no emission from the 2-1 S(1) line was seen, to a 3$\sigma$ detection limit equivalent of a {\it v}=2-1 S(1)  / {\it v}=1-0 S(1) line ratio of 0.03.  Thus, assuming the H$_2$ population traces denser gas in LTE, the temperature must be less than $\sim$1450K based on the non-detection of 2-1 S(1) flux. 

If stellar UV and X-ray heating of moderate density gas stimulates the H$_2$ at extended $\sim$70+ AU distances from DoAr 21, then the LTE gas temperature would be significantly below our measured line ratio limit.  With a {\it v}=2-1 S(1) / {\it v}=1-0 S(1) line ratio of 0.03, FUV pumping of low density gas is not the primary excitation mechanism of the H$_2$.   DoAr 21 is the only system in our survey sample that exhibits appreciable H$_2$ emission, but has no measurable line flux associated with the central stellar position.  Moreover, ALMA dust measurements of DoAr 21 detected no continuum emission from a dust disk in this system (referred to as $\rho$ Oph 6; \cite{cox17}).   \cite{jens09} mentioned that the extended dust and PAH flux flux might be illumination of nearby cloud material by DoAr 21, rather than material associated with the system in a circumbinary disk distribution.  Our results agree that heating of ambient local cloud material is a possible origin for the H$_2$ seen around DoAr 21.

\subsubsection{GG Tau A} 

GG Tau A is a sub-arcsecond triple system in the Taurus-Aurigae association (ensemble average distance of 140pc; Table~7).  It is comprised of the M0 primay star GG Tau Aa and the nearly equal mass Ab1+Ab2 binary \citep{whit99,difo14}.  GG Tau A is famous for it's spectacular and massive circumsystem ring of material that encompasses the three stars.  A distribution of dust has been seen near the north of the stars in the scattered light maps of GG Tau A, revealing that material exists at locations that should be dynamically cleared in this complex multiple system \citep{kris02,duch04}.  The GG Tau system also includes GG Tau Ba and Bb at a $\sim$10$\arcsec$ projected separation to the south \citep{whit99}.  Thus, GG Tau is a very young quintuple star system.  The GG Tau Bb component has a very late spectral type, making it a likely brown dwarf.

The {\it v} = 1-0 S(1) line emission was measured in GG Tau A by \cite{bary03} with a flux of 6.9$\times$10$^{-15}$\ergcms.  They interpreted the mission as arising from gas in the inner disks from one or more of the stars.  \cite{beck12} presented the H$_2$ maps from Figure~4 of this study and showed that the extended H$_2$ emission is strongest from the northestern region $\sim$40AU away from the stars.  Since the time of that publication, improved maps of the dust and CO gas from ALMA have revealed that the region of brightest H$_2$ is colocated with the CO streamer connecting the outer circumsystem disk and the inner distribution of dust \citep{dutr16}.  This H$_2$ emission may be excited through stellar heating from the young stars, though the 1700K LTE gas temperature measured from the H$_2$ line ratios is high for this distance from the stars.  \cite{beck12} postulated that the streamers of gas in H$_2$ could be excited through shocks associated with infalling accreting material in this dynamically complex young triple.



\subsubsection{GM Aur} 

GM Aur is a spectral type K5 star at 158.9 pc distance in the Taurus-Aurigae association \citep{espa11}.  It is a system in transition from the CTTS phase with a dense inner disk to the more evolved and X-ray luminous WTTS phase.  GM Aur has a $\sim$35 AU cavity in the optically thick dust in the inner disk, though modeling of it's SED and strong mass accretion suggest that gas and small particles persist within this cavity region.   The hot inner H$_2$ disk gas was detected in the UV electronic transitions \citep{fran12}, with an emission profile showing an origin within the inner disk region.  

Our high resolution spectrum presented in Figure~2 reveals the {\it v}=1-0 S(1) H$_2$ line emission measured in GM Aur.  This served as the basis for the follow up spectral imaging, where we spatially resolve the disk-like H$_2$ line emission to $\sim$70AU distances from the star.  The emission is strongest at the position of the star, and the $\sim$40~AU dust-cleared inner disk cavity \citep{maci18} has significant H$_2$ gas within it.  The H$_2$ distribution declines as a function of distance, in a manner consistent with the models of \cite{nomu05}, though the models were generated for an optically thick inner dust disk. \cite{maci16} measured a radio jet perpendicular to the disk of GM Aur.  The measured H$_2$ is consistent with what is expected from a disk origin, but it could also conceivably arise from this inner outflow; the H$_2$ does exhibit a slight north~-~south extension in the direction of the jet and our limits on the 2 - 1 S(1)/ 1-0 S(1) line ratio do not rule out shock excitation.   \cite{maci16} also found a spatially continuous photoevaporative wind component to the disk from GM Aur, so the measured H$_2$ from GM Aur may also arise from this region.  However, the H$_2$ is not appreciably shifted from the stellar velocity (Figure~2), as might be expected from a photoevaporative disk wind. 




\subsubsection{LkCa 15: A Null H$_2$ Detection}  

The LkCa 15 system is a K5 spectral type single star at 158.1pc distance in the Taurus-Aurigae association.  It has a transition disk with a large ($\sim$50AU), cleared inner cavity.  The system has garnered recent attention because of the detection of potential proto-planets orbiting within $<$20A region of the cavity, inside the dust ring \citep{sall15,thal16}.  The cool gaseous disk was found to extend to 900AU from the star in CO \citep{piet07}.   The hot inner H$_2$ disk gas was detected in the UV electronic transitions \citep{fran12}, with an emission profile showing an origin within $<$3AU from the star.  

The {\it v}=1-0~S(1) (2.12$\mu$m) ro-vibrational H$_2$ emission was measured in LkCa 15 by \cite{bary03} with a flux of 1.7$\times$10$^{-15}$~\ergcms.   LkCa 15 is the only star out of our nine source sample that not only does not show spatially extended ro-vibrational H$_2$ (Figure~6), we do not detect the line emission at all.  Our 3$\sigma$ flux detection limit for the H$_2$ integrated across the full spatial field was 1.2$\times$10$^{-15}$~\ergcms in the 1-D spectrum extracted over the full IFU field (1.\arcsec12 aperture; Table~4).  Measurement of spatially extended H$_2$ is easier than detection of central emission over the bright continuum.   Figure~6 revealed that the encircled energy in the continuum subtracted H$_2$ image had no extended flux beyond the PSF shape.  Still, tests were carried out extracting 1D spectra over different aperture widths and spatial locations within the IFU field, measuring the line flux limit, and then the line spread function(LSF) was scaled and added as an emission line with the flux measured by \cite{bary03}.  From this analysis, detection of any ro-vibrational H$_2$ emission from LkCa 15 at the level measured by \cite{bary03} was not successful.  If the system still had warm H$_2$ gas emission at the level reported previously, we should have detected it over the bright continuum flux from the star, especially if it was spatially extended.  Our failure to detect the H$_2$ here indicates that the emission from LkCa 15 might be time variable.

\subsubsection{LkH$\alpha$ 264}  

LkH$\alpha$ 264 is a K5 spectral type single star at 246.7pc distance in the sparse MBM 12 association of young stars \citep{hear00}.  The {\it v} = 1-0 S(1) (2.12$\mu$m) ro-vibrational H$_2$ emission was measured in LkH$\alpha$ 264 by \citet{itoh03, carm08b}, the latter study reported a flux of 3.0$\times$10$^{-15}$\ergcms.  Both investigations found the line emission to be spatially compact with a very narrow kinematic profile, and they interpret the H$_2$ as arising from gas in the inner cirumstellar disk.  LkH$\alpha$ 264 has optical spectroscopic evidence of accretion driven outflow and wind activity \cite{game02} but no apparent knowledge of a spatially resolved or collimated jet.  We find LkH$\alpha$ 264 to have disk-like near-IR H$_2$ consistent with these prior results, and a slightly skewed spatial emission profile with resolved low level linear extensions that extend to the west of the system.  Overall, the radial distribution of the H$_2$ is consistent with prior disk models \citep{nomu05}.  The  {\it v} = 2-1 S(1) / {\it v} = 1-0 S(1) near-IR H$_2$ line ratio limit measured in LkH$\alpha$ 264 appears to rule out significant grain growth to 10~cm sized grains in the disk, from comparison with the models of \cite{nomu07}.

\subsubsection{UY Aur} 

UY Aur is an 0.$\arcsec$9 separation binary at 154.9 pc distance in the Taurus-Aurigae association.  The optical primary UY Aur A is an K7 spectral type star and UY Aur B is an infrared luminous companion (IRC; \citet{kore97}) characterized by optical faintness, a large bolometric luminosity, and significant historical flux variability.   Like GG Tau A, UY Aur is one of the few young star systems that has a spatially resolved circumbinary ring surrounding the central stars \citep{clos98, hiok07, tang14}.  The stars both have optically thick circumstellar disks that are resolved by ALMA \citep{tang14}.  \cite{berd10} and \cite{tang14} proposed additional companions in the system to UY Aur A and B, respectively.  However, additional stars have not yet been resolved.  The PSF shapes of the stars in all of our continuum spectral images are round, symmetric and compact ($<$0.$\arcsec$1).  In the continuum image of UY Aur (Figure~3), the B component appears to be elongated in to west-northwest direction.  We considered that we might be spatially resolving the putative $\sim$20~AU separation UY Aur B binary system, as proposed by \cite{tang14}.  However, \cite{witt17} presented adaptive optics imaging from the W. M. Keck Telescope that  had superior sensitivity to a $\sim$20~AU separation binary, but they failed to resolve the UY Aur B system.  As a result of their non-detection, we interpret the apparent shape of UY Aur B in Figure~3 as likely an instrumental effect since the PSF is elongated in the direction across the larger 0.$\arcsec$1 wide spatial slices in the IFU optics.

Molecular hydrogen emission from UY Aur B was first measured by \cite{herb95}, and it was studied in both stars by \cite{ston14}.  We find extensive ro-vibrational H$_2$ gas surrounding the UY Aur binary within the circumbinary ring, and interpret the origin to include emission from the disks and central collimated and wide-angle outflows.  See Figure~14 and \S4.1 for discussion and comparison with the geometrical results of \cite{pyo14}.  UY Aur exhibits three extended arcs of H$_2$ surrounding the inner stars.  Based on the known system geometry of UY Aur, we propose that H$_2$ excitation in binary and higher order multiple stars can also result from gas colliding in extended outflows from slightly mis-aligned disk+outflow systems.   

\subsubsection{V773 Tau} 

V773 Tau is a WTTS system at 127.7 pc distance in the Taurus-Aurigae association.   It is proposed to be a young sub-arcsecond quintuple \cite{ghez93, lein93, duch03, bode12}.  The V773 A+B system is a spatially unresolved quadruple system and the brightest component in our IFU spectral images in Figure~3.  The primary, V773 Tau A, is a spectroscopic binary with a measured spectral type of K1-K3 \citep{welt95, bode12}.  The companion that we resolve is designated V773 Tau C.  V773 Tau exhibited one of the strongest X-ray flares ever measured in a young star system, with a peak luminosity L$_X$$\sim$10$^{32.4}$~ergs/s \citep{skin97}.  With five stars located within a projected separation of less than 60~AU, the V773 system is dynamically complex and highly active. 

We included this young multiple star system in our survey in part because of it's nature as an infrared companion (IRC) multiple system \citep{kore97, duch03}.  All of the IRCs that have been surveyed for H$_2$ line emission exhibit appreciable flux with complex character (e.g. T Tau, XZ Tau, UY Aur \citep{herb95, herb07, beck08}).  The brightest 2.12$\mu$m {\it v}=1-0 S(1) H$_2$ emission we see from the environment of V773 Tau is compact and centered on the location of V773 Tau C.  It is likely that the H$_2$ arises from X-ray or UV heating of gas in the inner circumstellar disk around this star.  However, the dynamical complexity of the V773 Tau system means that the disk of the C component is likely truncated within less than $\sim$50AU of the star.  There is also clear velocity structure in extended H$_2$ shock-excited knots of emission from a bi-polar outflow oriented north-south.  These shocked knots are located $>$70~AU from the central stars, and it's not clear which component originates the outflow.  The V773 Tau system also has diffuse H$_2$ line emission that nearly fills the NIFS field of view (e.g. Figure~15b).  This extended halo distribution of H$_2$ might be stimulated in part by strong high energy X-ray flares from the system\citep{skin97}.  Hence, V773 Tau exhibits evidence for up to three different stimulation mechanisms exciting the H$_2$ gas at multiple spatially extended locations.  The measurement of ro-vibrational H$_2$ in V773 Tau emphasizes the fact that there may be more than one simple excitation mechanism exciting the molecular gas in the environments of young stars.

\subsection{Accessing the Data Presented in this Study} 

The IFU data presented in this study is complex.  Figure~4 presents a single analysis method across all T Tauri star systems, to investigate extended H$_2$ emission per spatial element in the data cube.   The 3-D IFU data structure allows for tailored, specific analysis methods to investigate the sensitivity to measured H$_2$ depending on characteristics of the system (e.g., Figure~8 for AB Aur).  There are a multitude of ways to analyze the IFU H$_2$ line emission data, and each young star system presented and discussed here is worthy of it's own full-length article that delves into significantly more detail, analysis and modeling.   As a result, we recognize these IFU data need to be publicly available for future complementary studies and comparison.  The raw NIFS data presented in this study is available in the Gemini Observatory data archive.  However, the integral field spectroscopic observing modes require training and careful attention to process the data correctly, and it is very time consuming to do so.  As a result, the {\it v}=1-0 S(1) H$_2$ emission line maps, 2.1$\mu$m continuum images and the fully reduced and flux calibrated 3-Dimensional IFS datacubes ($\sim$2.01-2.45$\mu$m) used for this project are all available for public download at:  http://www.stsci.edu/$\sim$tbeck/data/2019/NIFSH2.  These files and any of the other line emission maps presented in this study can also be made available following request to the first author.  The emission line maps that we present here provide a valuable demonstration of the types of datasets expected on CTTSs from the James Webb Space Telescope Near-IR spectrograph (NIRSpec) and Mid-IR Instrument (MIRI) IFU observing modes.

\section{Summary}

We have presented an adaptive optics integral field spectroscopy survey for disk-like molecular hydrogen in the inner environments of nine Classical T Tauri Stars acquired using the Near-infrared Integral Field Spectrograph at the Gemini North Observatory.  Key findings from our study include:

1) Spatially extended {\it v}=1-0 S(1) 2.12$\mu$m molecular hydrogen emission is detected in eight of the nine Classical T Tauri Stars surveyed here.

2)  The resolved H$_2$ emission in these eight stars shows a wide range of spatial morphologies including knots, arcs, and spatially resolved flux centered at the position of the star.
 
3)  Only LkCa 15 did not show any appreciable molecular hydrogen emission, although our 3$\sigma$ detection limit was below the previously reported line flux measurement \citep{bary03}.

4)  In AA Tau, GG Tau A, UY Aur and V773 Tau, the {\it v} = 2-1 S(1) feature at 2.24$\mu$m is also detected.  In these cases, the {\it v} = 2-1 S(1) /  {\it v}=1-0 S(1) line ratio is typical of LTE gas with excitation temperatures in the range of 1700 - 2100K.

5)   The brightest knots of H$_2$ flux in AA Tau and UY Aur are consistent with little or no extinction along the line of site to the emitting region.  Structure seen in the spatial maps of the {\it v}=1-0 Q(3) / {\it v}=1-0 S(1) line ratio suggests that some differences in obscuration toward the H$_2$ emitting regions may exist in the overall environment of these two stars.

6) UY Aur and V773 Tau exhibit spatially resolved kinematics indicating clear shock excitation from their inner outflows.  Comparison of the H$_2$ maps of UY Aur with published [Fe~II] observations \citep{pyo14} reveals that the wide angle outflows from UY Aur A and B may be interacting and exciting the extended arcs of H$_2$ emission.

7) We find an apparent anti-correlation between stellar X-ray luminosity and ro-vibrational H$_2$ line luminosity.  This confirms that H$_2$ emission declines as CTTSs evolve into the more revealed WTTS and indicates that X-ray ionization and heating of the inner gas is not the dominant excitation mechanism for the near-IR H$_2$ from CTTSs.

8) The surveyed multiple stellar systems have greater H$_2$ emission line luminosity and overall emission area compared to single stars, and we postulate that two additional excitation mechanisms exist for young multiples: shocks from system dynamics and mass accretion infall, and interacting outflows in inclined systems.

9)  The ro-vibrational H$_2$ emission from GM Aur, AB Aur and LkH$\alpha$ 264 is consistent with an origin from their inner disks.  The disk-like H$_2$ is spatially resolved in all three systems, including the spiral arms encircling AB Aur, as reported in CO by \cite{tang17}.

10)  The ro-vibrational H$_2$ emission from GM Aur, AB Aur and LkH$\alpha$ 264 is compared with H$_2$ disk models of \cite{nomu05, nomu07}.  In GM Aur and LkH$\alpha$ 264, the distribution of the H$_2$ disk emission as a function of distance from the star is consistent with the models of \cite{nomu05}.  We calculate detection limits of the {\it v} = 2-1 S(1) /  {\it v}=1-0 S(1) line ratio in these disks and find that the dust evolution models to 10cm grain sizes of \cite{nomu07} can be ruled out for LkH$\alpha$ 264.

\acknowledgments

TLB acknowledges friend, mentor, and NIFS instrument Principal Investigator, Peter J. McGregor (deceased).  This project would not have been possible without his excellent instrument fabrication talent and project leadership.  Rest in peace, Peter.  We thank our anonymous referee, who provided a timely and constructive report that improved our manuscript.  We are very grateful to Michal Simon for providing comments and suggestions on a late draft of the paper.  We thank Tom Kerr for his help processing the UKIRT CGS4 data, Chad Bender for sharing his CSHELL reduction pipeline, and P. C. Schneider for information on the X-ray brightness of GG Tau A.  We are grateful to John Rayner and friends at the NASA Infrared Telescope Facility (IRTF) who acquired the CSHELL spectrum of GM Aur during directors discretionary observatory (DDO) time.  We also thank the support of Gemini Observatory staff for their assistance preparing our programs for data acquisition, and to staff observers for executing our program observations during the assigned queue time.  The data for this program was acquired under Gemini Program IDs:  GN-2007B-Q-40, GN-2009A-Q-100 and GN-2009B-Q-40.  Based on observations obtained at the Gemini Observatory, which is operated by the Association of Universities for Research in Astronomy, Inc., under a cooperative agreement with the NSF on behalf of the Gemini partnership: the National Science Foundation (United States), National Research Council (Canada), CONICYT (Chile), Ministerio de Ciencia, Tecnolog'a e Innovaci—n Productiva (Argentina), MinistŽrio da Cincia, Tecnologia e Inova‹o (Brazil), and Korea Astronomy and Space Science Institute (Republic of Korea). 


%
\vspace{2mm}
\facilities{Gemini--North}


\software{STARLINK data software suite \citep{curr14} \url{http://starlink.eao.hawaii.edu/starlink}  
 Gemini IRAF package:  \url{https://www.gemini.edu/sciops/data-and-results/processing-software}
 }



\section{Appendix 1: Measured Ghost in the NIFS Instrument}

When trying to achieve high contrast observations with a general use instrument, it is very important to understand the character and magnitude of optical ghosting that appears on the detector.  During the analysis for the project, it became apparent that a previously unknown ghost was present in the NIFS instrument.  The ghost is seen in the data on all sources and appears as a set of knots of emission that move through multiple velocity slices in the data at wavelengths of 2.023~$\mu$m to 2.026~$\mu$m.  The features appear to be $\sim$0$\farcs$9 away from the observed stars in the blue-shifted channels and slowly become closer to the star in the red-ward channels.  The ghosts are to the south-east of the stars in a north up, east left configuration.   The contrast of individual channel slice fluxes is at the $\sim$4$\times$10$^{-4}$ level in blue-ward channels, becoming brighter to the $\sim$10$^{-3}$ level in red channels compared to the flux of the central stars.  Figure~20 presents the character of this ghost from the GM~Aur spectral imaging data.  These ghosts are at the blue-ward edge of the spectral range sampled with NIFS in the standard K-band grating setting.   The blue to red-shifted character loosely resembles emission from the jets of young stars, at a wavelength that is just short-ward of the $\it v$~$=$~1-0 S(2) emission at 2.03~$\mu$m. Having characterized the behavior of this ghost, we have found that it does not impact the high contrast emission line detection limits we have achieved for this project and in no way affects the results and conclusion of our study. 

\begin{figure}[ht!]
\plotone{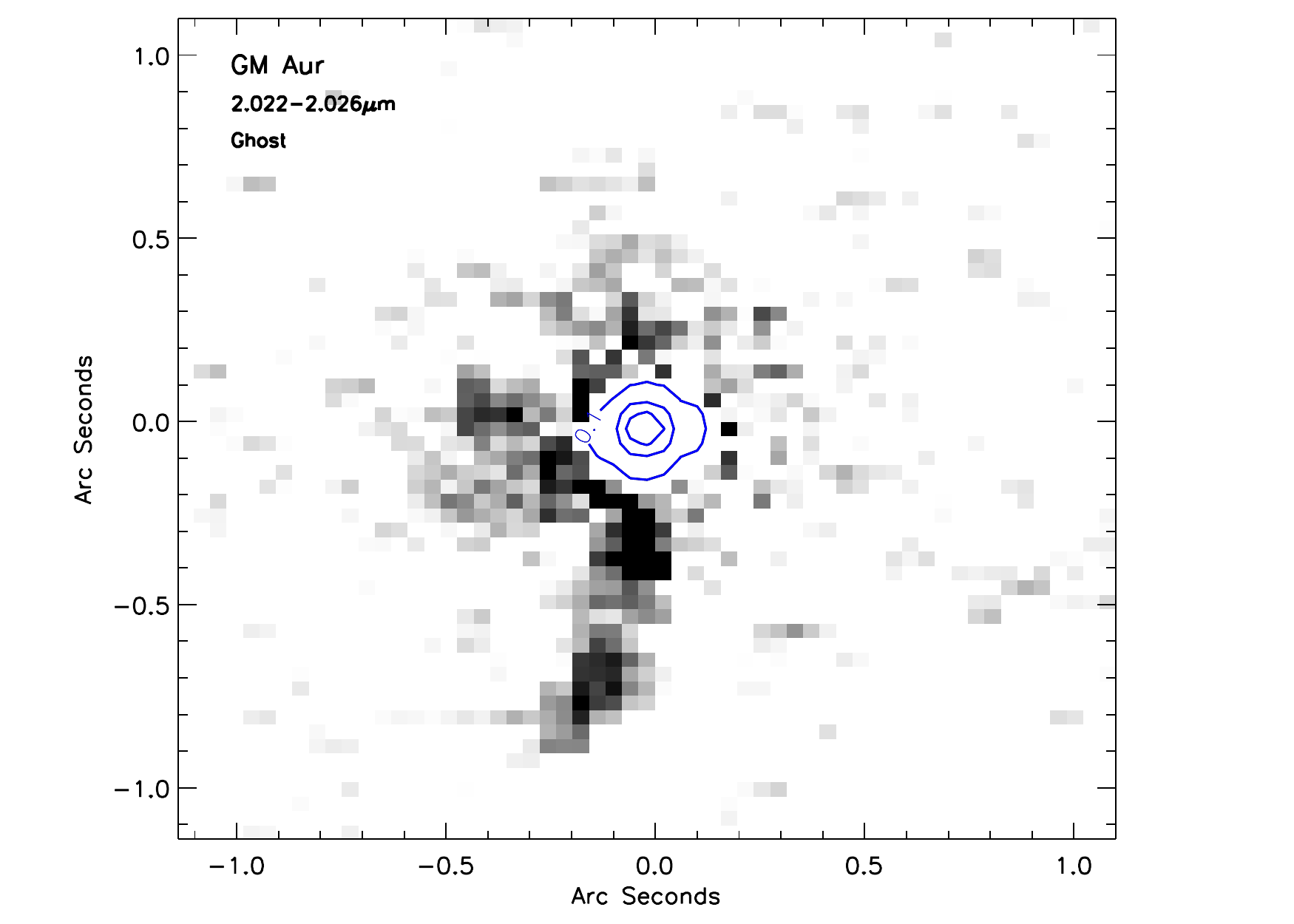}
\caption{A velocity integrated image of the previously unknown ghost detected in all of the IFU data at wavelengths from 2.023 to 2.026 $\mu$m.  Presented here is the ghost measured in the data on GM Aur.   \label{fig:ghostl}}
\end{figure}



\bibliographystyle{aasjournal}
\bibliography{references}



\end{document}